\documentclass[%
reprint,
showpacs,
showkeys,
amsmath,amssymb,
aps,
prd,
]{revtex4-1} 

\usepackage{diagbox}
\usepackage{graphicx}
\usepackage{pgffor}
\usepackage{capt-of}
\usepackage{soul}
\usepackage{lipsum}

\newcommand{\diff}{\mathrm{d}}
\def\beq{\begin{equation}}
\def\eeq{\end{equation}}
\def\bea{\begin{eqnarray}}
\def\eea{\end{eqnarray}}

\let\phi=\varphi

\let\phi=\varphi
\let\rho=\varrho

\newcommand{\uvozovky}[1]{``#1''}

\begin{document}

\frenchspacing

\author{Martin Blaschke}
\email{martin.blaschke@fpf.slu.cz}
\author{Zden\v{e}k Stuchl\'{\i}k}
\email{zdenek.stuchlik@fpf.slu.cz}
\affiliation{%
Institute of Physics and Research Centre of Theoretical Physics and Astrophysics, Faculty of Philosophy and Science, Silesian University in Opava,\\ Bezru\v{c}ovo n\'am.~13, CZ-746\,01 Opava, Czech Republic%
}

\title[Naked singularity mining instability] {Efficiency of the~Keplerian accretion in the~braneworld Kerr--Newman spacetimes and mining instability of some naked singularity spacetimes}

\begin{abstract}
We show that the~braneworld rotating Kerr--Newman black hole and naked singularity spacetimes with both positive and negative braneworld tidal charge parameter can be separated into fourteen classes according to properties of circular geodesics governing the~Keplerian accretion. We determine efficiency of the~Keplerian accretion disks for all braneworld Kerr--Newman spacetimes. We demonstrate occurrence of an~infinitely deep gravitational potential in Kerr--Newman naked singularity spacetimes having the~braneworld dimensionless tidal charge $b \in (1/4,1)$ and the~dimensionless spin $a \in (2\sqrt{b}-\sqrt{b(4b-1)},2\sqrt{b}+\sqrt{b(4b-1)})$, implying unbound efficiency of the~Keplerian accretion and possibility to extract the~whole naked singularity mass. Therefore, we call them braneworld \uvozovky{mining-unstable} Kerr--Newman naked singularity spacetimes. Fundamental restriction on the~relevance of the~extraordinary but fully classical phenomenon of the~mining instability is given by validity of the~assumption of geodesic motion of the~accreting matter. 
\end{abstract}


\maketitle 

\section{Introduction}

In recent years, one of the~most interesting and promising approaches to the~force-unification theory is represented by the~higher-dimensional String theory and particularly M-theory \cite{Hor:Wit:1996b:,Hor:Wit:1996:}. In the~String and M-theory, gravity is described as a~truly higher-dimensional interaction becoming effectively 4D at low enough energies. These theories inspired the~so called braneworld models, in which the~observable universe is a~3D-brane on which the~standard particle-model fields are confined, while gravity enters the~extra spatial dimensions \cite{Ark:Dim:Dva:1998:}. The~braneworld models provide an~elegant solution to the~hierarchy problem of the~electroweak and quantum gravity scales, as these scales could become of the~same order (TeV) due to the large scale extra dimensions \cite{Ark:Dim:Dva:1998:}. In fact, gravity can be localized near the~D3 brane in the~bulk space with a~non-compact, infinite size extra dimension with the~warped spacetime satisfying the~5D Einstein equations \cite{Ran:Sun:1999b:} - the~non-compact dimension can be related to the~M-theory. Future collider experiments can test the braneworld models quite well, including even the~hypothetical mini black hole production \cite{Dim:Lan:2001:}. 

The~5D Einstein equations at the~bulk space can be constrained to the~3D brane implying thus modified 4D Einstein equations \cite{Shi:Mae:Sas:2000:}. Solution of these constrained 4D Einstein equations is quite complex in the~presence of the~matter stress energy tensor, e.g., in the~case of models of neutron stars \cite{Ger:Maa:2001:,Hla-Stu:2011:JCAP:,Stu-Hla-Urb:GRG:2011:}. However, it can be relatively simple in the~case of vacuum solutions related to braneworld black holes. For both, the~spherically symmetric and static black holes that can be described by the Reissner--Nordstr{\" o}m geometry \cite{Dad:2000:}, and the axially symmetric and stationary rotating black holes that can be described by the Kerr--Newman geometry \cite{Ali-Gum:2005:CLAQG:}, the influence due to the~tidal effects from the~bulk is simply represented by a~single parameter that is called tidal charge because of the~similarity of the~effective stress-energy tensor of the~tidal effects of the~bulk space and the~stress-energy tensor of the~electromagnetic field \cite{Dad:2000:}. 
 
The~rotating braneworld black hole spacetimes, and the~related naked singularity spacetimes, are thus represented by the~Kerr--Newman geometry, but without the~associated electromagnetic field occurring in the~standard general relativity \cite{Mis-Tho-Whe:1973:Gra:}. The~tidal charge parameter can be both positive and negative \cite{Dad:2000:,Ali-Gum:2005:CLAQG:}, while in the~standard general relativity only positive parameter corresponding to the~square of the~electric charge occurs. 

The standard studies of the~Reissner--Nordstr{\" o}m or Kerr--Newman black hole and naked-singularity geodesic motion \cite{Stu:Cal:1991:,Stu:Hle:2000:,Stu-Hle:2002:ActaPhysSlov:,Pug-Que-Ruf:2011:PHYSR4:,Pug:Que:Ruf:2013:} can thus be directly applied for the~braneworld black holes and naked singularities with positive tidal charge. The~astrophysically relevant implications of the~geodesic motion were extensively studied for the~braneworld black holes (with both positive and negative tidal charges) in a~number of papers related to the~optical effects \cite{Sche:Stu:2009:,Sche:Stu:2009b:,Nun:2010:PHYSR4:,Eir:2005:,Ata-Abd-Ahm:PRD:2013:,Jia-Bam-Ste:2016:PRD:,Liu-Li-Bam:2015:JCAP:,Li-Kong-Bam:2014:APJ:}, or the~test particle motion \cite{Boh-etal:2008:CLAQG:,Kot-Stu-Tor:2008:CLAQG:,Stu:Kot:2009:,Ali-Tal:2009:prd:,Abd-Ahm:PRD:2010:,Rah-Abd-Ahm:APSS:2011:,Mo-Ahm-APSS:2011:,Sha-Ahm-Abd:PRD:2013:}. 

Here we study circular motion of test particles and photons in the~braneworld Kerr--Newman spacetimes and give classification of the~braneworld black hole and naked singularity spacetimes according to the~properties of the~radial profiles of specific angular momentum and specific energy of sequences of corotating and retrograde circular orbits. We give the~classification for both the~positive and negative values of the~dimensionless tidal charge parameter. We also determine efficiency of the~Keplerian accretion disks that is related to the astrophysically relevant accretion from infinity (large distance) downwards to the~first limit on existence of stable circular geodesics. 

A very detailed analysis of the~circular motion of~electrically neutral test particles in the~standard Kerr--Newman spacetimes has been presented in \cite{Pug:Que:Ruf:2013:}, where both black hole and naked singularity spacetimes were discussed. The results of this study are relevant in the~braneworld spacetimes with positive tidal charges; we do not repeat them, focusing our study to the~phenomena  related to the~Keplerian accretion, its efficiency, and the~phenomenon of a new special instability of the~naked singularity spacetimes that were not considered in the~paper \cite{Pug:Que:Ruf:2013:}. Along with the~standard classical instability due to the~Keplerian accretion occurring in the~Kerr naked singularity spacetimes, leading to their conversion to a~Kerr black hole \cite{deF:1974:aap,Stu:1980:BAC:,Stu:Hle:Tru:2011:,Stu-Sche:2012:CLAQG:}, we have found a~special class of classical instability, called here \uvozovky{mining} instability, as this instability is related to an~unlimitedly deep gravitational potential well, occuring in the~class of the~Kerr--Newman naked singularity spacetimes with appropriately restricted values of their dimensionless spin $a$ and dimensionless tidal charge $b$. We briefly discuss the~limits on the~applicability of the~Keplerian accretion in relation to the~mining instability. 

\section{Braneworld Kerr--Newman geometry}

Using the~standard Boyer-Lindquist coordinates $(t,r,\theta,\varphi)$ and the geometric units $(c=G=1)$, we can write the~line element of a~rotating (Kerr--Newman) black hole or naked singularity, representing solution of the Einstein equations constrained to the~3D-brane, in the~form \cite{Ali-Gum:2005:CLAQG:,Dad:2000:} 

\begin{widetext}
\begin{eqnarray}
\mkern-106mu
\mathrm{d}s^2= - \left(1-\frac{2Mr - b}{\Sigma }\right)\mathrm{d}t^2 &-& \frac{2a(2Mr - b)}{\Sigma}\,\mathrm{sin}^2\theta\,\mathrm{d}t\,\mathrm{d}\phi + \frac{\Sigma}{\Delta}\,\mathrm{d}r^2
\ \nonumber \\
 &+& \Sigma\, \mathrm{d}\theta^2 + \left(r^2 + a^2 + \frac{2Mr - b}
{\Sigma}\,a^2\mathrm{sin}^2\theta\right)
\mathrm{sin}^2\theta\, \mathrm{d}\phi^2\, , \label{Metrika}
\end{eqnarray}
\end{widetext}
where 
\begin{eqnarray}
\Delta = r^2 -2Mr +a^2 + b\, ,\\ 
\Sigma = r^2 + a^2\mathrm{cos}^2\theta\, .
\end{eqnarray}
$M$ is the~mass parameter of the~spacetime, $a=J/M$ is the~specific angular momentum of the~spacetime with internal angular momentum $J$, and the~braneworld parameter $b$, called \uvozovky{tidal charge}, represents imprint of the~non-local (tidal) gravitational effects of the~bulk space \cite{Ali-Gum:2005:CLAQG:}. 

The~form of the~metric (\ref{Metrika}) is the~same as that of the~standard Kerr--Newman solution of the~4D Einstein--Maxwell equations, with squared electric charge $Q^2$  being replaced by the~tidal charge $b$ \cite{Mis-Tho-Whe:1973:Gra:}. 
We can separate three cases: 
\begin{itemize}
\item[a) ] $b=0$ in which we are dealing with the standard Kerr metric.  
\item[b) ] $b>0$ in which we are dealing with the standard Kerr--Newman metric. 
\item[c) ] $b<0$ in which we are dealing with the non-standard Kerr--Newman metric.    
\end{itemize}
Notice that in the~braneworld Kerr--Newman spacetimes the~geodesic structure is relevant also for the~motion of electrically charged particles, as there is no electric charge related to these spacetimes. On the~other hand, the case (b) can be equally considered for the~analysis of the~uncharged particle motion in the~standard electrically charged Kerr--Newman spacetime. 

For simplicity we put in the~following considerations $M=1$. Then the~spacetime parameters $a$ and $b$, and the~time $t$ and radial $r$ coordinates become dimensionless. This is equivalent to the~redefinition when we express all the~quantities in units of $M$: $a/M \to a$, $b/M^2 \to b$, $t/M \to t$ and $r/M \to r$. 

Separation between the~black hole and naked singularity spacetimes is given by the~relation of the~spin and tidal charge parameters in the~form  
\beq
     a^2 + b = 1 
\eeq
determining the~so called extreme black hole with coinciding horizons. 
The~condition $0< a^2+b < 1$ governs black hole spacetimes with two distinct event horizons, while the~condition $a^2+b < 0$ governs black hole spacetimes with only one distinct event horizon at $r>0$. For $a^2+b > 1$, the~spacetime describes a~naked singularity. 

For positive tidal charges the~black hole spin has to be $a^2<1$, as in the~standard Kerr--Newman spacetimes, but for negative tidal charges there can exist black holes violating the~well know Kerr limit, having $a^2>1$ \cite{Stu:Kot:2009:}. 
 
Using substitutions 
\begin{eqnarray}
dt &=& dx^0 + \left(\frac{r^2+a^2}{\Delta}-1\right) dr\, , \\
d\phi &=& d\tilde \phi + \frac{a}{\Delta}dr\, ,\\
x &=& \left(r \cos(\tilde\phi)+ a \sin(\tilde\phi)\right)\sin\theta\, ,\\
y &=& \left(r\sin(\tilde \phi)-a \cos(\tilde\phi)\right)\sin\theta\, ,\\
z &=& r\cos\theta\, ,
\end{eqnarray}
the~braneworld Kerr--Newman geometry can be transformed into the~so called Kerr-Schild form using the~Cartesian coordinates: 
\begin{widetext}
\begin{equation}\label{metric2}
ds^2 = -(dx^0)^2 + (dx)^2 + (dy)^2 + (dz)^2 + \frac{(2Mr-b)r^2}{r^4+a^2z^2}\left\{dx^0-\frac{1}{r^2+a^2}\left[r(xdx+ydy)+a(xdy-ydx)-\frac{1}{r}z dz\right]\right\}^2 \, , 
\end{equation}
\end{widetext}

where $r$ is defined, implicitly, by
\[
r^4 - r^2(x^2+y^2+z^2-a^2) - a^2z^2=0\, .
\] 
\subsection{Singularity}\label{ssingular}
The~metric (\ref{metric2}) is analytical everywhere except at points satisfying the~condition 
\begin{equation}
x^2+y^2+z^2 = a^2 \hspace{5mm} \mathrm{and}\hspace{5mm}z=0\, .
\end{equation}
This condition is the~same as in the~case of the~Kerr black holes or naked singularities, so we clearly see that the braneworld parameter $b$ has no influence on the~position of the~physical singularity of the~space-time. The~physical \uvozovky{ring} singularity of the~braneworld rotating black holes (and naked singularities) is located at $r=0$ and $\theta = \pi/2$, as in the~Kerr spacetimes. 

\begin{figure}[t]
\begin{center}
\centering
\includegraphics[width=1\linewidth]{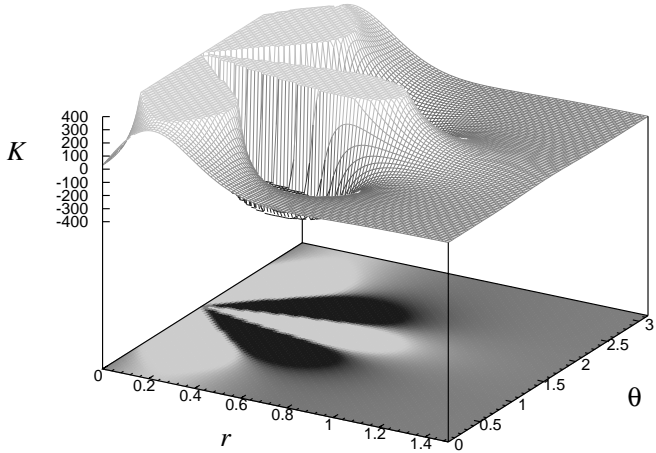}
\caption{\label{Fkre} Example of the~behaviour of Kretschmann's scalar $K$ for $a=0.8$ and $b=-0.8$ to illustrate  it's similarity to Kerr--Newman case.} 
\end{center}
\end{figure}

We describe the~influence of the braneworld tidal charge parameter $b$ on the~Kerr-like ring singularity at $r=0\, ,\theta=\pi/2$ using the Kretschmann scalar $K=R_{\alpha\beta\gamma\delta} R^{\alpha\beta\gamma\delta}$ representing an~appropriate tool to probe the~structure of spacetimes singularities. Using (\ref{Metrika}) we obtain 
\begin{widetext} 
\begin{equation}\label{K}
K =\frac{8}{(r^2 + a^2 y^2)^6}\left(r^4 A -  2 a^2 r^2 B y^2 + a^4 C y^4 - 6a^6 M^2 y^6\right)\, ,
\end{equation}
\end{widetext}
where 
\begin{eqnarray}
y &=& \cos \theta\, ,\\
A &=&(7 b^2 - 12 b M r + 6 M^2 r^2)\, , \\
B &=& (17 b^2 - 60 b M r + 45 M^2 r^2)\, , \\
C &=&(7 b^2 - 60 b M r + 90 M^2 r^2)\, .
\end{eqnarray}

The Kretschmann scalar is formally the same as in the~case of the~Kerr--Newman metric with $Q^2 \rightarrow b$\, \cite{Hen:2000:apj:}. Naturally, the~negative values of brane parameter would have some effect onto $K$, but as we can see from denominator of (\ref{K}), it does not influence the~location of the~singularity. As an~example we demonstrate behaviour of the~scalar $K$ for $a=0.8$ and $b=-0.8$ near the~ring singularity in the~Figure (\ref{Fkre}). 

For completeness we give also the~Ricci tensor whose components take the~form:  
\begin{eqnarray}
R_{tt}&=&4b\frac{a^2 + 2\Delta - a^2 \cos(2 \theta)}{\left(a^2 + 
  2 r^2 + a^2 \cos(2 \theta)\right)^3}\, , \\
R_{t\phi} &=& -8 a b \frac{(a^2 + \Delta) \sin^2\theta}{\left(a^2 + 2 r^2 + a^2 \cos(2 \theta)\right)^3}\, , \\
R_{\phi t} &=& R_{t\phi}\, ,\quad R_{rr} = -\frac{R_{\theta\theta}}{\Delta}\, , \\
R_{\theta\theta} &=& \frac{2 b}{a^2 + 2 r^2 + a^2 \cos(2 \theta)}\, , \\
R_{\phi\phi} &=& 4 b \sin^2(\theta) \frac{3 a^4 + 2 r^4 + a^2 \left(b - 2 M r + 5 r^2\right)}{\left(a^2 + 2 r^2 + a^2 \cos(2\theta)\right)^3} \\ 
&-&  \frac{a^2 \Delta\cos(2\theta)}{\left(a^2 + 2 r^2 + a^2 \cos(2\theta)\right)^3}\, . 
\end{eqnarray}

Ricci scalar is zero automatically by construction of the~braneworld Kerr--Newman solution \cite{Dad:2000:}. 

\subsection{Ergosphere}
\begin{figure*}[ht]
\begin{center}
\begin{minipage}{.5\linewidth}
\centering
\includegraphics[width=\linewidth]{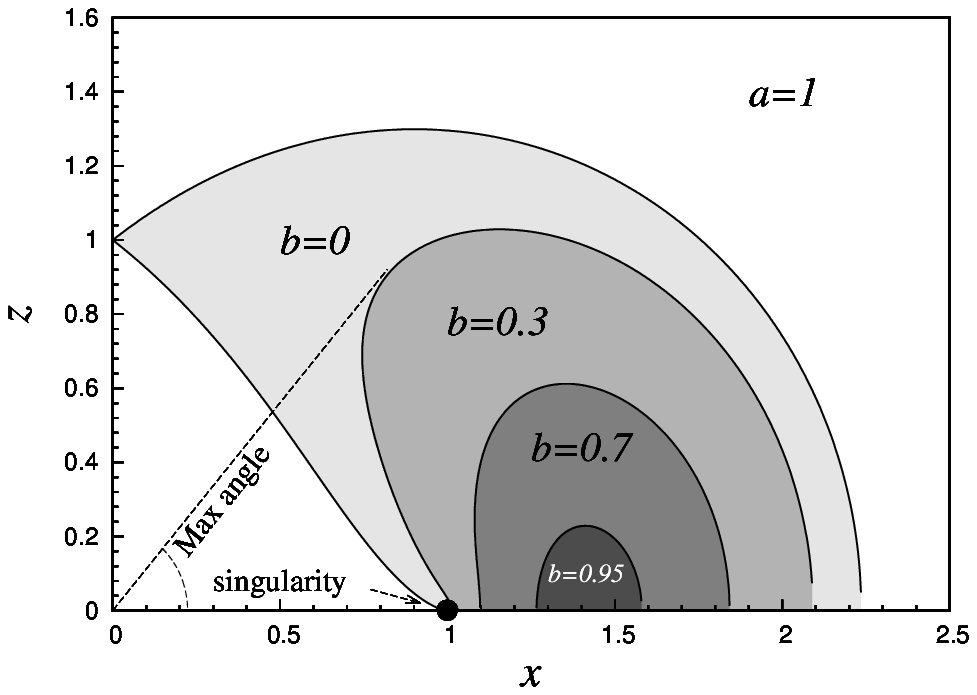}
\end{minipage}\hfill
\begin{minipage}{.5\linewidth}
\centering
\includegraphics[width=\linewidth]{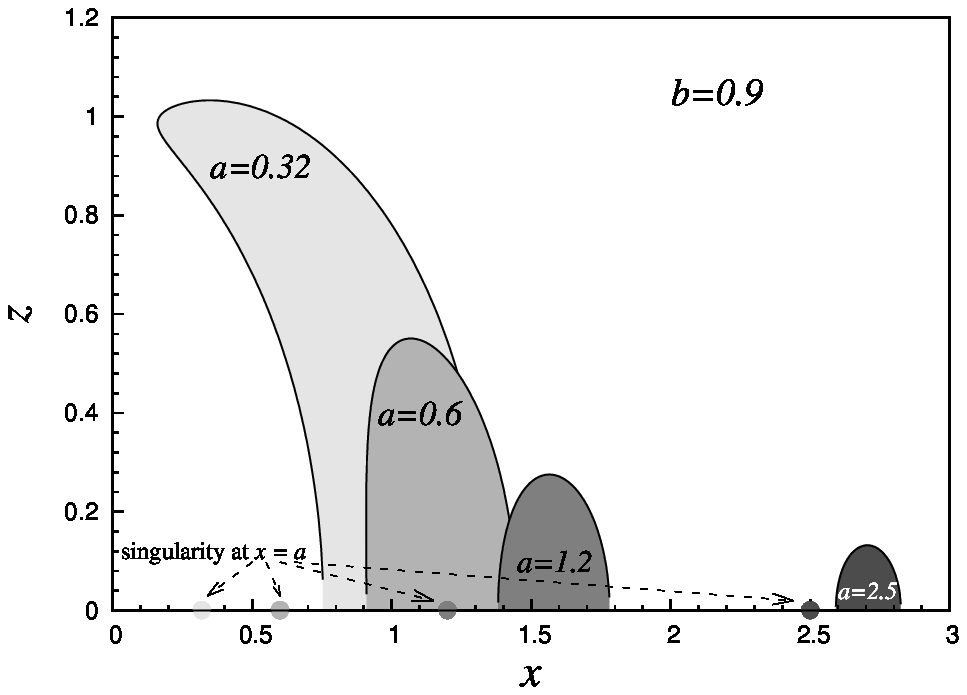}
\end{minipage}\hfill
\begin{minipage}{.5\linewidth}
\includegraphics[width=\linewidth]{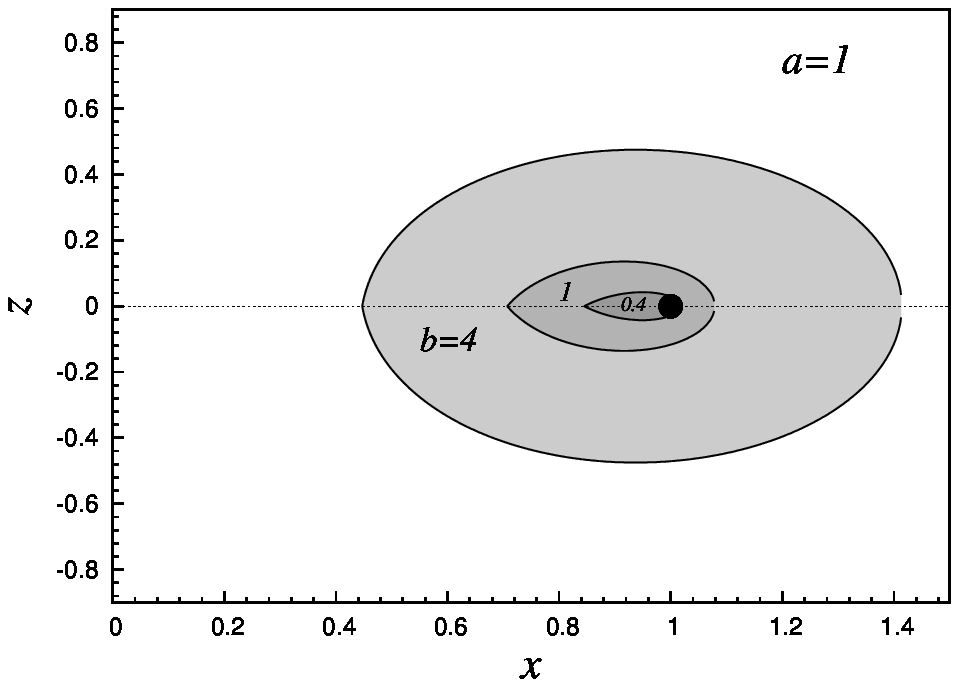}
\end{minipage}\hfill
\begin{minipage}{.5\linewidth}
\centering
\includegraphics[width=\linewidth]{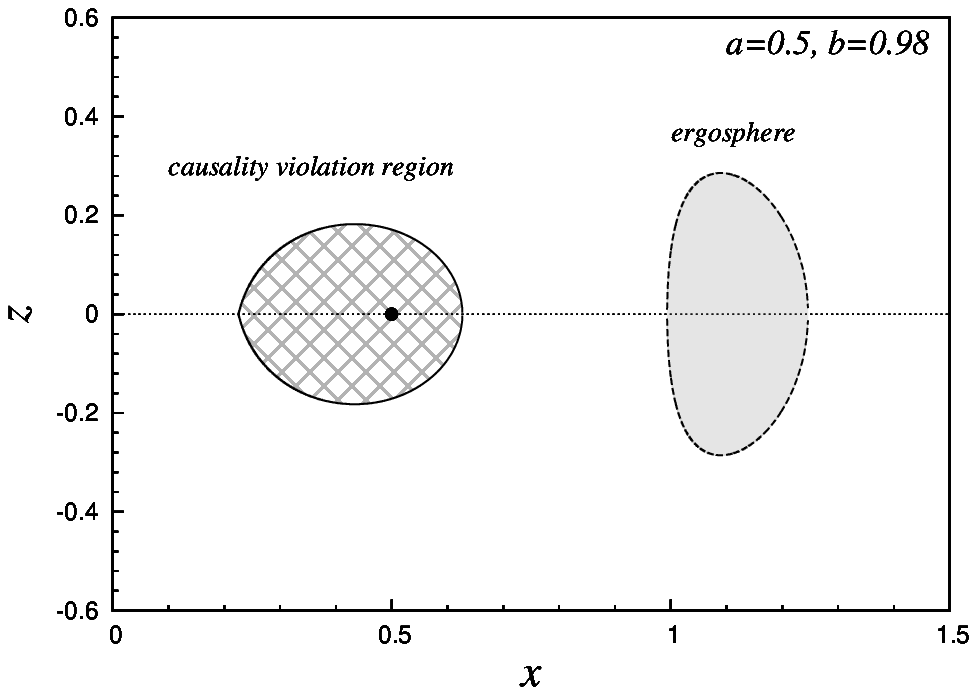}
\end{minipage}
\caption{\label{ejkio} {\bf Upper Left:}Polar slice through the braneworld Kerr--Newman spacetime in the Cartesian Kerr–Schild
coordinates. Dimensionless spin parameter $a$ is fixed to value 1 and the braneworld parameter $b$ is appropriately chosen to demonstrate its influence on the~ergosphere. {\bf Upper Right:} Polar slice through the braneworld Kerr--Newman spacetime in the Cartesian Kerr–Schild
coordinates. The braneworld parameter $b$ is fixed to value 0.9 and the spin parameter $a$ is appropriately chosen to demonstrate its influence on the~ergosphere. {\bf Lower Left:} Causality violation region. {\bf Lower Right:} Ergosphere and causality violation region.} 
\end{center}
\end{figure*}

Here we demonstrate influence of the~braneworld tidal charge parameter $b$ on the~ergosphere whose boundary is defined by the~condition:
\begin{equation}
g_{tt} = r^2-2Mr+a^2cos^2\theta+b = 0\, .
\end{equation}
Extension of the~ergosphere in the~latitudinal coordinate $\theta$ is determined by the~maximal latitude given by the~relation 
\begin{equation}
    \cos^2\theta_{\mathrm{max}} = \frac{1-b}{a^2} . 
\end{equation}
We can see that existence of the~ergosphere is limited by the~condition  
\begin{equation}
         b < 1 .
\end{equation}
We can infer that the~ergosphere extension increases as the~tidal charge parameter $b$ decreases. 

It is convenient to represent location of the~ergosphere in the~Kerr-Schild coordinates (\ref{metric2}). Using the~spacetime symmetry we can focus only on the~polar slices with $y=0$. In this case the~condition for the~static limit surface governing the~border of the~ergosphere is simply given by \cite{Car:1973:BlaHol:}
\begin{eqnarray}
x^2 &=& \frac{\left(a^2+r^2\right)\Delta}{a^2}\, ,\nonumber \\
z^2 &=& \frac{(2r-b)r^2-r^4}{a^2}\, .
\end{eqnarray}
In Figure (\ref{ejkio}) we illustrate influence of the~braneworld tidal charge $b$ on the~ergosphere extension. 

The~ergosphere does not always completely surrounds the~ring singularity. To illustrate this phenomenon, we give also dependence of the~maximal allowed latitudinal angle of the~ergosphere on the dimensionless spin and dimensionless tidal charge. 

For $b<1$, the~ergosphere exists for each dimensionless spin $a>0$, covering all values of the~latitudinal angle for the Kerr--Newman black holes. However, as the~spin $a$ increases for the~Kerr--Newman naked singularities, the~ergosphere extension shrinks -- the~maximal angle $\alpha$ decreases. 

\subsection{Causality violation region}
\begin{figure}[t]
\begin{center}
\centering
\includegraphics[width=1\linewidth]{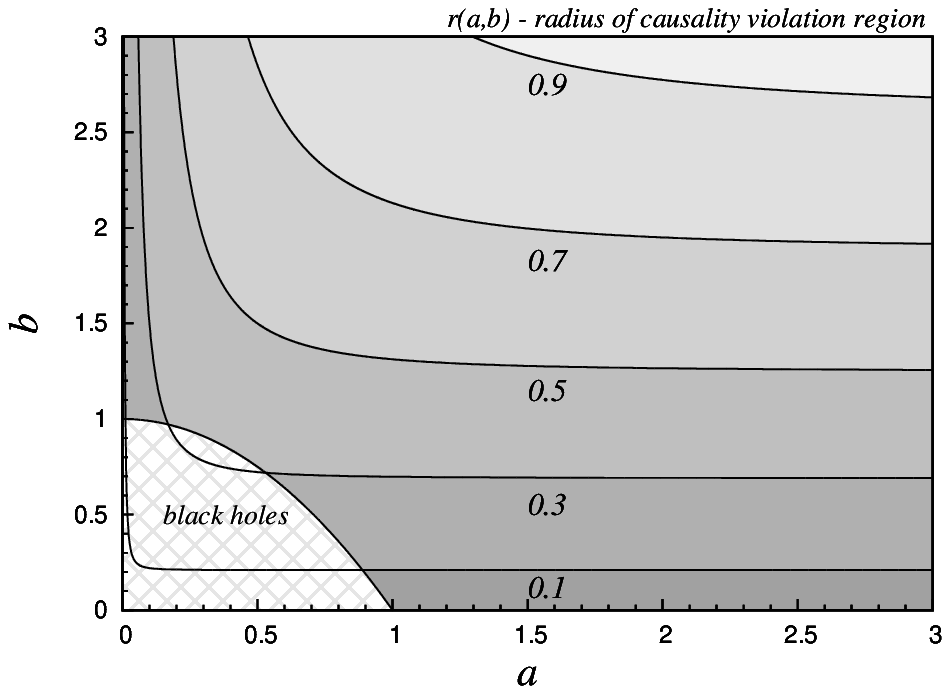}
\caption{\label{causal} Contour plot for radii of boundary of the~causality violation region in the~equatorial plane.} 
\end{center}
\end{figure}

In the \uvozovky{causality violation region} (sometimes called time-machine region) the~axial coordinate $\phi$ takes time-like character implying possible existence of closed time-like curves. The causality violation region is defined by the~condition 
\begin{equation}
g_{\phi\phi}<0.
\end{equation}
In the equatorial plane the~boundary of the~causality violation region is determined by the~condition 
\begin{equation}\label{blb}
r^4+a^2(r^2+2r-b)=0\, .
\end{equation}
The boundary of the~causality violation region can be expressed by the~relation 
\begin{equation}\label{dis3}
b = b_{\mathrm{CV}} \equiv \frac{r\left(2a^2+a^2 r + r^3\right)}{a^2}\, . 
\end{equation}
At Figure (\ref{causal}) we give some examples of the~extension of the~causality violation region. We see that for this region to exist above the~ring singularity, the tidal charge has to be positive. With increasing values of the~parameters $b>0$ and $a$, the causality violation region expands. 

The~equation (\ref{blb}) gives us maximal possible extension of causal violation region located at
\begin{equation}
r_{\mathrm{Max}}=\sqrt{1+b}-1\, .
\end{equation}
For positive $b$ the~value of $r_{\mathrm{Max}}$ is less than $b$ and therefore, as we shall see later, the causality violation region cannot reach the~region where the~circular geodesics exist.  

In the~Kerr--Schild coordinates the~boundary of the~causality violation region is given by the~relations 
\begin{eqnarray}
x^2 &=& \frac{\left(a^2+r^2\right)^3}{a^2 \Delta}\, ,\\
z^2 &=& \frac{r^2(a^2(b-2r-r^2)-r^4)}{a^2 \Delta}
\end{eqnarray}
It can be proved that the~causality violation region never overlaps with ergosphere and its extension is influenced by the braneworld parameter $b$ in an opposite way. While causality violation region increases with increasing $b$, the~ergosphere extension is getting smaller. This phenomenon is illustrated in Figure (\ref{ejkio}) where the Kerr--Schild coordinates are used. 

In the~following, we consider geodesic motion only in the~regions above the causality violation region. For astrophysical phenomena occuring in the~naked singularity spacetimes, it is usually assumed that above the~boundary of the causality violation region the~Kerr or Kerr--Newman spacetime is removed and substituted by different solution that could be inspired by the String theory -- such objects are called superspinars \cite{Gim-Hor:2009:PhysLetB:,Stu-Sche:2010:CLAQG:,Stu-Sche:2013:CLAQG:}. Therefore, it is quite natural to assume that in the~braneworld model framework, the~inner boundary of the~superspinars is located at radii larger than those related to the~boundary of the~causality violation region. 

\subsection{Locally non-rotating frames}

In the~rotating Kerr--Newman spacetimes physical processes can be most conveniently expressed in the~family of locally non-rotating frames (LNRF), corresponding to zero angular momentum observers (ZAMO), with tetrad vectors given by the relations \cite{Bar-Pre-Teu:1972:ApJ:} 
\beq\label{LNRF}
\mathrm{\bf e}^{(t)} = \left ( \omega^2g_{\phi\phi} - g_{tt}\right ) ^{\frac{1}{2}}{\bf d}t\ ,
\end{equation}
\begin{equation}
\mathrm{\bf e}^{(\phi)} = (g_{\phi\phi})^{\frac{1}{2}}({\bf d}\phi - \omega{\bf d}t)\, ,
\end{equation}
\begin{equation}
\mathrm{\bf e}^{(r)} = \left(\frac{\Sigma }{\Delta }\right)^{\frac{1}{2}}{\bf d}r\ ,
\end{equation}
\begin{equation}
\mathrm{\bf e}^{(\theta)} = \Sigma^{\frac{1}{2}}{\bf d}\theta\ ,
\eeq
where $\omega$ is the~angular velocity of the~LNRF relative to distant observers and reads
\beq\label{omega}
\omega =-\frac{g_{t\phi}}{g_{\phi\phi}}= \frac{a(2r-b)}{\Sigma(r^2 + a^2) + (2r - b)\,a^2\,\mathrm{sin^2\theta}}\, .
\eeq
Convenience of the~LNRF can be demonstrated, e.g., in the~case of the~free fall of particles from infinity, which is purely radial only if related to the~family of LNRFs \cite{Stu-Bic-Bal:1999:GRG:}. 

\subsection{Geodesic motion and Carter's equations}

Using the~Hamilton-Jacobi method, Carter found separated first order differential equations of the~geodesic motion \cite{Car:1968:PRD:,Car:1973:BlaHol:}, which in the~case of  the braneworld Kerr spacetimes take the~form 

\begin{eqnarray}
 	\Sigma\frac{\diff r}{\diff w}&=&\pm\sqrt{R(r)},\label{ce7}\\
 	\Sigma\frac{\diff \theta}{\diff w}&=&\pm\sqrt{W(\theta)},\label{ce8}\\
 	\Sigma\frac{\diff \varphi}{\diff w}&=&-\frac{P_W}{\sin^2\theta}+\frac{a P_R}{\Delta},\label{ce9}\\
 	\Sigma\frac{\diff t}{\diff w}&=&-a P_W + \frac{(r^2+a^2)P_R}{\Delta},\label{ce10}
\end{eqnarray}
where 

\begin{eqnarray}
  R(r)&=&P^2_R-\Delta(m^2 r^2 + \tilde{K}),\label{ce11}\\
  W(\theta)&=&(\tilde{K}-a^2m^2\cos^2\theta)-\left(\frac{P_w}{\sin\theta}\right)^2,\label{ce12}\\
  P_R(r)&=&\tilde{E}(r^2+a^2)-a\tilde{\Phi},\label{ce13}\\
  P_W(\theta)&=&a\tilde{E}\sin^2\theta - \tilde{\Phi}.\label{ce14}
\end{eqnarray}
Along with the~conservative rest energy $m$, three constants of motion related to the~spacetime symmetries has been introduced: $\tilde{E}$ is the~energy (related to the time Killing vector field), $\tilde{\Phi}$ is the~axial angular momentum (related to the axial Killing vector field) and $\tilde{K}$ is the~constant of motion related to total angular momentum (related to the Killing tensor field) that is usually replaced by the~constant $\tilde{Q}=\tilde{K}-(a\tilde{E}-\tilde{\Phi)}^2$, since for the motion in the~equatorial plane ($\theta=\pi/2$) there is $\tilde{Q}=0$. 

Note that the separable Eqs (38)-(41) are quaranteed in the Petrov type D spacetimes, in particular when the metric in the Boyer-Lindquist coordinates can be expressed in the Kerr-like form by replacing the mass parameter $M$ by a function $M(r)$ independent of latitude $\theta$. In the braneworld rotating black hole spacetimes there is $M(r) = M - \frac{b}{2r}$. Generally, these equations can be integrated and expressed in terms of the~hyper-elliptic integrals \cite{Mis-Tho-Whe:1973:Gra:,Kra:2005:CLAQG:,Kra:2007:CLAQG:}. The Carter equations can be also generalized to the~motion in the~Kerr--Newman-de Sitter spacetimes \cite{Car:1973:BlaHol:,Stu:1983:BAC:,Stu-Hle:1999:PHYSR4:,Kra:2005:CLAQG:,Kra:2014:GRG:}. 

For the~geodesic motion of photons, we put $m=0$ in the~Carter equations. Analysis of the~photon motion in the~standard~Kerr--Newman spacetimes \cite{Stu:1981:BULAI:NullGeoKN:,Bal-Bic-Stu:1989:BAC:,Pug:Que:Ruf:2013:} can be directly applied to the~case of photon motion in the~braneworld Kerr--Newman spacetimes. It is has been done in \cite{Hle-Stu:2007:RAGtime8and9:Proceedings:,Sche:Stu:2009b:}; we use results of these works in the following discussions. 

We have to construct classification of the braneworld Kerr--Newman spacetimes according to the~properties of circular geodesics governing the Keplerian accretion that can be related not only to the~standard accretion discs, but also to the~quasicircular motion of gravitationally radiating particles. We give classification according to properties of the~circular null geodesics and stability of the~circular geodesics that is the~critical attribute of the~Keplerian accretion. Finally, we combine the~effects given by these two classifications. Of course, we have to include into the~classification as relevant criteria also existence of the~event horizons and existence of the~ergosphere. 

\section{Circular geodesic motion}

In general stationary and axially symmetric spacetime with the Boyer-Lindquist coordinate system $(t,r,\theta,\phi)$ and $\displaystyle (-+++)$ signature of the~metric tensor, the~line element is given by 
\begin{equation}
ds^{2}=g_{tt}dt^{2}+2g_{t\phi }dtd\phi +g_{rr}dr^{2}+g_{\theta \theta
}d\theta ^{2}+g_{\phi \phi }d\phi ^{2}\;.  \label{ds2rcoappr}
\end{equation}%
The~metric (\ref{ds2rcoappr}) is adapted to the~symmetries of the~spacetime, endowed with the~Killing vectors $(\partial/ \partial t)$ and $(\partial/ \partial \phi)$ for time translations and spatial rotations, respectively. For geodesic motion in the~equatorial plane ($\theta =\pi /2$), the~metric functions $g_{tt}$, $g_{t\phi }$, $g_{rr}$, $g_{\theta\theta }$ and $g_{\phi \phi }$ in Eq.~(\ref{ds2rcoappr}) depend only on the~radial coordinate $r$. So except the~rest energy $m$, two integrals of the motion are relevant as $\tilde{Q}=0$: 
\begin{equation}
U_t = -\,E\, , \; U_\phi =\,L\, ,
\end{equation}
where the 4-velocity $U_\alpha = g_{\alpha\nu} dx^\nu/d\tau\,$, with $\tau $ being the~affine parameter.
In the~case of asymptotically flat spacetime, we can identify at infinity the~motion constant $E=\tilde{E}/m$ as the specific energy, i.e., energy related to the~rest energy, and the~motion constant $L=\tilde{\Phi}/m$ as the~specific angular momentum. 

The geodesic equations of the~equatorial motion take the~form (see, e.g., \cite{Kov-Har:2010:PRD:}) 
\begin{equation}\label{eq14}
 \frac{dt}{d\tau }=\,\frac{E g_{\phi \phi }+%
 L g_{t\phi }}{g_{t\phi }^{2}-g_{tt}g_{\phi \phi }},
\;\; \frac{d\phi }{d\tau }=-\,\frac{E g_{t\phi }+%
 L g_{tt}}{g_{t\phi }^{2}-g_{tt}g_{\phi \phi }},
\end{equation}%
and
\begin{equation}\label{geodeqs3}
g_{rr}\left( \frac{dr}{d\tau }\right)^{2}=R(r),
\end{equation}%
where the~radial function $R(r)$ is defined by
\begin{equation}\label{Rrr}
R(r)\equiv -1+\frac{E^{2}g_{\phi \phi }+2E%
Lg_{t\phi }+L^{2}g_{tt}}{g_{t\phi
}^{2}-g_{tt}g_{\phi \phi }}\, .
\end{equation} 

\subsection{Energy, angular momentum and angular velocity of circular geodesics}

For circular geodesics in the~equatorial plane, the conditions  
\begin{equation}\label{podminky}
R(r)=0\,\quad \mathrm{and}\,\quad \partial_r R(r)=0\, 
\end{equation}
must be satisfied simultaneously. These conditions determine the~specific energy $E$, the~specific angular momentum $L$ and the angular velocity $\Omega=d\phi/dt$ related to distant observers, for test particles following the~circular geodesics, as functions of the~radius and the~spacetime parameters in the~form 
\begin{eqnarray}
E &=&\pm\frac{g_{tt}+g_{t\phi }\Omega }{\sqrt{-\left(g_{tt}+2g_{t\phi}\Omega +g_{\phi \phi }\Omega ^{2}\right)}}\;,  \label{tildeE} \\
L &=&\mp\frac{g_{t\phi }+g_{\phi \phi }\Omega }{\sqrt{-\left(g_{tt}+2g_{t\phi }\Omega +g_{\phi \phi }\Omega ^{2}\right)}},  \label{tildeL} \\
\Omega  &=& \frac{-g_{t\phi ,r}\pm\sqrt{(g_{t\phi
,r})^{2}-g_{tt,r}g_{\phi \phi ,r}}}{g_{\phi \phi ,r}}\;,\label{Omega}
\end{eqnarray}
where the~upper and lower signs refer to two families of solutions. To avoid any misunderstanding, we will refer to these two families as the~upper sign family, and the~lower sign family. At large distances in the~asymptotically flat spacetimes, the~upper family orbits are corotating, while the~lower family orbits are counterrotating with respect to rotation of the~spacetime. This separation holds in the~whole region above the~event horizon of the~Kerr--Newman black hole spacetimes, but it is not necessarily so in all the~Kerr--Newman naked singularity spacetimes -- in some of them the~upper family orbits become counterrotating close to the~naked singularity as demonstrated in \cite{Stu:1980:BAC:}. 

Using the~spacetime line element of the~braneworld rotating spacetimes given by (\ref{Metrika}) \cite{Ali-Gum:2005:CLAQG:,Dad-Kal:1977:}, with the~assumption of $M=1$, we obtain the~radial profiles of the specific energy, specific axial angular momentum and the~angular velocity related to infinity of the~circular geodesics in the~form: 
\begin{eqnarray}
E &=& \frac{r^2-2r + b \pm a \sqrt{r-b}}{r\sqrt{r^2-3r +2b \pm 2a\sqrt{r-b}}}\, ,\label{E} \\
L &=& \pm\frac{\sqrt{r-b}\left(r^2+a^2 \mp 2a\sqrt{r-b}\right)\mp ab}{r\sqrt{r^2-3r +2b\pm 2a\sqrt{r-b}}}\, ,\label{L} \\
\Omega  &=& \pm \frac{1}{\frac{r^2}{\sqrt{r-b}}\pm a}\, .\label{Om}
\end{eqnarray}

From equations (\ref{E})--(\ref{Om}) we immediately see that two restrictions on the~existence of circular geodesics have to be satisfied: 
\begin{eqnarray}
r^2-3 r +2b\pm 2a\sqrt{r-b} \geq 0\, ,\\
r\geq b\, .
\end{eqnarray}
The equality in the~first condition determines the~photon circular geodesics -- this demonstrates that positions of circular orbits of test particles are limited by the~circular geodesics of massless particles. The~second, reality condition is relevant in the~Kerr--Newman spacetimes with positive tidal charge $b$ only, if we restrict attention to the~region of positive radii. 

\subsection{Effective potential}

Instead of the~radial function $R(r,a,b,E,L)$, the~equatorial motion of test particles can be conveniently treated by using the~so called effective potential $V_{\mathrm{Eff}}(r,a,b,L)$ that is related to the~particle specific energy and depends on the~specific angular momentum of the~motion and the~spacetime parameters. The equation $E = V_{\mathrm{Eff}}$ determines the~turning points of the radial motion of the~test particle. 

The notion of the~effective potential is useful in treating the Keplerian (quasigeodesic) accretion onto the~central object that is directly related to the~circular geodesic motion \cite{Nov-Tho:1973:BlaHol:,Pag-Tho:1974:ApJ:}. The~circular geodesics are governed by the~local extrema of the~effective potential; the~accretion process is possible in the~regions of stable circular geodesics corresponding to the~local minima of the~effective potential. 

The~effective potential can be easily derived using the~normalization condition for the~test particle motion  
\begin{equation}
U_\alpha U^\alpha = -1 
\end{equation}
that implies for the equatorial motion relation 
\begin{equation}
  g_{rr}\left(\frac{dr}{d\tau}\right)^2 = (E - V_{\mathrm{Eff+}})(E - V_{\mathrm{Eff-}}) ,  
\end{equation}
and in the~general stationary and axisymmetric spacetimes the~effective potential can be expressed in the~form  
\begin{equation}
V_{\mathrm{Eff}\pm}(r,a,b,L)= \frac{\beta \pm \sqrt{\beta^2-\alpha\gamma}}{\alpha}\, ,
\end{equation}
where
\begin{eqnarray}
\alpha &=& \frac{g_{\phi\phi}}{g^2_{\phi t}-g_{\phi\phi}g_{tt}}\, ,\quad \beta = \frac{-Lg_{t\phi}}{g^2_{\phi t}-g_{\phi\phi}g_{tt}}\, , \\ 
\gamma &=& \frac{L^2g_{tt}}{g^2_{\phi t}-g_{\phi\phi}g_{tt}}-1\, . 
\end{eqnarray}
This form can be simplified into 
\begin{equation}
V_{\mathrm{Eff}\pm}=\frac{-L g_{t\phi}\pm\sqrt{\left(L^2+g_{\phi\phi}\right)\left(g_{t\phi}^2-g_{\phi\phi}g_{tt}\right)}}{g_{\phi\phi}}\, .
\end{equation}

We have to choose the~upper (plus) sign of the~general expression of the~effective potential, as this case represents the~boundary of the~motion of particles in the~so called positive root states having positive locally measured energy and future-oriented time component of 4-velocity. The lower (minus) sign expression of the~effective potential is irrelevant here, as it determines in the~regions of interest particles in the~so called negative-root states having negative locally measured energy and past-oriented time component of the 4-velocity, being thus related to the~Dirac particles -- for details see \cite{Mis-Tho-Whe:1973:Gra:,Bic-Stu-Bal:1989:BAC:}. 

The physically relevant condition of the~test particle motion reads 
\begin{equation}
E\geq V_{\mathrm{Eff}+}\, .
\end{equation}
For particles with the non-zero rest mass, $m>0$, the~explicit form of the~effective potential in the~braneworld Kerr--Newman spacetimes reads
\begin{widetext} 
\begin{equation}\label{efect}
V_{\mathrm{Eff}}(r,a,b,L)= \frac{aL(2r-b) + r\sqrt{\Delta}\sqrt{L^2r^2 + r^4+a^2(r^2+2r-b)}}{r^4+a^2(r^2+2r-b)}\,. 
\end{equation}
\end{widetext}
For massless particles, $m=0$, we formally obtain 
\begin{equation}\label{efect2}
 \frac{V_{\mathrm{Effp}}(r,a,b)}{L}= \frac{a(2r-b) \pm r^2\sqrt{\Delta}}{r^4+a^2(r^2+2r-b)}\, .
\end{equation}
Here the $+$ sign is valid if $L>0$ and the $-$ sign is valid if $L<0$. 
Of course, we know that the~photon geodesic motion is independent of the~photon energy, being dependent of the~impact parameter $l=L/E$ for the~equatorial motion \cite{Bar:1973:BlaHol:,Stu-Sche:2010:CLAQG:}. 

The~effective potential is symmetric under transformation $a\rightarrow-a,\, L\rightarrow -L$, therefore, we will only study the~Kerr--Newman braneworld spacetimes with non-negative values of the~spin parameter $a$. 

The~effective potential has a~discontinuity (divergence) at radii determined by the~conditions: 
\begin{eqnarray}
g^2_{\phi t}-g_{\phi\phi}g_{tt} = 0\, ,\label{dis0}\\
r^4+a^2(r^2+2r-b)=0\, .\label{dis2}
\end{eqnarray}
At the~equatorial plane, the~quantity $g^2_{\phi t}-g_{\phi\phi}g_{tt}\equiv\Delta$, and the~condition (\ref{dis0}) implies that the~effective potential diverges at the~event horizons. The second condition (\ref{dis2}) for possible divergence of the~effective potential can be transformed to the~relation 
\begin{equation}\label{dis4}
b = b_{\mathrm{s}} \equiv \frac{r\left(2a^2+a^2 r + r^3\right)}{a^2}\, . 
\end{equation}

Notice that the~functions $b_{\mathrm{s}}(r,a)$ and $b_{\mathrm{CV}}(r,a)$ are equivalent -- therefore, the divergence could occur just at the boundary of the~causality violation region. In the~limit of $b\rightarrow b_S$, the numerator of (\ref{efect}) reads 
\begin{equation}
-\frac{r}{a}\left(L-|L|\right)\left(a^2+r^3\right)\, .
\end{equation}
Thus if $L\geq0$, both numerator and denominator of (\ref{efect}) are zero and we have to use the~L'Hopital rule to obtain
\begin{equation}
\mathrm{lim}_{b\rightarrow b_\mathrm{S}} V_{\mathrm{Eff}} = \left\{
\begin{array}{cc}
\frac{r^4+a^4 + 2 a^2(L^2+r^2)+L^2 r^2}{2aL(a^2+r^2)} & L\geq 0 \\
\infty & L<0 \\
\end{array}\, ,\right.
\end{equation}
Therefore, for the~specific angular momentum $L<0$ the~effective potential approaches at the~discontinuity the~positive infinity, creating thus for the~test particles with $L< 0$ an impenetrable barrier. 
On the~other hand, for particles with $L \geq 0$, the~effective potential takes at the~boundary of the~causality violation region a~finite value that depends on the~specific angular momentum. 

The last square root in equation (\ref{efect}) is negative for 
\begin{equation}
b>\frac{r\left(2a^2+a^2 r + L^2 r+ r^3\right)}{a^2}\, ,
\end{equation}   
therefore the~effective potential can be undefined for small values of $r$. However, this could happen only in the~causality violation region where the~effective potential looses its relevance because of the~modified meaning of the~axial coordinate that has time-like character in this region. 

\subsection{Energy measured in LNRF}

It is useful to determine for particles on the~circular geodesics the~locally measured energy, related to some properly defined family of observers. The specific energy related to the LNRF ($E_{\mathrm{LNRF}}$) is given by the~projection of the 4-velocity on the time-like vector of the~frame: 
\begin{widetext}
\begin{eqnarray}
E_{\mathrm{LNRF}}=U^{(t)} &=& U^\mu\mathbf{e}^{(t)}_{\mu} = \left(\frac{dt}{d\tau}\right) \mathbf{e}^{(t)}_{t}  \\ \nonumber
&=& \frac{r^2\pm a\sqrt{r-b}}{\sqrt{r^4+a^2\left(r^2+2r-b\right)}}\frac{\sqrt{\Delta}}{\sqrt{r^2-3r+2b\pm 2a\sqrt{r-b}}}\, .
\end{eqnarray}
\end{widetext}
The~locally measured particle energy must be always positive for the~particles in the~positive-root states assumed here -- while it is negative for the~negative-root states that are physically irrelevant in the~context of our study \cite{Bic-Stu-Bal:1989:BAC:}.  

The LNRF energy of particle following the circular geodesics diverges on the~photon circular orbit as well as the~covariant energy $E$. It also diverges for circular orbits approaching the~boundary of the~causality violation region given by equation (\ref{dis2}).

\subsection{Future-oriented particle motion}

For the~positive-root states the~time evolution vector has to be oriented to future, i.e., $dt/d\tau > 0$. On the~other hand, the negative-root states have past oriented time vectors, $dt/d\tau < 0$, being thus physically irrelevant for our study. To be sure that we are using the~solutions related to the~proper effective potential $V_{\mathrm{eff}}$ with the~correct upper sign, we have to check that the~considered geodesics have proper orientation $dt/d\tau > 0$. 

Using the~metric (\ref{Metrika}) and relations for the~specific energy (\ref{E}) and specific angular momentum (\ref{L}) in equation (\ref{eq14}), we obtain the~time component of the 4-velocity for both the~upper and lower family circular geodesics in the~form 
\begin{equation}  
\frac{dt}{d\tau} =\frac{r^2\pm a \sqrt{r-b}}{r\sqrt{r^2-3 Mr +2b\pm 2a\sqrt{Mr-b}}}\, .
\end{equation}
We see from this equation that the~time component is always positive for both the~orbits of the~upper and lower family, so we always have the~positive-root states and no mixing with the~negative-roots states occurs. 

\section{Circular geodesics of photons}

We first study motion of photons, as the~photon circular orbits represent a~natural boundary for existence of the~circular geodesic motion \cite{Bar:1973:BlaHol:,Bal-Bic-Stu:1989:BAC:}.

\begin{table}[t]
\begin{center}
\scalebox{0.9}{
\begin{tabular}{|l||c|c|c|c|c|c||c|}
\hline
\hline
\backslashbox{$r$}{$b$} & $(-\infty;0)$ & $\left(0;\frac{3}{4}\right)$ & $\left(\frac{3}{4};1\right)$ & $1$ & $\left(1;\frac{9}{8}\right)$ & $\left(\frac{9}{8};\infty\right)$ & $a_{\mathrm{ph}}(r,b)$ \\
\hline
\hline
1 & max & max & min & $\displaystyle -$ & $\displaystyle -$ & $\displaystyle -$& $\sqrt{1-b}$ \\
\hline
$r_1\, ,r_2$ & min & min & min & min & min & $\displaystyle -$ & 0 \\
\hline
$\frac{4}{3}b$ & $\displaystyle -$ & min & max & max & max & min &$ \frac{\sqrt{b}}{3\sqrt{3}}\left(8b-9\right) $\\
\hline
\hline
$a_{\mathrm{phMin}}$ & $-b$ & $\infty$ & $\infty$ & 0 & $\infty$ & $\infty$ & $\displaystyle -$ \\
\hline
\hline
\end{tabular}}
\caption{\label{table} All kinds of extrema of function $a_{\mathrm{ph}}(r,b)$. We show the (+) sign part only because of the~symmetry corresponding to interchangeability between signs $(\pm)$ and nature of local minima (max/min). Function $a_{\mathrm{phMin}}$ is value of function $a_{\mathrm{ph}}(r,b)$ at lowest possible $r$, which is $r=0$ for non-positive values of $b$ and $r=b$ otherwise.}
\end{center}
\end{table}  

The~general photon motion in the~braneworld Kerr--Newman black hole spacetimes was studied in \cite{Sche:Stu:2009b:}. Here we concentrate on the~equatorial photon motion and especially on the~existence of the~photon circular orbits. In case of the~equatorial photon orbits, the~radial function $R(r)$ is determined by the~equation (\ref{Rrr}) with removed term $-1$ (as the~rest energy $m=0$) that can be transformed into the form \cite{Sche:Stu:2009b:}:                                                           
\begin{equation}
\frac{R}{E^2}=\frac{\left[r^2-a(\lambda-a)\right]^2-\Delta(\lambda-a)^2}{r^2\Delta}\, ,
\end{equation}
where the~impact parameter $\lambda$ is defined by the relation 
\begin{equation}
\lambda=\frac{L}{E}\, .
\end{equation} 
Notice that photon orbits depend only on the~impact parameter $\lambda$. 

Applying conditions for the~circular motion (\ref{podminky}), we find that the~equatorial photon circular orbits are given by the~equations 
\begin{eqnarray}
\left[r^2-a(\lambda-a)\right]^2-\Delta(\lambda-a)^2=0\, , \\
2r(r^2+a^2-a\lambda)-(r-1)(\lambda-a)^2=0\, .\label{cosi2}
\end{eqnarray}
These two conditions imply that the~radii of circular photon orbits are determined by~equation
\begin{equation}\label{cosi}
r^2-3r+2a^2+2b\pm2a\sqrt{\Delta}=0\, , 
\end{equation}
and the~impact parameter $\lambda$ is given by the~equation 
\begin{equation}\label{lambda}
\lambda =-a\frac{r^2+3r-2b}{r^2-3r+2b}\, .
\end{equation} 
Furthermore, the~equation (\ref{cosi}) can by transformed into the form 
\begin{equation}\label{jedno}
r^2-3r+2b\pm 2a\sqrt{r-b}=0\, 
\end{equation}
that implies the~same reality condition on the~radius of the~photon orbit $r_{ph}$ as the~one that follows from equations (\ref{E})-(\ref{Om}): 
\begin{equation}
r_{\mathrm{ph}}\geq b\, .
\end{equation} 

\begin{figure}[t]
\begin{center}
\centering
\includegraphics[width=1\linewidth]{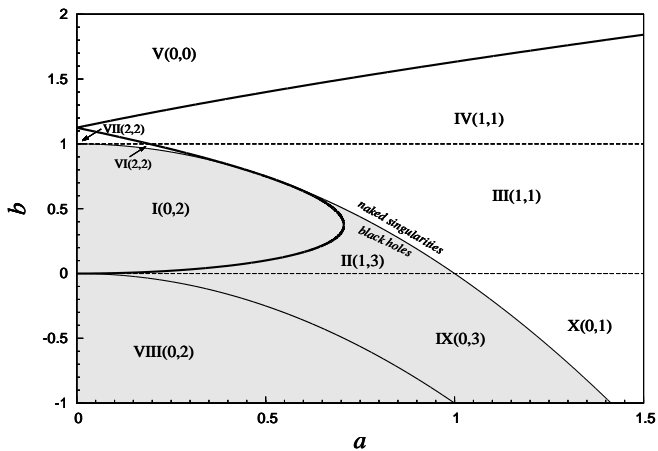}
\caption{\label{reg} Braneworld Kerr--Newman black holes and naked singularities can be divided into ten distinguish classes according to the~properties of circular photon geodesics. Curve $a_\mathrm{ph}(4b/3,b)$ (full line), given by (\ref{aphbb}), plays the main role in the classification. The corresponding regions of $b$--$a$ plane are denoted by I--X; the numbers in brakets denote the number of the~circular photon orbits in the respective class. The first number determines number of the stable circular photon geodesics, the second number determines number of the unstable circular photon geodesics.}
\end{center}
\end{figure}

Due to the~reality condition the~numerator in equation (\ref{lambda}) is positive, while the~$(\pm)$ sign of the~denominator is determined by the~sign in equation (\ref{jedno}). Thus we obtain corotating orbits $(\lambda>0)$ for the~upper sign in (\ref{jedno}) and counterrotating orbits $(\lambda<0)$ in the~other case. 

The solution of the~equation (\ref{jedno}) can be expressed in the~form 
\begin{equation}
a= a_{\mathrm{ph}}(r,b) \equiv \pm \frac{\left(3r-r^2-2b\right)}{2\sqrt{r-b}}\, .
\end{equation}
For given $a$ and $b$ the~points of a~line $a=\mathrm{const}$ crossing the~function $a_{\mathrm{ph}}(r,b)$ determine radius $r_{\mathrm{ph}}$ of the~photon circular orbits. We restrict our discussion to the~solutions corresponding to $a>0$, giving both corotating and counterrotating orbits.  
The zeros of the~function $a_{\mathrm{ph}}(r,b)$ are located at 
\begin{equation}
r_{\mathrm{ph\pm}}=\frac{1}{2}\left(3\pm\sqrt{9-8b}\right)\, .
\end{equation}
Note that these solutions represent radii of photon circular orbits in the~Reissner--Nordstr{\" o}m spacetimes \cite{Stu-Hle:2002:ActaPhysSlov:,Pug-Que-Ruf:2011:PHYSR4:}. Since
\begin{equation}
\frac{\partial a_{\mathrm{ph}}}{\partial r}=\pm\frac{(r-1)(3r-4b)}{4\left(r-b\right)^{3/2}}\, ,
\end{equation}
the~extrema of the~curves $a_{\mathrm{ph}}(r,b)$ are located at $r=1$ and at $r=4b/3 \,$ . The value of the~function $a_{\mathrm{ph}}(r,b)$ at the~point $r=1$ reads (recall that we consider the~positive values of spin)   
\begin{equation}\label{aphbb}
a_{\mathrm{ph-ex}}(r=1,b)=\sqrt{1-b}\, ,
\end{equation} 
corresponding to the~extreme Kerr--Newman black holes, while at $r=4b/3$ it reads
\begin{equation}\label{aphbb2}
a_{\mathrm{ph-ex}}(r=4b/3,b) = \pm\frac{\sqrt{b}}{3\sqrt{3}}\left(8b-9\right)\, .
\end{equation} 

The~position, value and kind of the~extrema of the~function $a_{\mathrm{ph}}(r,b)$ are listed in Table (\ref{table}). The results are summarized in Figure (\ref{reg}). We see that all curves drawn there -- the~curve $a_{\mathrm{ph-ex}}(r=4b/3,b)$, the~line $a^2+b=1$ (corresponding to the extremal black holes), the~line $a^2+b=0$, and the~line $b=1$ (separating the~braneworld Kerr--Newman naked singularities with the~ergosphere from those without it) -- divide the~$b$ -- $a$ plane into ten regions.
In this sense, the~braneworld Kerr--Newman spacetimes can be divided into ten different classes, characterized by:
\begin{enumerate}
\item existence of the~horizon
\item existence of the~ergosphere
\item the number of stable and unstable circular photon orbits. 
\end{enumerate} 
The~situation is summarized in Table (\ref{tabnej}) and is also visualized in Figure (\ref{reg}), in accord with the~analysis of circular photon orbits in the~standard Kerr--Newman spacetimes \cite{Bal-Bic-Stu:1989:BAC:}. In the~case of the braneworld Kerr--Newman black holes new regions VIII,IX and X corresponding to negative values of the~tidal charge, $(b<0)$, occur in addition to the~standard Kerr--Newman spacetimes. 

\begin{table}[t]
\begin{center}
\begin{tabular}{|c|c|c|c||c|c|c|c|}
\hline
\hline
Class & Horiz. & Ergo. & Orbits & Class & Horiz. & Ergo. & Orbits \\
\hline
\hline
I & yes & yes & 0,2 & VI & no & yes & 2,2 \\ \hline
II & yes & yes & 1,3 & VII & no & no & 2,2 \\ \hline
III & no & yes & 1,1 & VIII & yes & yes & 0,2\\ \hline
IV & no & no & 1,1 & IX & yes & yes & 0,3 \\ \hline
V & no & no & 0,0 & X & no & yes & 0,1 \\ \hline
\hline
\end{tabular}
\caption{\label{tabnej} Ten possible divisions of braneworld Kerr--Newman spacetimes with respect to existence of the~horizon, existence of the~ergosphere and the~number of stable and unstable circular photon orbits. The~first number in the~column orbits corresponds to amount of stable circular photon orbits, the second corresponds to amount of unstable circular photon orbits.}
\end{center}
\end{table}
    
\section{Stable circular geodesics}

It is well known that character of the~test particle (geodesic) circular motion governs structure of the~Keplerian (geometrically thin) accretion disks orbiting a~black hole \cite{Nov-Tho:1973:BlaHol:,Pag-Tho:1974:ApJ:} or a~naked singularity (superspinar) \cite{Stu:Hle:Tru:2011:,Stu-Sche:2012:CLAQG:}; similarly, it can govern also motion of a~satellite orbiting the~black hole or the~naked singularity (superspinar) along a~quasicircular orbit slowly descending due to gravitational radiation of the~orbiting satelite \cite{Ruf:1973:BlaHol:}. The Keplerian accretion, starting at large distances from the~attractor, is possible in the~regions of the~black hole or naked singularity spacetimes where local minima of the~effective potential exist and the~energy corresponding to these minima decreases with decreasing angular momentum \cite{Mis-Tho-Whe:1973:Gra:}. In other words, in terms of the~radial profiles of the~quantities characterizing circular geodesics, the~Keplerian accretion is possible where both specific angular momentum and the~specific energy of circular geodesics decrease with decreasing radius. In the~standard model of the~black hole accretion disks, the~inner edge of the~accretion disk is located in the~so called marginally stable circular geodesic where the~effective potential has an~inflexion point \cite{Nov-Tho:1973:BlaHol:}, but the~situation can be more complex in the~naked singularity spacetimes \cite{Stu-Hle:2002:ActaPhysSlov:,Stu-Sche:2014:CLAQG:}. 

We study stability of the~circular geodesic motion of test particles relative to the~radial perturbations in the~braneworld Kerr--Newman spacetimes. Note that the~equatorial circular motion is then always stable relative to the~latitudinal perturbations perpendicular to the~equatorial plane \cite{Bic-Stu:1976:BAC:}. We show that the~most interesting and, in fact, unexpected result occurs for test particles orbiting the~special class of the~braneworld mining-unstable Kerr--Newman naked singularities demonstrating an~infinitely deep gravitational well enabling (formally) unlimited energy mining from the~naked singularity spacetime. Of course, such a~mining must be limited by violation of the~assumption of the~test particle motion.  

\subsection{Marginally stable circular geodesics}

The~loci of the~stable circular orbits are given by the~condition related to the~radial motion $R(r)$ function 
\begin{equation}\label{R2}
\frac{\partial^2 R(r,a,b,E,L)}{\partial r ^2}\leq 0\, ,
\end{equation} 
or the~relation
\begin{equation}\label{R3}
\frac{\partial^2 V_{\mathrm{Eff}}(r,a,b,L)}{\partial r ^2}\leq 0\, ,
\end{equation} 
related to the~effective potential $V_{\mathrm{Eff}}(r)$, where the~case of equality corresponds to the~marginally stable circular orbits at $r_{\mathrm{ms}}\,$ with $L=L_{\mathrm{ms}}$, corresponding to the~inflexion point of the~effective potential -- for lower values of the~specific angular momentum $L$ the~particle cannot follow a~circular orbit. Such marginally stable circular orbit represents the~innermost stable circular orbit and the~inner edge of the~Keplerian disks in the~Kerr black hole and naked singularity spacetimes. 
 
Using the~relations (\ref{E}) and (\ref{L}), we obtain for the~braneworld Kerr--Newman spacetimes \cite{Stu:Kot:2009:}\footnote{Formally same results relevant for Kerr--Newman spacetime can be found in \cite{Ali-Gal:1981:}} 
\begin{widetext}
\begin{eqnarray}\label{rms}
r(6r - r^2 - 9b +3a^2) + 4b(b-a^2)\mp 8a(r-b)^{3/2}=0\, .
\end{eqnarray}
\end{widetext}
In the~previous studies, only the~braneworld black hole spacetimes were usually considered \cite{Kot-Stu-Tor:2008:CLAQG:,Stu:Kot:2009:}. Standard Kerr--Newman naked singularity spacetimes were discussed in \cite{Bic-Stu-Bal:1989:BAC:,Pug:Que:Ruf:2013:}. Here we consider whole family of the~braneworld Kerr--Newman spacetimes, with both positive and negative tidal charges. The~solution of equation (\ref{rms}) can be express in the~form 
\begin{equation}
a_{\mathrm{ms}}=\mp\frac{4\left(r-b\right)^{3/2}\pm\sqrt{3 b r^2 - (2 + 4 b) r^3 + 3 r^4}}{4b-3r}\, ,
\end{equation} 
where the $\mp$ signs correspond to the~upper and lower family of the~circular geodetics. The $\pm$ signs correspond to the~two possible solutions of Eq. (\ref{rms}). The~local extrema of the~function $a_{\mathrm{ms}}(r,b)$ are given by the~relation 
\begin{equation}\label{extr}
a_{\mathrm{ms(extr)}} \equiv \mp\left(2 \sqrt{b}\pm\sqrt{b\left(4b-1\right)}\right) .   
\end{equation}  
Thus it can be shown that there is no solution for equation (\ref{rms}) related to the~lower family of circular geodesics when
\begin{equation}\label{con}
b>\frac{5}{4} \quad \wedge \quad a < -2\sqrt{b}+\sqrt{b\left(4b-1\right)}\, 
\end{equation}
and there is no solution for the~upper sign family of circular geodesics when $1 > b > 1/4 $ and 
\begin{equation}\label{con2}
2\sqrt{b}-\sqrt{b\left(4b-1\right)} < a < 2\sqrt{b}+\sqrt{b\left(4b-1\right)}\, .
\end{equation}
The existence of the~marginally stable circular geodesics in dependence on the~dimensionless parameters of the~braneworld Kerr--Newman spacetimes is represented in Figure (\ref{pod2}). This figure will be crucial for construction of the~classification of the~Kerr--Newman spacetimes according to the~Keplerian accretion, but it is not sufficient, as the~classification of the~photon circular geodesics plays a crucial role too. 

\begin{figure}[t]
\begin{center}
\centering
\includegraphics[width=1\linewidth]{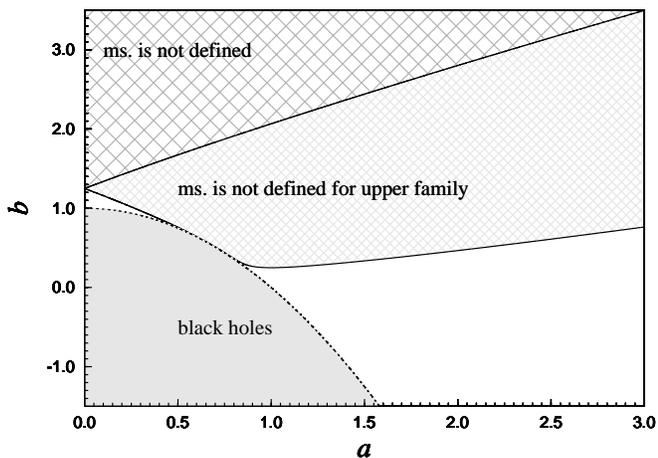}
\caption{\label{pod2} Mapping of existence of the~marginally stable circular geodesics in the~parameter space of the~braneworld Kerr--Newman spacetimes.}
\end{center}
\end{figure}

The~function $a_{\mathrm{ms}}(r,b)$ determines in a~given Kerr--Newman spacetime location of the~marginally stable circular geodesics that are usually considered as the~boundary of Keplerian accretion discs determined by the~quasi-geodesic motion. 

\subsection{Innermost stable circular geodesics}

The~standard treatment when the~inner edge of the~Keplerian accretion discs is located at the~marginally stable orbits defined by the~inflexion point of the~effective potential (this point is also the~innermost stable circular orbit (ISCO) \cite{Nov-Tho:1973:BlaHol:}) works perfectly in the~braneworld Kerr--Newman black hole spacetimes, but the~situation is more complex in the~braneworld naked singularity spacetimes, as the~innermost stable circular orbits (ISCO) do not always correspond to the~marginally stable orbit defined by equation (\ref{rms})\footnote{See for example \cite{Stu-Hle:2002:ActaPhysSlov:,Fav:2011:PRD:}}. The ISCOs that are not coinciding with a marginally stable circular geodesic, related to an inflexion point of the effective potential, correspond to the orbits with lowest radius in sequence of stable circular geodesics. Contrary to the case of the marginally stable circular orbits from which the particles can move inwards, in the case of ISCO representing the inner limit of stable circular geodesics the particle remains captured at this orbit or in its vicinity. Such ISCO orbits were found for the first time in the Reissner-Nordstrom(-de Sitter) naked singularity spacetimes when they correspond to orbits with vanishing angular momentum (particles at static positions) \cite{Stu-Hle:2002:ActaPhysSlov:,Pug-Que-Ruf:2011:PHYSR4:}. Here we demonstrate existence of a new class of this kind of ISCO representing the limit of stable circular geodesics located at the stable photon circular geodesic. In order to alow for the standard accretion with decreasing $E$ and $L$ with decreasing radius of the stable orbits, we have consider the stable circular geodesics with $E \to -\infty$ and $L \to -\infty$. 

The ISCO can be formally determined, if we consider the function of the radius of the circular geodesic $r_{c}(L_{c};a,b)$, given implicitly by Eq.(56), or $r_{c}(E_{c};a,b)$, given implicitly by Eq.(55). Then the $r_{\mathrm{ISCO}}$ can be defined in a given spacetime, with fixed parameters $a,b$, by the relations $dr_c/dL_{c} = 0$, $dr_c/dE_{c} = 0$ that can be expressed as $dL_{c}/dr_{c} \to -\infty$ and $dE_{c}/dr_{c} \to -\infty$, related to the standard accretion with decreasing energy and angular momentum of accreting matter. Note that the conditions $dL_{c}/dr_{c} \to \infty$ and $dE_{c}/dr_{c} \to \infty$ can determine the outermost stable circular geodesics from which the accretion could start, but such a situation is not related to plausible astrophysical situation as discussed in detail in \cite{Stu-Sche:2014:CLAQG:}. 

There are two relevant cases when the~conditions $dL_{c}/dr_{c} \to \pm\infty$ and $dE_{c}/dr_{c} \to \pm\infty$ can be satisfied: 
\begin{eqnarray}
\mathrm{a)}\qquad  0&=&r^2-3 r +2b\pm 2a\sqrt{r-b}\, , \\
\mathrm{b)}\qquad  r&=&b\, .
\end{eqnarray}
The case a) tells us that the~innermost circular geodesics corresponds to the~photon circular geodesics that can be also stable with respect to radial perturbations so that this condition is also applicable as a~limit on the~stable circular orbits of test particles, as demonstrated in \cite{Stu-Hle:2002:ActaPhysSlov:}. Then the~specific energy and the~specific angular momentum tend asymptotically to $E \to \pm \infty$ and $L \to \pm \infty$ but the~impact parameter $\lambda=L/E$ remains finite. The~condition $r>b$ could restrict the~condition implied by the~photon circular geodesics. Here we consider the case of $dL_{c}/dr_{c} \to -\infty$ and $dE_{c}/dr_{c} \to -\infty$. 

\begin{figure*}[t]
\begin{center}
\centering
\includegraphics[width=\linewidth]{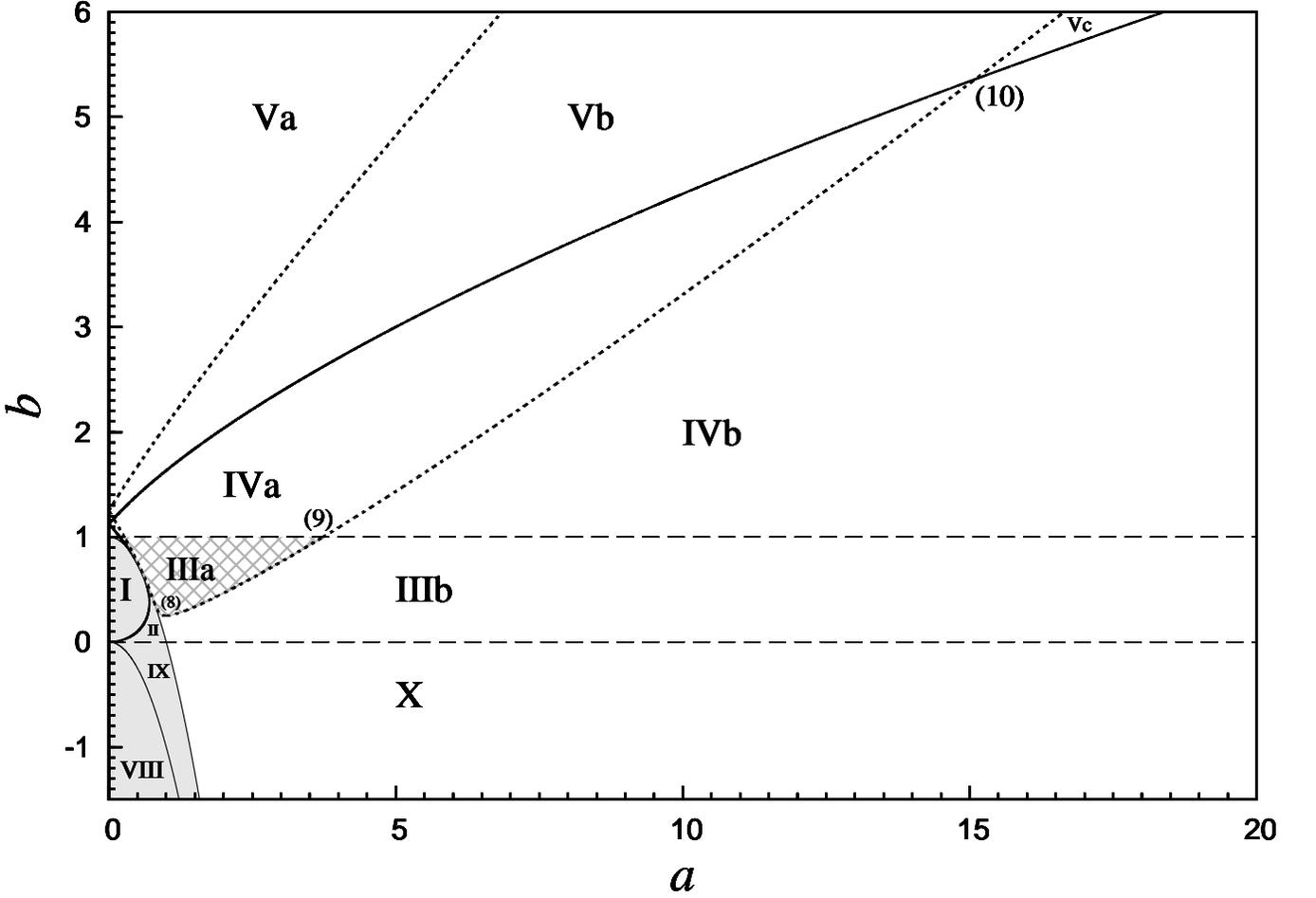}
\caption{\label{regionab2} Classification of the~braneworld Kerr--Newman spacetimes according to the~properties of circular geodesics relevant for the~Keplerian accretion. The parameter space $b-a$ is separated by curves governing the~extrema of the~functions determining the~photon circular orbits (thick lines) and the~marginally stable orbits (dashed lines). Point ($\sqrt{0.5}$,0.5) is an intersection of dashed line and curve separating black holes from naked singularities ($b=1-a^2$). This two curves are tangent at the~common point. }   
\end{center}
\end{figure*}

The~case b) can be relevant for the~braneworld spacetimes with positive braneworld parameter $b$, as demonstrated in \cite{Stu-Hle:2002:ActaPhysSlov:} where the~effective potential $V_{\mathrm{Eff}}(r,b,L)$ for the~Reissner--Nordstr{\" o}m naked singularity spacetimes clearly demonstrates that the~inner edge of the~Keplerian disc is located at  $r=b$, having $L=0$, while no marginally stable circular orbit, corresponding to an~inflexion point of the~effective potential, exist. Note that in some spacetimes the sequence of the stable circular geodesics can start at the outermost stable circular geodesic with $dL_{c}/dr_{c} \to +\infty$ and $dE_{c}/dr_{c} \to +\infty$ corresponding to the stable circular geodesic. 

\subsection{Effective potential dependence on specific angular momentum and analytical proof that Keplerian accretion with infinite efficiency can exist in approximation of geodesic motion}

The~Keplerian accretion works if there exist a~continuous sequence of local minima of the~effective potential with decreasing values of angular momentum $L$. In terms of the~effective potential (\ref{efect}), the~conditions for existence of Keplerian accretion disks can be express in the~form 
\begin{eqnarray}
\frac{\partial V_{\mathrm{Eff}}(r,a,b,L)}{\partial r}&=&0\, ,\nonumber \\
\frac{\partial^2 V_{\mathrm{Eff}}(r,a,b,L)}{\partial r ^2}&\leq& 0\, ,\nonumber\\
\frac{\partial V_{\mathrm{Eff}}(r,a,b,L)}{\partial L}&<&0\, .\label{cond2}
\end{eqnarray} 
Except the~possibility to stop this procedure by the~inflexion point of the~effective potential, there exist other possible way to stop validity of these conditions: 
\begin{equation}
\frac{\partial V_{\mathrm{Eff}}(r,a,b,L)}{\partial L}=0\,  
\end{equation}
that will be satisfied at a~turning point where 
\begin{equation}
  L = L_{\mathrm{T}}\equiv\frac{\pm a (2r-b)}{r\sqrt{r^2-2r+b}}\, . 
\end{equation} 
At this turning point the~minimum of effective potential, given by $\partial V_{\mathrm{eff}}(r,a,b,L_{\mathrm{T}})/\partial r=0$, is located where:
\begin{equation}\label{mreturn} 
 \frac{r-b}{r^2\sqrt{r^2-2r+b}}=0\, \Rightarrow\, r=b\, .
\end{equation}
In such situations, the~inner edge of the~Keplerian accretion disk is located at $r=b$. Putting this result into the~definition of the~function $L_{\mathrm{T}}$, we find 
\begin{equation}
L_{\mathrm{T}}(r=b)=\frac{\pm a b}{b\sqrt{b(b-1)}}\, \Rightarrow b>1\, .
\end{equation}  
We see that the~effect of existence of the~lowest possible value of angular momentum $L$ associated with local minima of effective potential occurs only for values of the~tidal charge $b> 1$ when equation (\ref{mreturn}) is well defined at $r>b$. 

\begin{figure*}[t]
\begin{center}
\centering
\includegraphics[width=\linewidth]{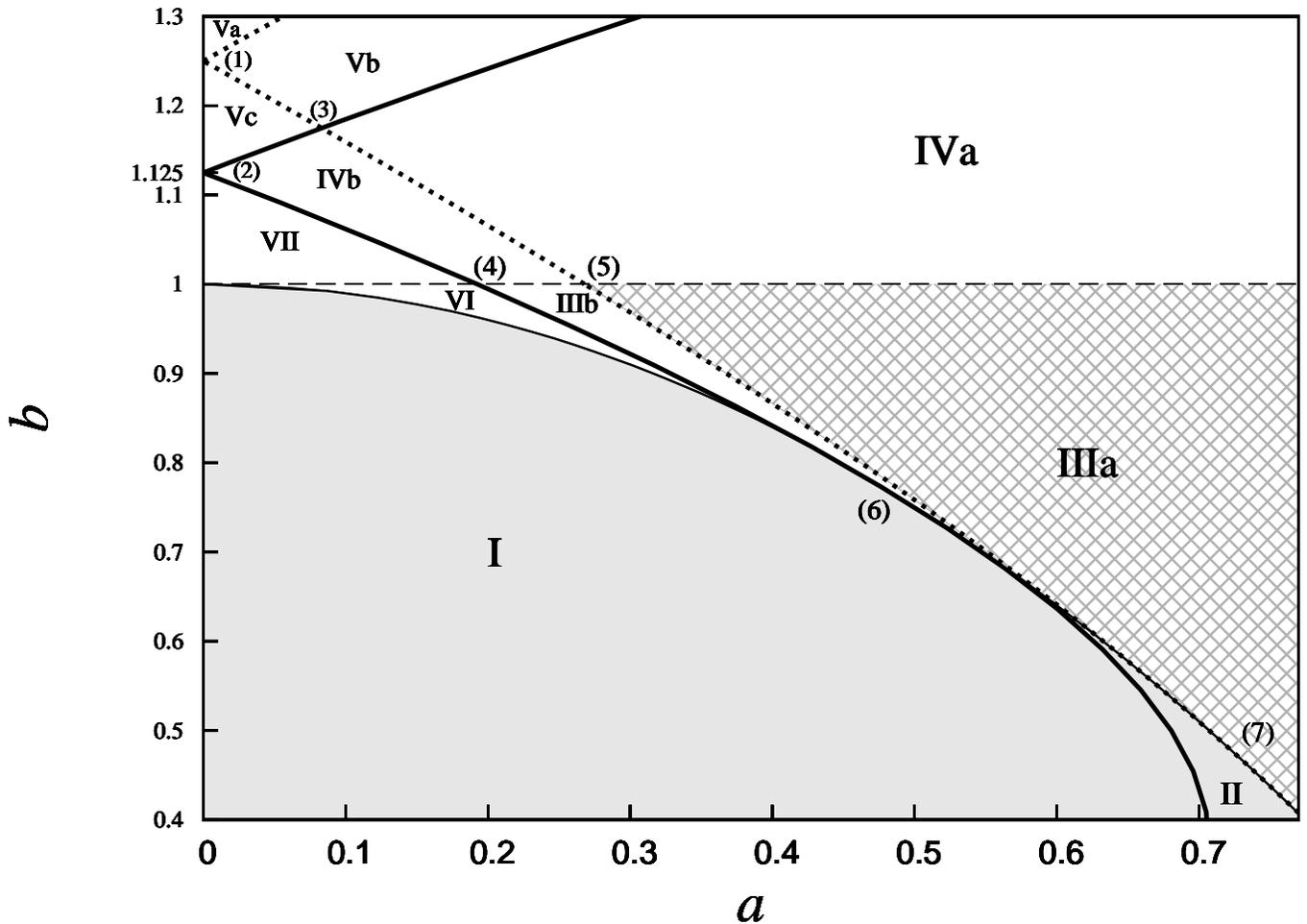}
\caption{\label{regionab3} Classification of the~braneworld Kerr--Newman spacetimes according to the~properties of circular geodesics relevant for the~Keplerian accretion. The parameter space $b-a$ is separated by curves governing the~extrema of the functions determining the~photon circular orbits (thick lines) and the~marginally stable orbits (dashed lines). Detailed structure for small values of spin $a$ and $b \sim 1$.}   
\end{center}
\end{figure*}

The~second possible way the~conditions (\ref{cond2}) are not well matched is related to the~situation when the~local minima of the~effective potential turn into an~inflexion point at $r$ defined by (\ref{rms}). 

So the~inflexion point of the~effective potential is not always defined, if the~tidal parameter $b>5/4$ for the~lower family solution of equation (\ref{rms}). In this case the~Keplerian accretion stops at the~point $r=b$ where the~minimum of the~effective potential starts to increase in energy level with decreasing $L$.

On the~other hand, for the~upper family solution of equation (\ref{rms}) situation becomes extraordinary and much more interesting for the~tidal charge in the~interval $1>b>1/4$, and appropriately tuned spin $a$; then the~inflexion point of the~effective potential does not exist -- for this reason, the~Keplerian accretion starting at large values of the~angular momentum of accreting matter cannot be stopped, and it continues with no limit to unlimitedly large negative values of the~angular momentum and unlimitedly large negative values of the~energy. 

Therefore, in this case, unrestricted mining of the~energy due to the~Keplerian accretion could occur. Of course, this mining has to be stopped at least when the~energy of the~accreting matter starts to be comparable to the~mass parameter of the~Kerr--Newman naked singularity, and the~approximation of the~test particle motion of matter in the~disk is no longer valid. 

\section{Classification of braneworld Kerr--Newman spacetimes according to radial profiles of circular geodesics}

\begin{figure*}[ht]
\begin{center}
\begin{minipage}{.5\linewidth}
\centering
\includegraphics[width=\linewidth]{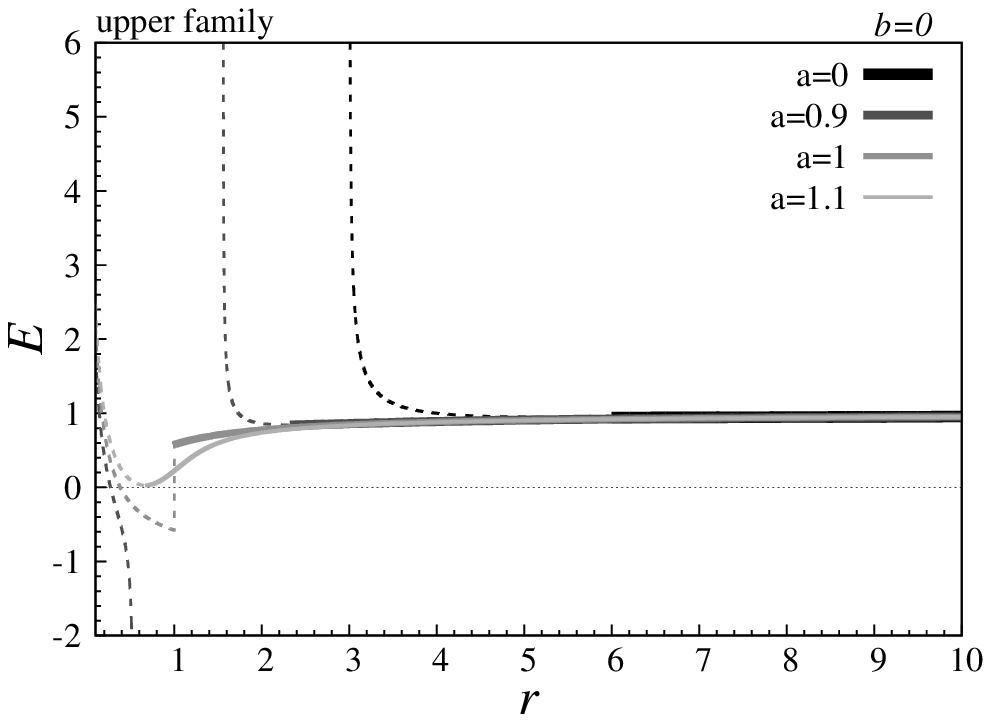}
\end{minipage}\hfill
\begin{minipage}{.5\linewidth}
\centering
\includegraphics[width=\linewidth]{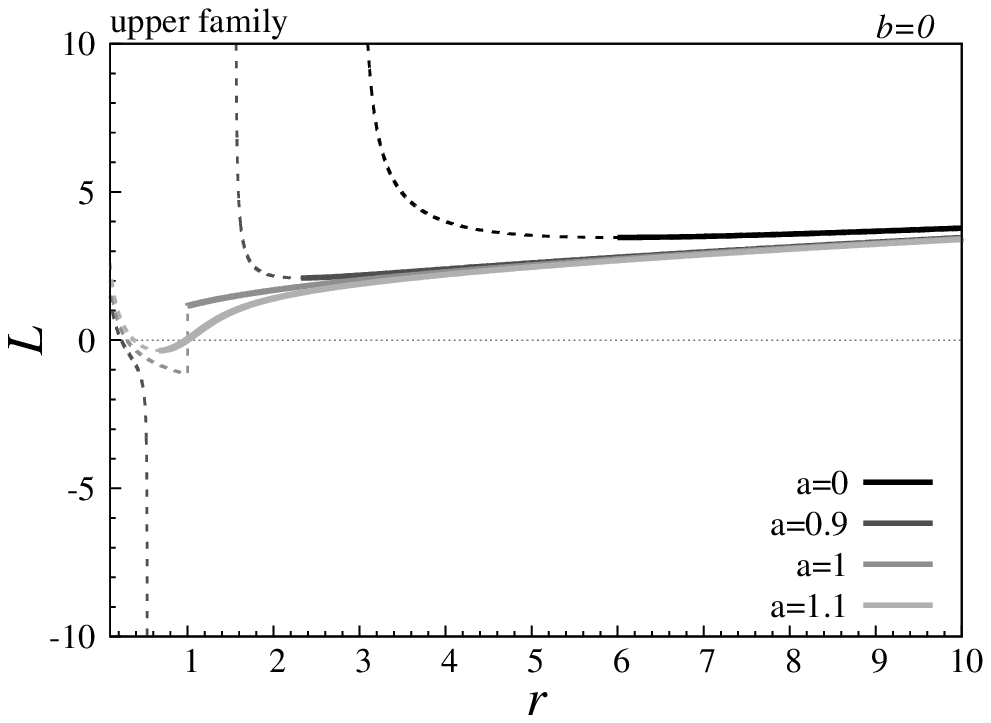}
\end{minipage}
\begin{minipage}{.5\linewidth}
\centering
\includegraphics[width=\linewidth]{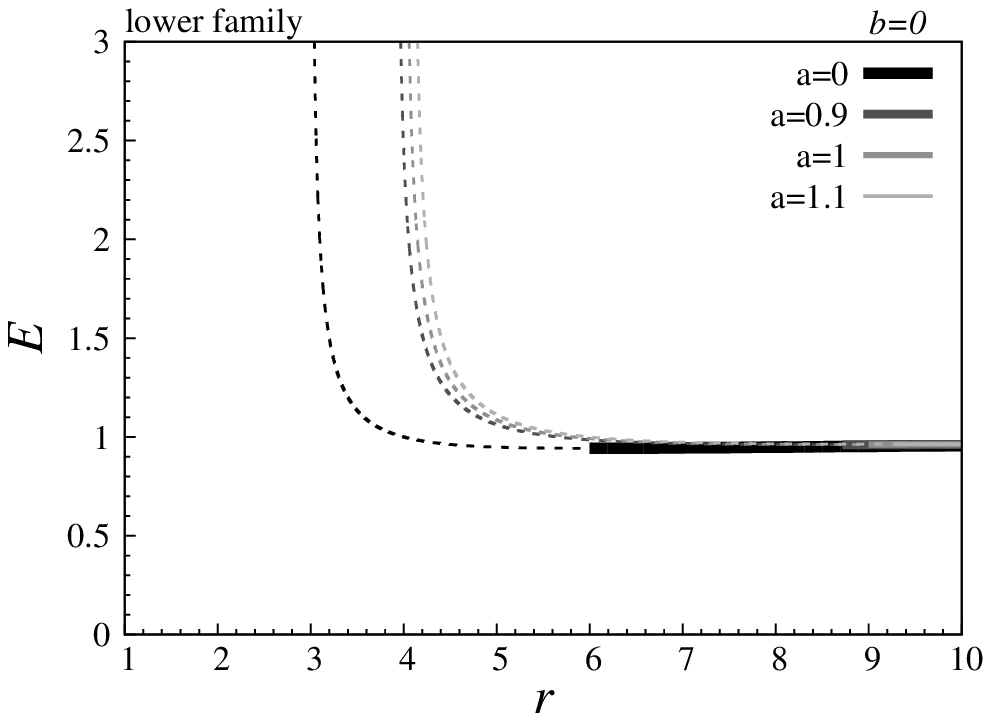}
\end{minipage}\hfill
\begin{minipage}{.5\linewidth}
\centering
\includegraphics[width=\linewidth]{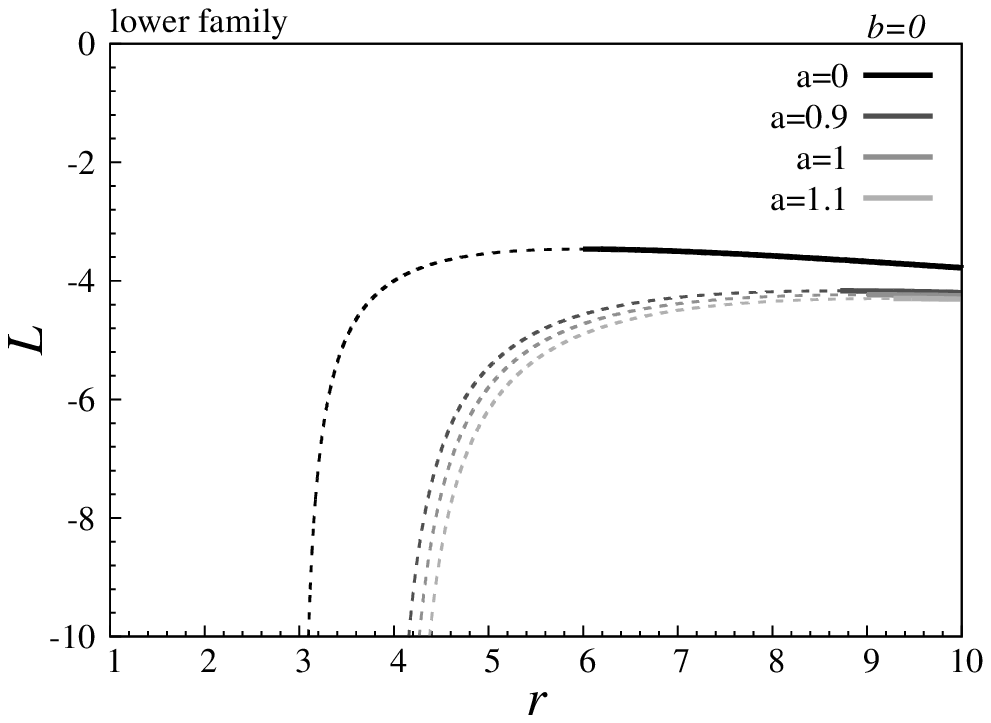}
\end{minipage}
\caption{\label{el0} $E$ and $L$ for Kerr black hole and naked singularities.} 
\end{center}
\end{figure*}

In order to create classification of the~braneworld Kerr--Newman black hole and naked singularity spacetimes according to possible regimes of the~Keplerian accretion, we consider existence of the~event horizons, existence of the~ergosphere, and we use the~characteristics of the~circular geodesics: existence of the~circular photon geodesics and their stability, existence of the~marginally stable circular geodesics related to inflexion points of the~effective potential, and relevance of the~limiting radius $r=b$. We use the~classification of the~braneworld Kerr--Newman spacetimes introduced for the~characterization of the~photon circular geodesics, and generate a~subdivision of the~introduced classes according to the~criteria related to the~marginally stable orbits. 

The~individual classes of the~Kerr--Newman spacetimes will be represented by typical radial profiles of the~specific angular momentum $L$, specific energy $E$, and the~effective potential $V_{\mathrm{Eff}}$ that enable understanding of the~Keplerian accretion and calculation of its efficiency. We first briefly summarize results of the~two special cases -- Kerr and Reissner--Nordstr{\" o}m spacetimes. In the~following classification of the~braneworld Kerr--Newman spacetimes the~characteristic types of the~behaviour of the~circular geodesics in the~special Kerr and Reissner--Nordstr{\" o}m spacetimes occur, but also some quite new and extraordinary situations arise. The results of the~circular geodesic analysis in the~braneworld Kerr--Newman spacetimes can be directly applied also for the~circular geodesics in the~standard Kerr--Newman spacetimes, if we make transformation $b \to Q^2$ where $Q^2$ represents the~squared electric charge parameter of the~Kerr--Newman background.

\subsection{Case b=0: Kerr black hole and naked singularity spacetimes}

The~limiting case of the~well known results of the~test particle circular orbits in the~Kerr spacetimes that were studied in detail in \cite{Bar-Pre-Teu:1972:ApJ:,Stu:1980:BAC:} demonstrates clearly the~necessity of very careful treating of the~families of circular orbits in the naked singularity spacetimes where the~simple decomposition of the~circular orbits to corotating and counterrotating (retrograde) is not always possible. Namely, in the spacetimes with $1 < a < a_{\mathrm{c}} = 1.3$ the~circular orbits that are corotating at large distances from the~ring singularity become retrograde near the~ring singularity, at the~ergosphere; moreover, in the~spacetimes with $1 < a < a_{0} = 1.089$ the~covariant energy of such orbits can be negative. The~specific energy and specific angular momentum of the~circular geodesics of the~Kerr black hole and naked singularity spacetimes are illustrated in Figure (\ref{el0}). Notice that the~unstable circular geodesics approach the~radius $r=0$ with unlimitedly increasing covariant energy and axial angular momentum, however, the~photon circular geodesic cannot exist at the~ring singularity. The~Kerr naked singularities are classically unstable as Keplerian accretion from both the~corotating and counterrotating disks inverts the~naked singularity into an~extreme Kerr black hole -- the~transition is discontinuous (continuous) for corotating (counterrotating) Keplerian disk \cite{deF:1974:aap,Stu:1980:BAC:,Stu:1981:BAC:,Stu:Hle:Tru:2011:}. We shall see latter that the~Keplerian accretion cannot be generally treated simply in the~Kerr--Newman naked singularity spacetimes due to the~complexities discussed in detail in \cite{Stu-Sche:2014:CLAQG:}. We expect addressing this issue in future work. 

\subsection{Case a=0: Reissner--Nordstr{\" o}m black hole and naked singularity spacetimes}

The~other limiting case of the~Reissner--Nordstr{\" o}m (RN) and Reissner--Nordtr{\" o}m-(anti-)de Sitter black hole and naked singularity spacetimes has been treated in \cite{Pug-Que-Ruf:2011:PHYSR4:,Stu-Hle:2002:ActaPhysSlov:}. It has been demonstrated that in the~Reissner--Nordstr{\" o}m naked singularity spacetimes even two separated regions of circular geodesics could exist. The doubled regions of the~stable circular motion occur in the~RN naked singularity spacetimes with the charge parameter $1<Q^2<5/4$, or even only stable circular geodesics could exist, if the~charge parameter $Q^2 > 5/4$. In the~RN naked singularity spacetimes also doubled photon circular geodesics can occur, with the~inner one being stable relative to radial perturbations, if the~charge parameter is in the~interval of $1<Q^2<9/8$ \cite{Stu-Hle:2002:ActaPhysSlov:,Pug:Que:Ruf:2013:}. The~same phenomena occur in the~naked singularity Kehagias--Sfetsos spacetimes of the~Ho{\v r}ava quantum gravity \cite{Stu-Sche:2014:CLAQG:,Stu-Sche-Abd:2014:PHYSR4:} or in the~no-horizon regular Bardeen or Ayon--Beato--Garcia spacetimes \cite{Stu-Sche:2015:IJMPD:,Sche-Stu:2015:JCAP:}. 

We shall see that in the~braneworld Kerr--Newman naked singularity spacetimes the~special naked singularity effects of the~Kerr and Reissner--Nordstr{\" o}m case are mixed in an~extraordinary way leading to existence of an~infinitely deep gravitational well implying the~new effect we call mining instability. 

\begin{table}[ht]
\begin{center}
\begin{tabular}{|c|c|c|c|c|c|c|}
\hline
\hline
\bfseries Class &
\bfseries ISCO &
\bfseries MSO(u) &
\bfseries MSO(l) &
\bfseries Hor./Erg. &
\bfseries SP &
\bfseries UP \\
\hline
\hline
I & =MSO & classic & classic & yes/yes & 0 & 2 \\ \hline
II & =MSO & classic & classic & yes/yes & 1 & 3 \\ \hline
IIIa & =Photon & $\displaystyle -$ & classic & no/yes & 1 & 1 \\ \hline
IIIb & =MSO & classic & classic & no/yes & 1 & 1 \\ \hline
IVa & at r=b & $\displaystyle -$ & classic & no/no & 1 & 1 \\ \hline
IVb & at r=b & classic & classic & no/no & 1 & 1 \\ \hline
Va & at r=b & $\displaystyle -$ & $\displaystyle -$ & no/no & 0 & 0 \\ \hline
Vb & at r=b & $\displaystyle -$ & classic & no/no & 0 & 0 \\ \hline
Vc & at r=b & classic & classic & no/no & 0 & 0 \\ \hline
VI & =MSO & classic & classic & no/yes & 2 & 2 \\ \hline
VII & at r=b & classic & classic & no/no & 2 & 2 \\ \hline
VIII & =MSO & classic & classic & yes/yes & 0 & 2 \\ \hline
IX & =MSO & classic & classic & yes/yes & 0 & 3 \\ \hline
X & =MSO & classic & classic & no/yes & 0 & 1 \\ \hline
\hline
\end{tabular}
\caption{\label{class} Classification of parameter space $b-a$ with respect to: ISCO - radius of innermost stable circular orbit; MSO(u) - radius of Marginally Stable Orbit for upper sign family; MSO(l) - radius of Marginally Stable Orbit for lower sigh family; SP - number of stable photon circular orbit; UP - number of unstable photon circular orbit. ISCO has only two possible outcomes. It can be identical with MSO or lies at $r=b$. Word \uvozovky{classic} in this context means that MSO is defined by equation (\ref{rms}).} 
\end{center}
\end{table}

\subsection{Characteristic points of the Kerr--Newman spacetime classification}

The classification of the~braneworld Kerr--Newman spacetimes according to the~character of the~circular geodesics and related effective potential are determined by the~functions governing the~local extrema of the~functions giving the photon circular geodesics and the~marginally stable circular geodesics corresponding to the~inflexion points of the~effective potential. In the~space of the spacetime parameters $b - a$ then exist fourteen regions corresponding to classes of the~braneworld Kerr--Newman spacetimes demonstrating different behaviour of the~circular geodesics and Keplerian accretion as demonstrated in Figure (\ref{regionab2}) and in Figure (\ref{regionab3}) giving details of the~regions of low values of the dimensionless parameters $a$ and $b$. These regions are governed by intersection points of the~curves (\ref{aphbb}) and (\ref{extr}) that give thirteen characteristic points in the~parameter space that are summarized in the~following way: the pairs $(a,b)$ are ordered gradually from the top to the bottom and from the left to the right,
\begin{eqnarray}
&(1)&\rightarrow(0,1.25)\, ,\nonumber \\
&(2)& \rightarrow(0,1.125)\, ,\nonumber \\
&(3)&\rightarrow\left(\frac{A_{-} \left(12+A_{-}\right)}{16 \sqrt{2}},\frac{3}{32} \left(12 + A_{-}\right)\right)=(0.0831,1.1748)\, ,\nonumber \\   
&(4)&\rightarrow\left(\frac{1}{3\sqrt{3}},1\right)=(0.19245,1)\, ,\nonumber  \\
&(5)&\rightarrow\left(2-\sqrt{3},1\right)=(0.268,1)\, ,\nonumber  \\
&(6)&\rightarrow\left(0.5,\frac{3}{4}\right)\, ,\nonumber \\
&(7)&\rightarrow\left(\sqrt{0.5},0.5\right)\, ,\nonumber \\
&(8)&\rightarrow\left(1,0.25\right)\, ,\nonumber \\
&(9)&\rightarrow\left(2+\sqrt{3},1\right)=(3.732,1)\, ,\nonumber \\
&(10)&\rightarrow\left(\frac{A_{+} \left(12+A_{+}\right)}{16 \sqrt{2}},\frac{3}{32} \left(12 + A_{+}\right)\right)=(15.0992,5.361)\, ,\nonumber   
\end{eqnarray}
where
\begin{equation}
A_{\pm}=\left(9+8 \sqrt{3}\pm\sqrt{3 \left(83+48 \sqrt{3}\right)}\right)
\end{equation}
The point (6) is the~crossing point of the~function of the extrema of the~photon circular orbit function $a_{\mathrm{ph-ex}}$, and the~curve separating black hole and naked singularity spacetimes, $b=1-a^2$. The point (7) is the~common point of the dotted curve, given by the~function $a_{\mathrm{ms(extr)}}$ and limiting the~spacetimes allowing for existence of the~marginally stable orbits, and the~curve separating black holes from naked singularities ($b=1-a^2$). These two curves are tangent at the~common point. 

\subsection{Character of circular geodesics in the Kerr--Newman spacetimes}

The~parameter space of the~braneworld Kerr--Newman spacetimes $b-a$ is divided into fourteen regions due to the~criteria reflecting basic properties of the~spacetimes and properties of their circular geodesics:
\begin{itemize}
\item Existence of event horizons and ergosphere
\item Existence of unstable and stable circular photon geodesics
\item Existence of the marginally stable geodesics or the innermost stable circular geodesics (ISCO)
\end{itemize} 
The classification is summarized in Table (\ref{class}). Basically, we combine Figure (\ref{pod2}) with Figure (\ref{reg}) to obtain Figures (\ref{regionab2},\ref{regionab3}) where properties of the~photon circular geodesics and properties of marginally stable geodesics or ISCO's are reflected. We will show that the~most surprising properties of the Keplerian accretion arise in the~spacetimes of Class IIIa.  

Now we give properties of the~circular geodesics in all the~fourteen classes of the~braneworld Kerr--Newman spacetimes, presenting and discussing typical radial profiles of their specific energy and specific angular momentum, complemented by sequences of the~effective potential. Classification of the~standard Kerr--Newman spacetimes according to properties of the~circular geodesics contains all the~classes except those related to $b<0$ -- therefore, classes VIII, IX and X are excluded. 

\subsubsection{Class I}

Class of the~black hole spacetimes with two horizons, two unstable photon circular orbits and ergosphere. Class border is given by line $b=a_{\mathrm{ph-ex}}(r=4b/3,b)\, ,$ $b=1-a^2$ with intersection at point $(0.5,3/4)$ (point number (6) in Figure (\ref{regionab2})) and line $a=0\, .$  

Marginally stable orbits for test massive particles are given by the~inflexion point of the~effective potential are defined by equation (\ref{rms}) and coincide with the ISCO's (this is the standard scenario of Keplerian accretion: shortly -- classic). 

\begin{widetext}
\begin{center}
\begin{minipage}{.33\linewidth}
\includegraphics[width=1\linewidth]{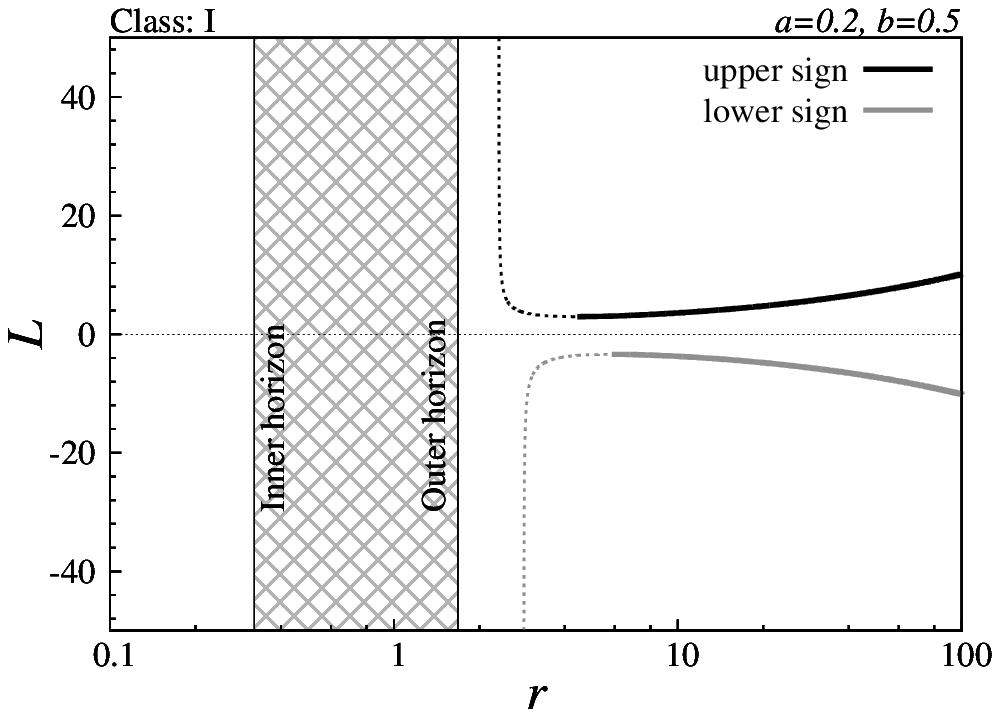}
\end{minipage}\hfill
\begin{minipage}{.33\linewidth}
\includegraphics[width=1\linewidth]{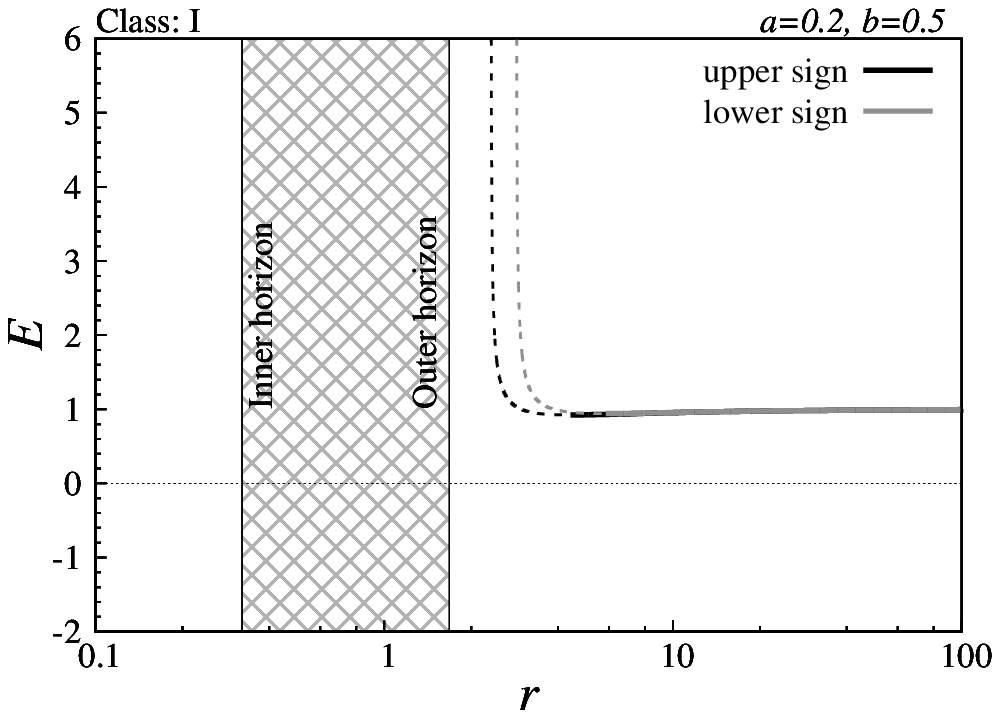}
\end{minipage}\hfill
\begin{minipage}{.33\linewidth}
\includegraphics[width=1\linewidth]{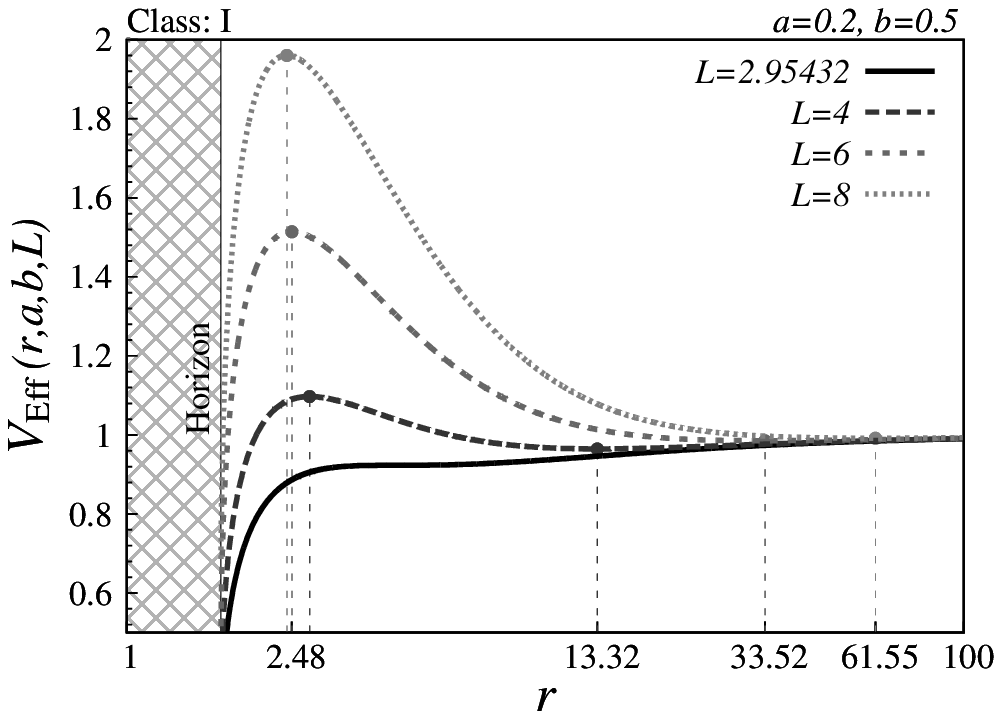}
\end{minipage}
\captionof{figure}{\label{ClI} $L$, $E$ and effective potential for class I.} 
\end{center}
\end{widetext}
\subsubsection{Class II}

Class of the~black hole spacetimes with two horizons, one stable and three unstable photon circular orbits and ergosphere. Notice that the~stable and unstable photon circular geodesic are located under the inner horizon, being thus irrelevant for the~Keplerian accretion. Border is given by line $b=a_{\mathrm{ph-ex}}(r=4b/3,b)$ and $b=1-a^2$ with intersection at point $(0.5,3/4)$ (point number (6) in figure (\ref{regionab2})) and line $b=0\, .$ 

Marginally stable orbits for test massive particles are given by the~inflexion point of the effective potential, coinciding with the ISCO's (classic).

\begin{widetext}
\begin{center}
\begin{minipage}{.33\linewidth}
\includegraphics[width=1\linewidth]{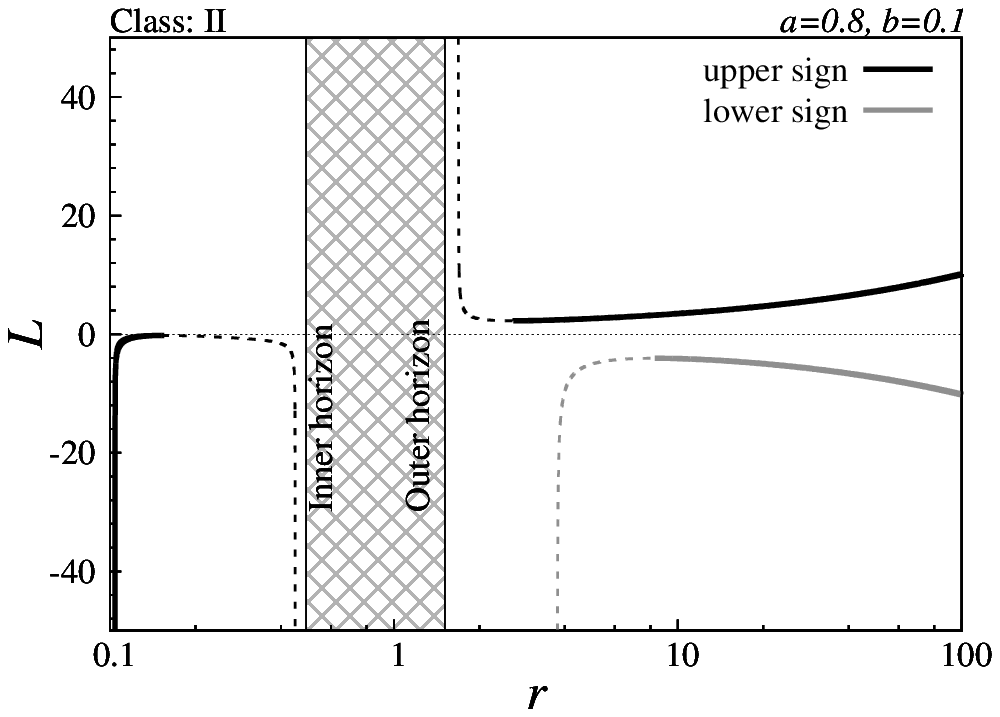}
\end{minipage}\hfill
\begin{minipage}{.33\linewidth}
\includegraphics[width=1\linewidth]{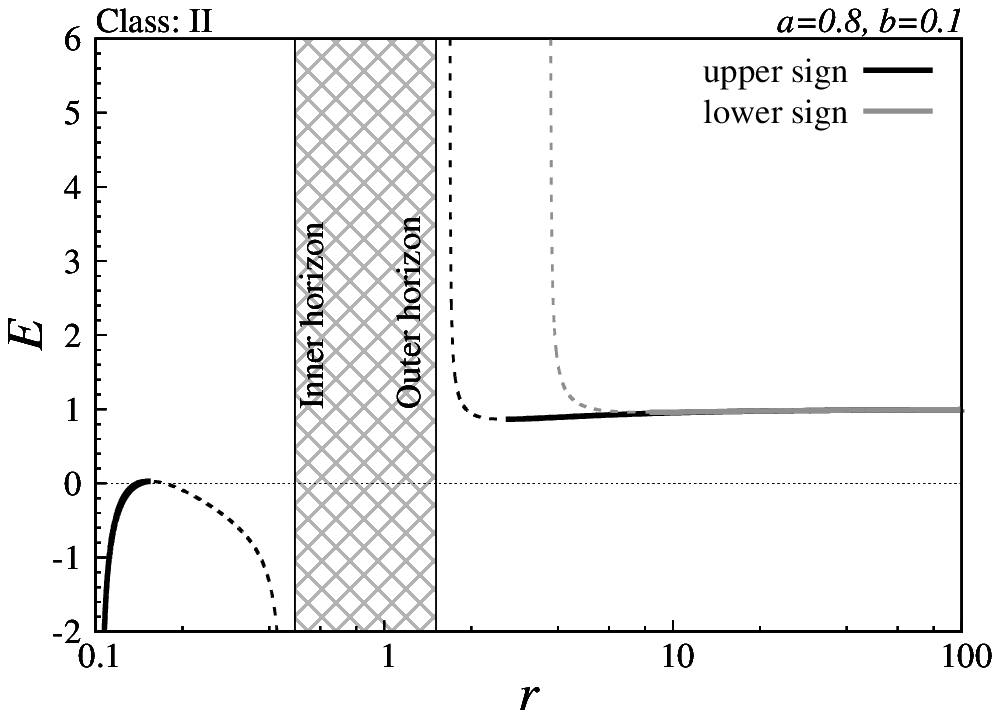}
\end{minipage}\hfill
\begin{minipage}{.33\linewidth}
\includegraphics[width=1\linewidth]{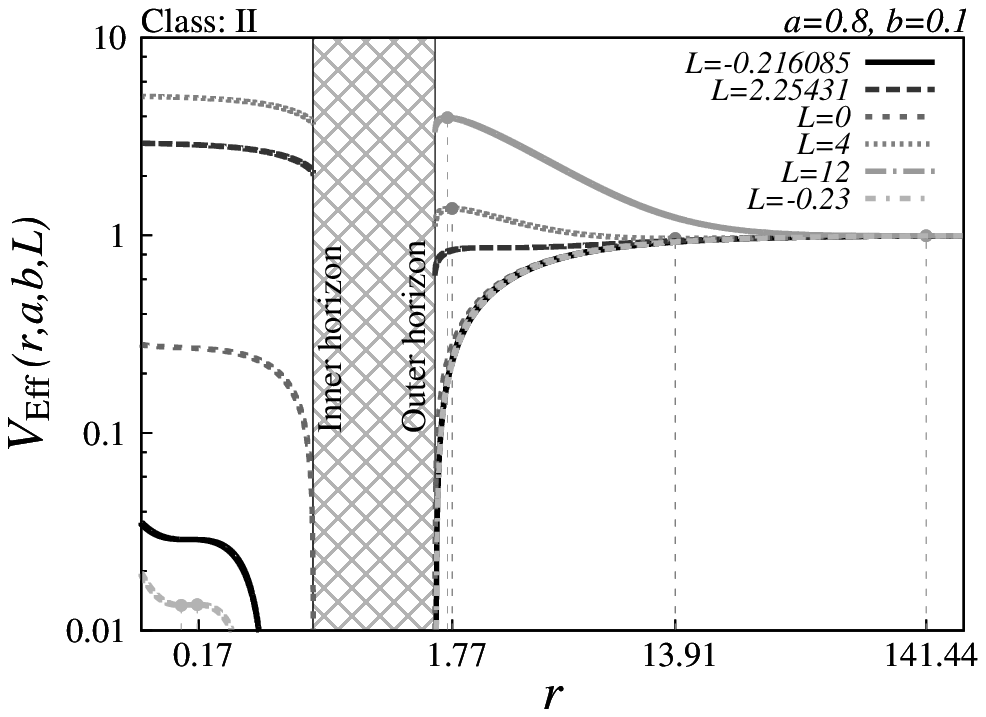}
\end{minipage}\hfill
\captionof{figure}{\label{ClII} $L$, $E$ and effective potential for class II.} 
\end{center}
\end{widetext}
\subsubsection{Class IIIa}

Class of the~naked singularity spacetimes with one stable and one unstable photon circular geodesics and ergosphere. The border of the~Class IIIa region is given by line $b=a_{\mathrm{ms(extr)}}$ and line $b=1$ with intersection points at $\left(2-\sqrt{3},1\right)=(0.268,1)$ (point number (5)) and $\left(2+\sqrt{3},1\right)=(3.732,1)$ (point number (9)). We have also marked the~point number (7) with coordinates $(\sqrt{0.5},0.5)$ where the~line $b=a_{\mathrm{ms(extr)}}$ and line $b=1-a^2$ touch and are tangent to each other. This theoretically means that effect of mining instability can be achieved for extremal Kerr--Newman black holes with spin parameter $a=\sqrt{0.5}$ and charge or braneworld tidal charge parameter $b=0.5$. But it occurs under the~event horizon. We have also marked the point number (8) with coordinates $(1,0.25)$ giving information on minimal amount of electric charge or braneworld tidal charge parameter $b$. 

Marginally stable orbit of massive test particles is in case of the~lower family circular geodesics given by inflexion point of the~effective potential and coincides with the ISCO (classic). 

\begin{widetext}
\begin{center}
\begin{minipage}{.33\linewidth}
\includegraphics[width=1\linewidth]{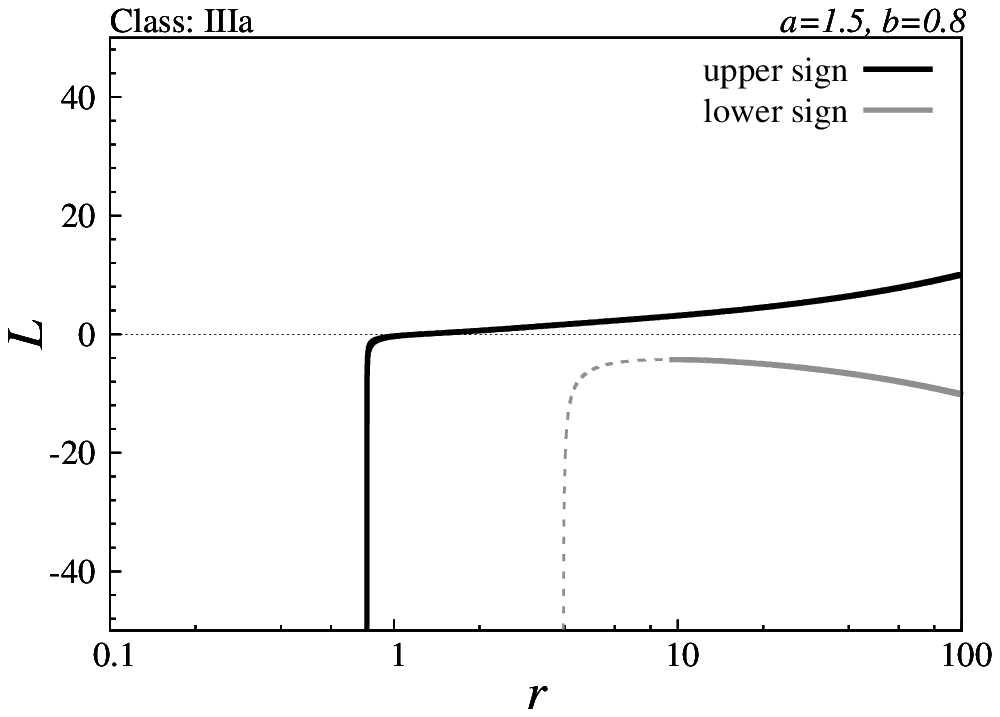}
\end{minipage}\hfill
\begin{minipage}{.33\linewidth}
\includegraphics[width=1\linewidth]{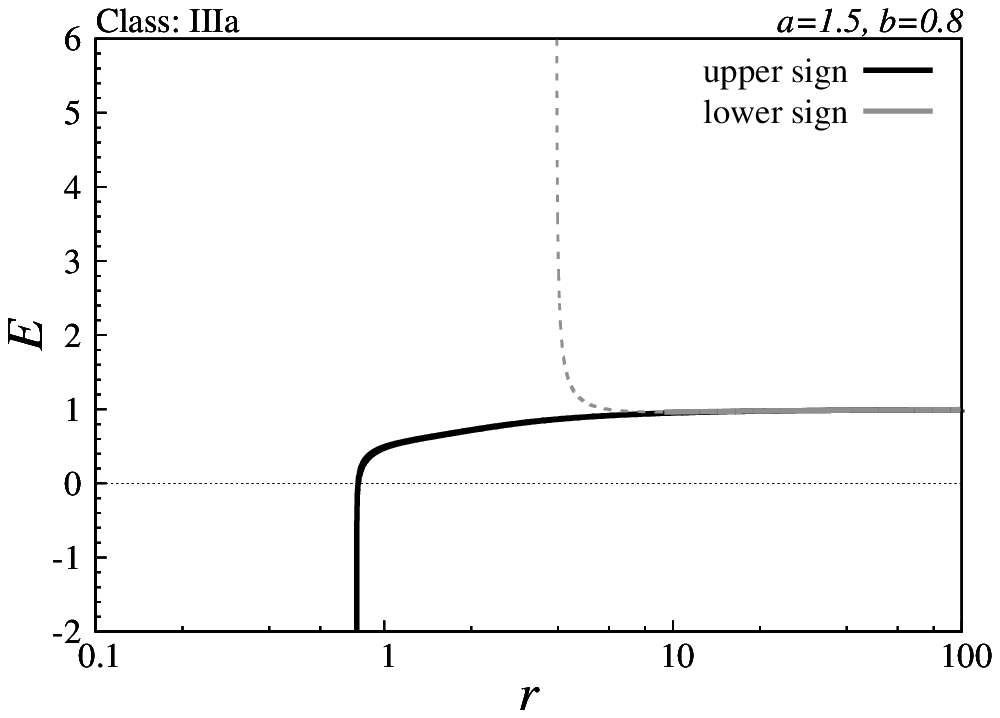}
\end{minipage}\hfill
\begin{minipage}{.33\linewidth}
\includegraphics[width=1\linewidth]{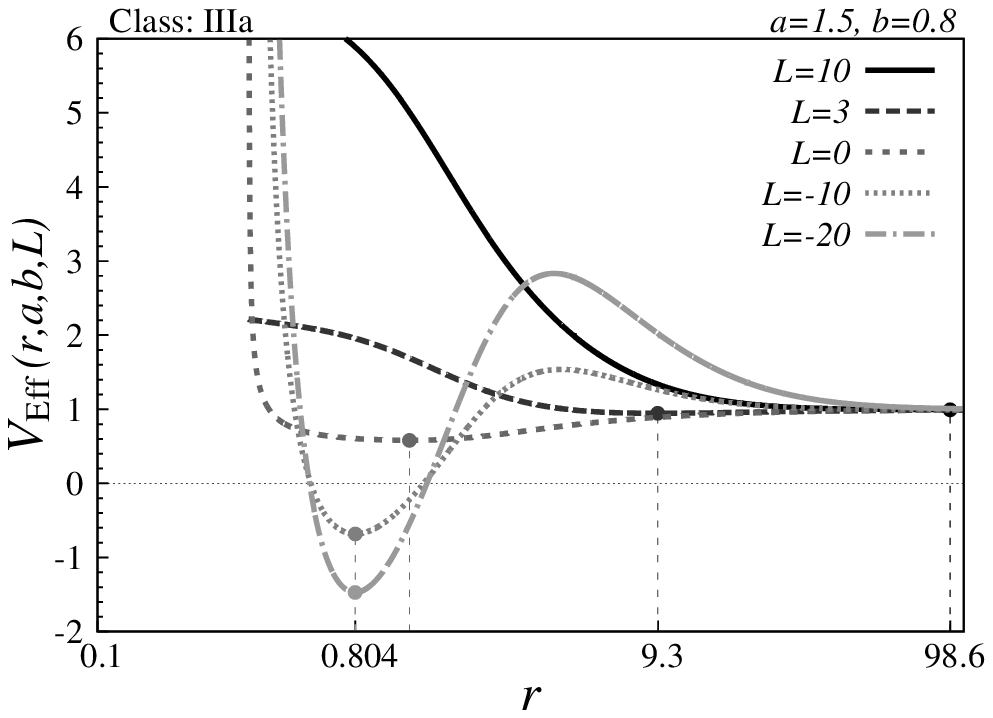}
\end{minipage}
\captionof{figure}{\label{ClIIIa} $L$, $E$ and effective potential for class IIIa.} 
\end{center}
\end{widetext}
In case of the~upper family circular geodesics, the~inflexion point of the~effective potential is not defined and the~sequence of minima of the~effective potential continues with decreasing specific energy and specific angular momentum of the~accreting matter down to the~stable photon orbit. This orbit can be therefore considered as ISCO of the~massive test particles. We have thus found an~infinitely deep gravitational well enabling theoretically an~unlimited mining of energy from the~naked singularity. In fact, such a mining instability could work only up to the~energy contained in the~naked singularity spacetime. We can expect that the~energy mining could work also in more realistic situations when the~naked singularity is removed and an~astrophysically more plausible superspinar is created by joining a regular (e.g. stringy) solution to the~Kerr--Newman spacetime at a radius overcoming the~outer radius of the~causality violation region \cite{Gim-Hor:2009:PhysLetB:,Stu-Sche:2012:CLAQG:}. The mining could work if the~matching radius of the~internal stringy spacetime and the~outer Kerr--Newman spacetime is smaller than the~radius of the~stable photon orbit related to the~mining instability of the~Class IIIa spacetimes. 

For completeness, we give in this case also the~locally measured (LNRF) specific energy of the~upper family circular geodesics. As shown in Figure \ref{ELNRF}, the specific energy $E_{LNRF}$ diverges, along with the covariant specific energy $E$ as the~orbit approaches the~limiting photon circular orbit. On the~other hand, Figure \ref{ElE} clearly demonstrates that the~ratio $|E|/E_{LNRF}$ remains finite while the~orbits approach the~location corresponding to the~stable photon circular orbit. 

\begin{widetext}
\begin{center}
\begin{minipage}{.49\linewidth}
\includegraphics[width=1\linewidth]{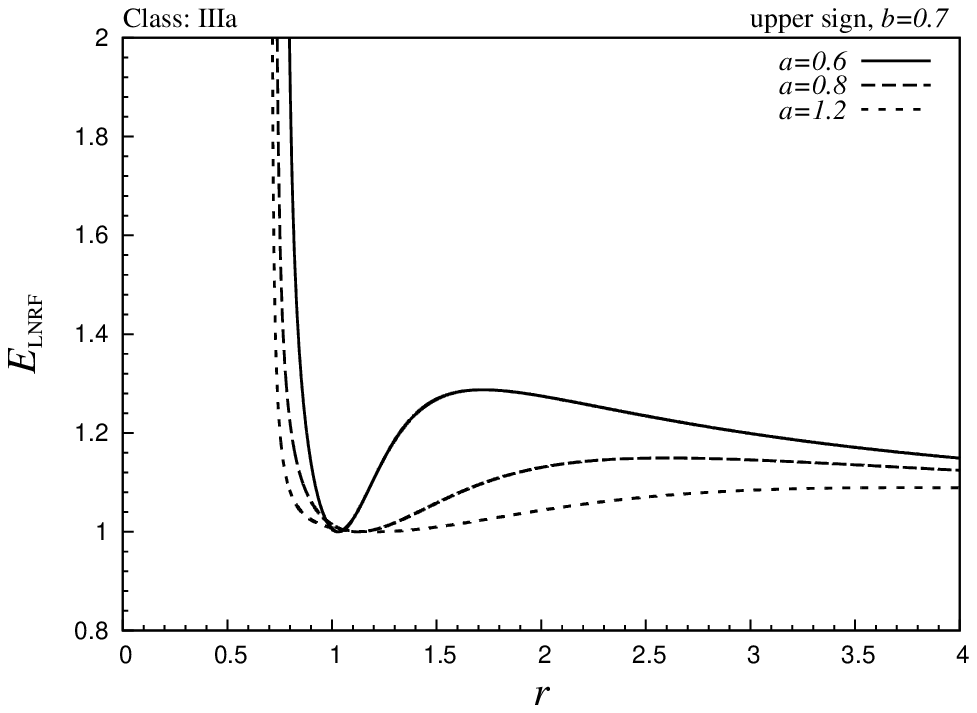}
\end{minipage}\hfill
\begin{minipage}{.48\linewidth}
\includegraphics[width=1\linewidth]{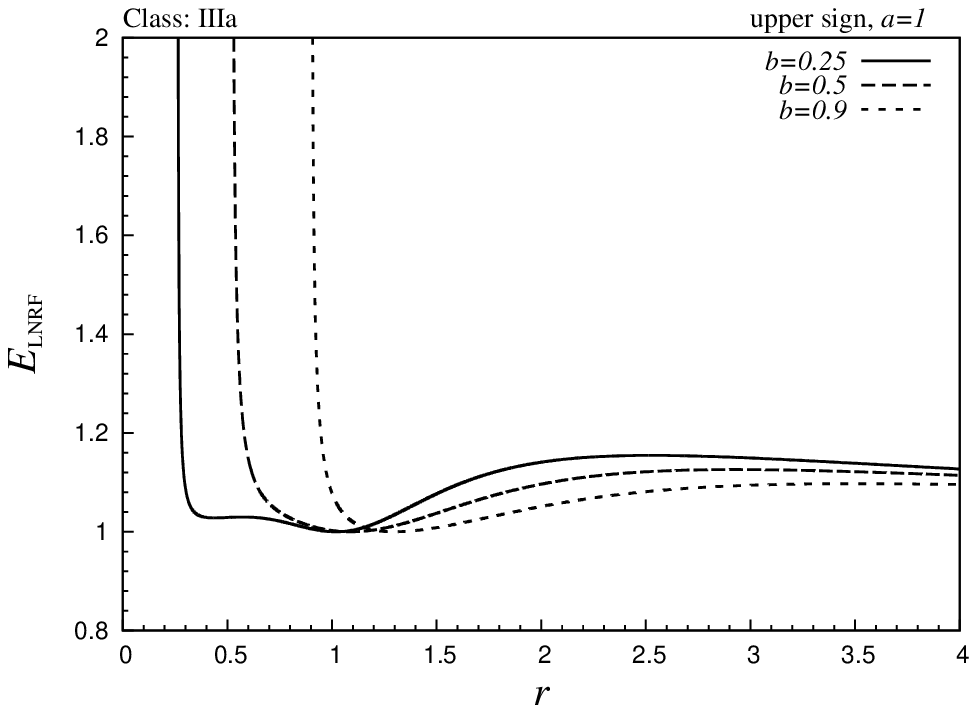}
\end{minipage}
\captionof{figure}{\label{ELNRF} Energy measured by the~LNRF observers (upper sign family orbits only) in the mining unstable Kerr-Newman spacetimes of class IIIa. The~energy diverges at the~radius of the~stable photon circular orbit.} 
\end{center}
\end{widetext}

Of course, the mining instability could work only if the~assumption of the~test particle motion of the~accreting matter is satisfied. Therefore, the~assumption requires validity of the~relation 
\begin{equation}
   |\tilde{E}| \ll M ,  
\end{equation}
the~covariant energy of the~particle (accreting matter) has to be much smaller than the~naked singularity mass parameter $M$. Of course, the~issue of the~mining instability and the~related interaction of the~mining unstable Kerr--Newman naked singularity (Kerr--Newman superspinar) and the~accreting mass is much more complex and deserves a~more detailed study. 

\begin{widetext}
\begin{center}
\begin{minipage}{.49\linewidth}
\includegraphics[width=1\linewidth]{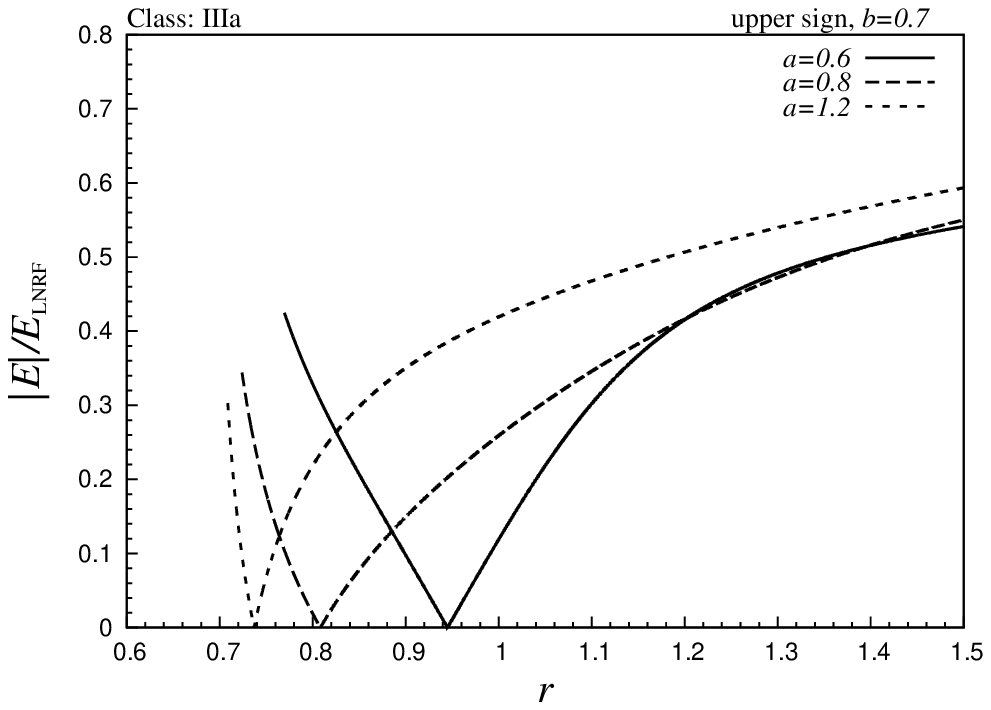}
\end{minipage}\hfill
\begin{minipage}{.48\linewidth}
\includegraphics[width=1\linewidth]{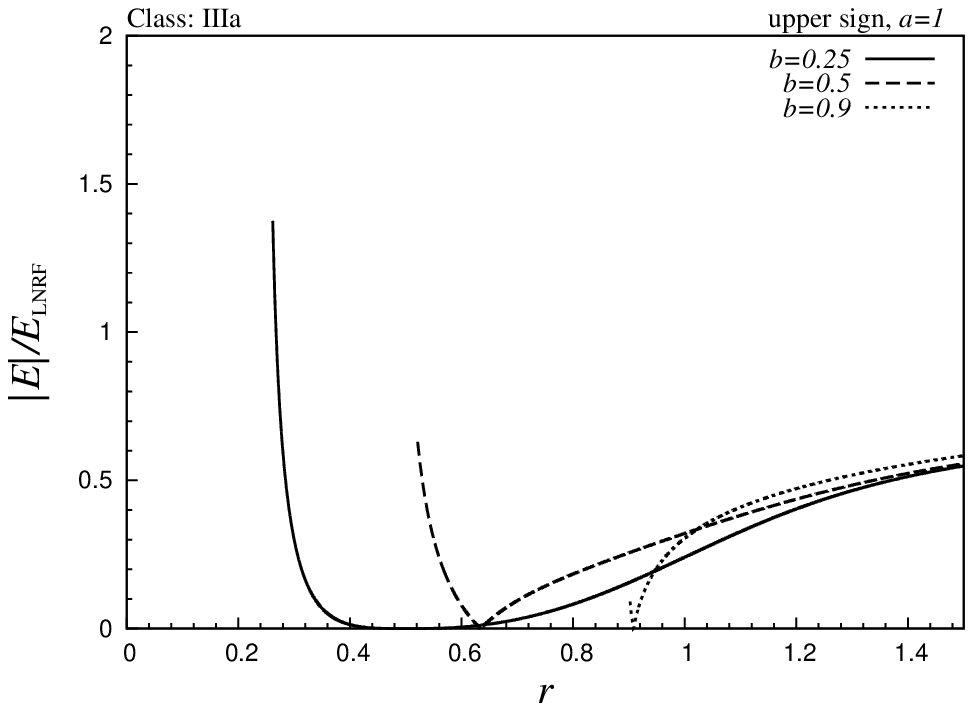}
\end{minipage}
\captionof{figure}{\label{ElE} Absolute value of covariant energy $/E/$ divided by LNRF energy (for the upper sign family orbits) in the mining unstable Kerr-Newman spacetimes of class IIIa. The~fraction is defined down to the radius of the stable circular photon orbit and at this point it has a finite value.} 
\end{center}
\end{widetext}

\subsubsection{Class IIIb}

Class of naked singularity spacetimes with one stable and one unstable photon circular orbit and ergosphere. 
In the~parameter space $b-a$ the area related to this class is not compact and disintegrates into two separated areas. The first area is infinitely large, its border is given by the~line $b=a_{\mathrm{ms(extr)}}$, line $b=1-a^2$, line $b=1$ and $b=0$ with intersections at the~point $(\sqrt{0.5},0.5)$ (point number (7)) and $\left(2+\sqrt{3},1\right)$ (point number (9)). The second area is compact and finite. Its border is given by the~line $b=a_{\mathrm{ms(extr)}}$, line $b=a_{\mathrm{ph-ex}}(r=4b/3,b)$, line $b=1-a^2$ and line $b=1$ with intersection points number (4), (5) and (7). It is not obvious from figures, but $b=a_{\mathrm{ms(extr)}}$ and $b=a_{\mathrm{ph-ex}}(r=4b/3,b)$ do not intersect. 

Marginally stable orbits of both the~lower and upper family of circular geodesics are given by the~inflexion point of the~effective potential and coincide with ISCO's (classic). Notice that in this case the~sequence of the~upper family orbits with descending specific energy $E$ and specific angular momentum $L$ is interrupted by a sequence where both $E$ and $L$ increase with decreasing radius, corresponding thus to the~unstable geodesics. In this case, the infinitely deep gravitational well still exists, but the Keplerian accretion sequence is interrupted and this gravitational well cannot be applied in an~astrophysically natural accretion process. Nevertheless, it is still possible to use this gravitational well, if matter with appropriate initial conditions (values of the motion constants), enabling starting of the mining instability, could appear close to the~naked singularity. 

\begin{widetext}
\begin{center}
\begin{minipage}{.33\linewidth}
\includegraphics[width=1\linewidth]{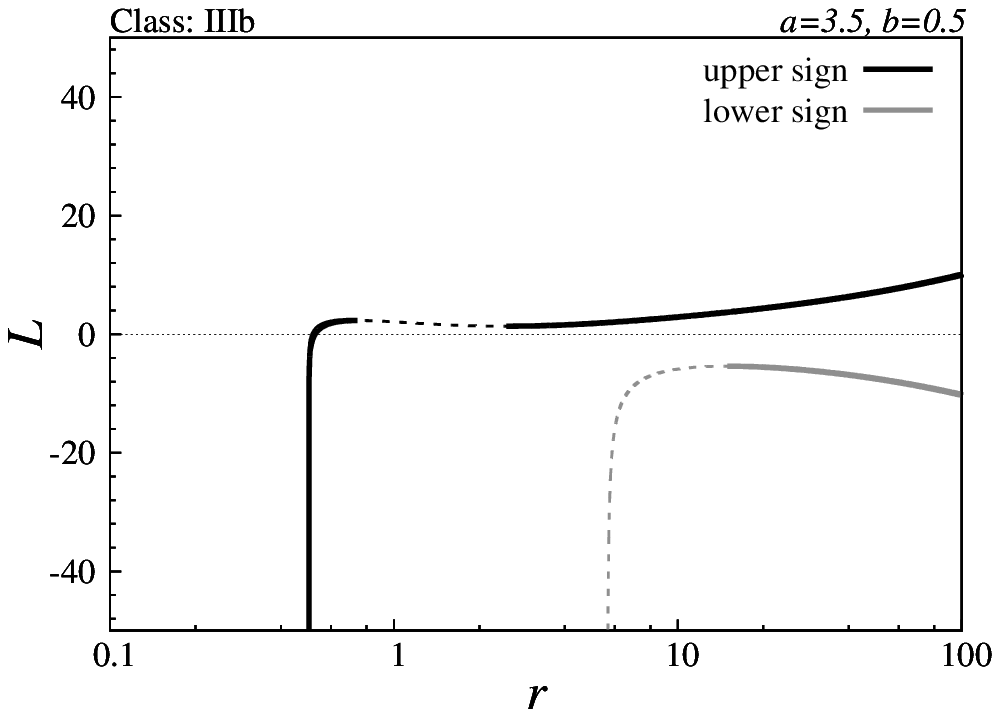}
\end{minipage}\hfill
\begin{minipage}{.33\linewidth}
\includegraphics[width=1\linewidth]{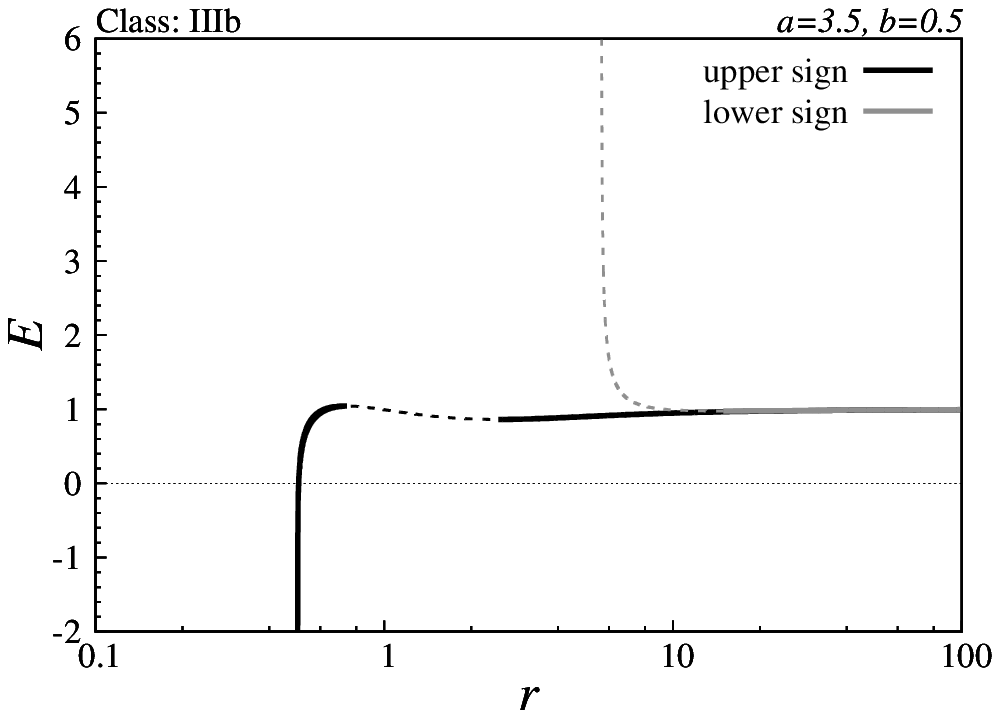}
\end{minipage}\hfill
\begin{minipage}{.33\linewidth}
\includegraphics[width=1\linewidth]{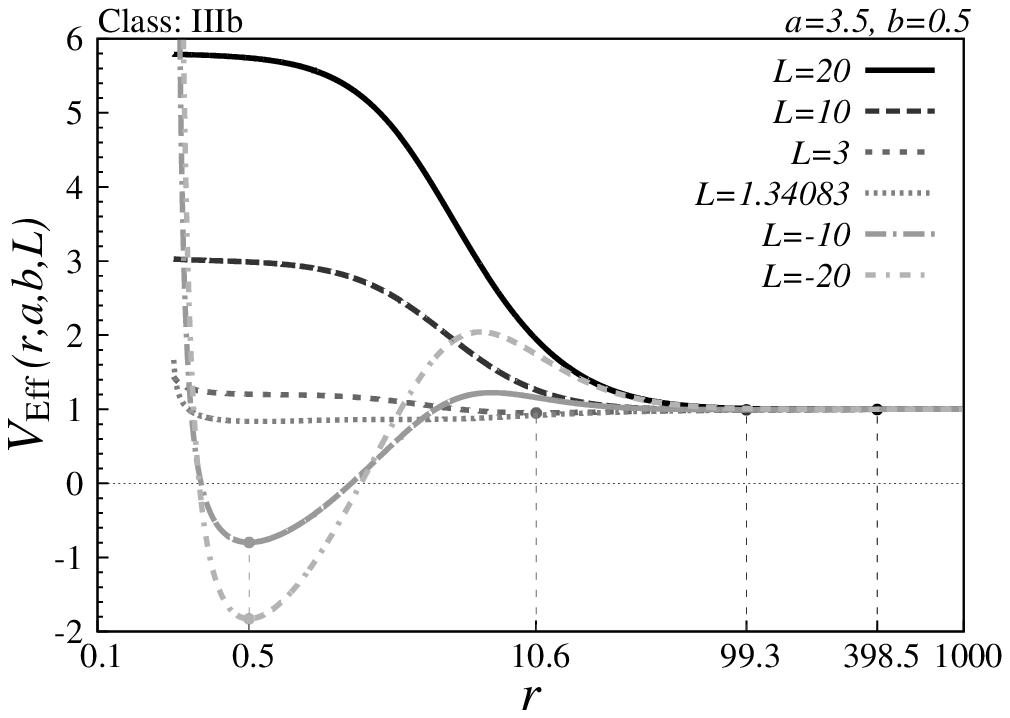}
\end{minipage}
\captionof{figure}{\label{ClIIIb} $L$, $E$ and effective potential for class IIIb.} 
\end{center}
\end{widetext}

\subsubsection{Class IVa} 

Class of naked singularity spacetimes with one stable and one unstable photon circular orbit. Class is without ergosphere. Border of this class in the~spacetime parameter space is given by line $b=a_{\mathrm{ph-ex}}(r=4b/3,b)$ and $b=a_{\mathrm{ms(extr)}}$ with intersection points (3),(5),(9) and (10).

For the lower family circular geodesics, the marginally stable orbits defined by the~inflexion point of the~effective potential occur (classic). For the~upper family circular geodesics marginally stable orbit is not defined, and the~ISCO is located at $r=b$ as it is always for all classes with $b>1$. A sequence of stable circular geodesics with sharply increasing specific energy occurs near (above) the radius $r=b$, approaching the stable photon circular orbit. (Such sequences of stable circular orbits were discussed in \cite{Stu-Sche:2014:CLAQG:}.)

Note that probability we are actually living in a~spacetime with the~braneworld tidal charge parameter greater than one is very small \cite{Kot-Stu-Tor:2008:CLAQG:,Boh-etal:2008:CLAQG:}. 

\begin{widetext}
\begin{center}
\begin{minipage}{.33\linewidth}
\includegraphics[width=1\linewidth]{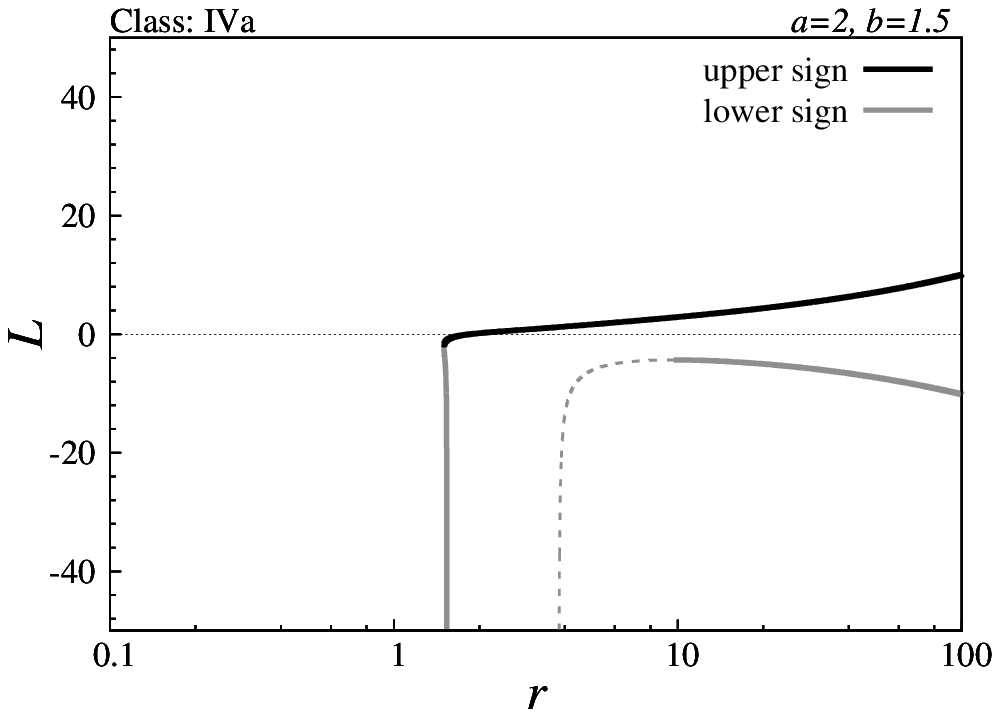}
\end{minipage}\hfill
\begin{minipage}{.33\linewidth}
\includegraphics[width=1\linewidth]{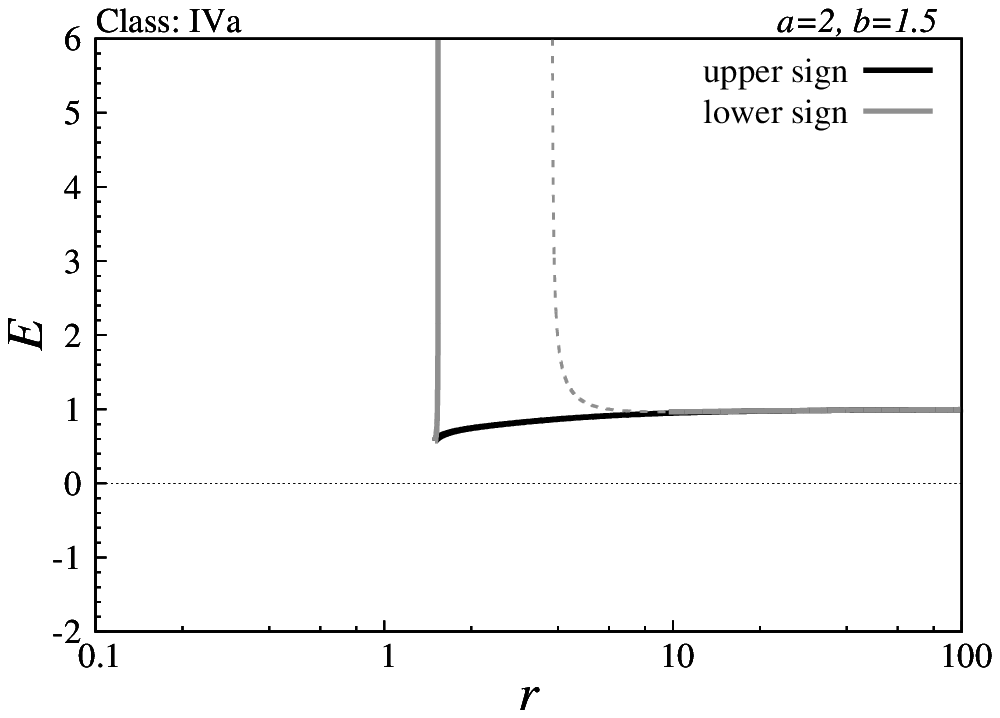}
\end{minipage}\hfill
\begin{minipage}{.33\linewidth}
\includegraphics[width=1\linewidth]{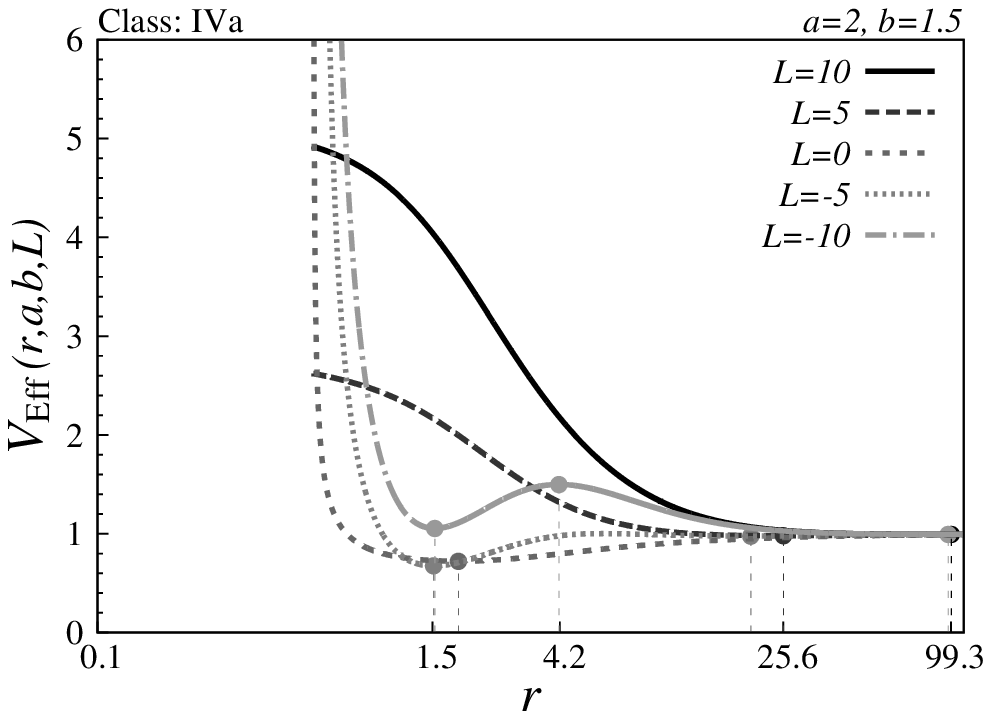}
\end{minipage}
\captionof{figure}{\label{ClIVa} $L$, $E$ and effective potential for class IVa.} 
\end{center}
\end{widetext}

\subsubsection{Class IVb}

Class of naked singularity spacetimes with one stable and one unstable photon circular orbit. These spacetimes are without ergosphere. In the~parameter space $b-a$, this class is not compact and disintegrates into two separated areas. 
The~first area is infinitely large, the~border is given by the~line $b=a_{\mathrm{ph}}(4b/3,b)$, line $b=a_{\mathrm{ms(extr)}}$ and line $b=1$ with intersection points (9) and (10). The~second area is finite and its border is given by same lines and intersection points (2), (3), (4) and (5). 

For both the upper and lower families of circular geodesics, the~marginally stable orbits of test massive particles are given by the~inflexion point of the~effective potential, governing the~sequence of geodesics related to the~standard Keplerian accretion (classic). There is additional internal sequence of stable circular geodesic, with the ISCO located at $r=b$, as it is always for all classes with $b>1$. This sequence approaches the stable photon circular geodesic at the outer edge. 

\begin{widetext}
\begin{center}
\begin{minipage}{.33\linewidth}
\includegraphics[width=1\linewidth]{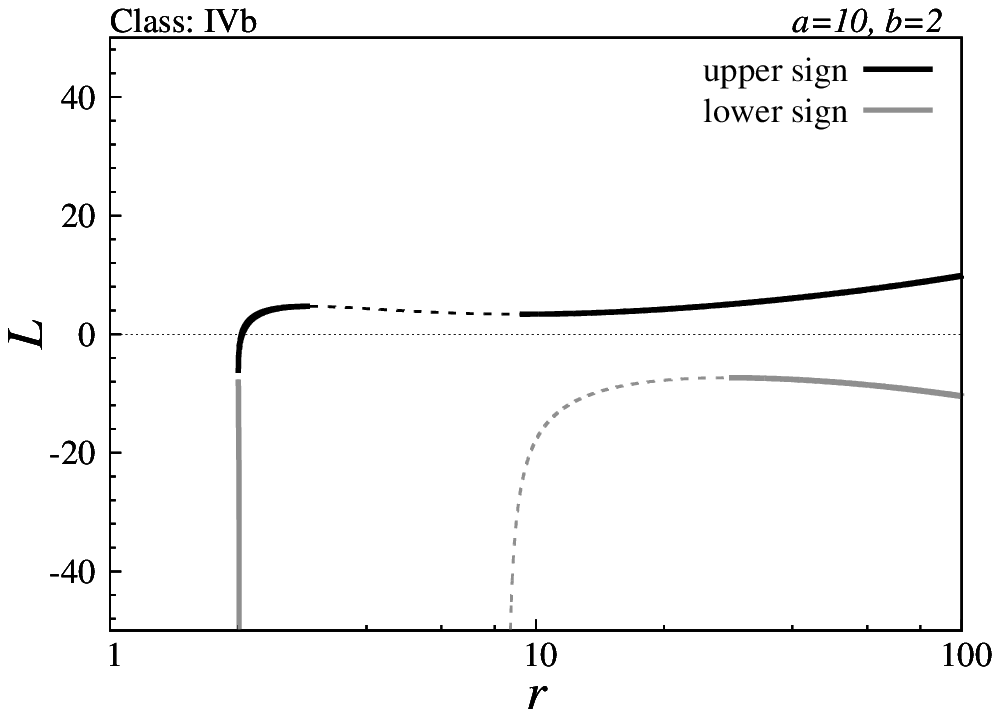}
\end{minipage}\hfill
\begin{minipage}{.33\linewidth}
\includegraphics[width=1\linewidth]{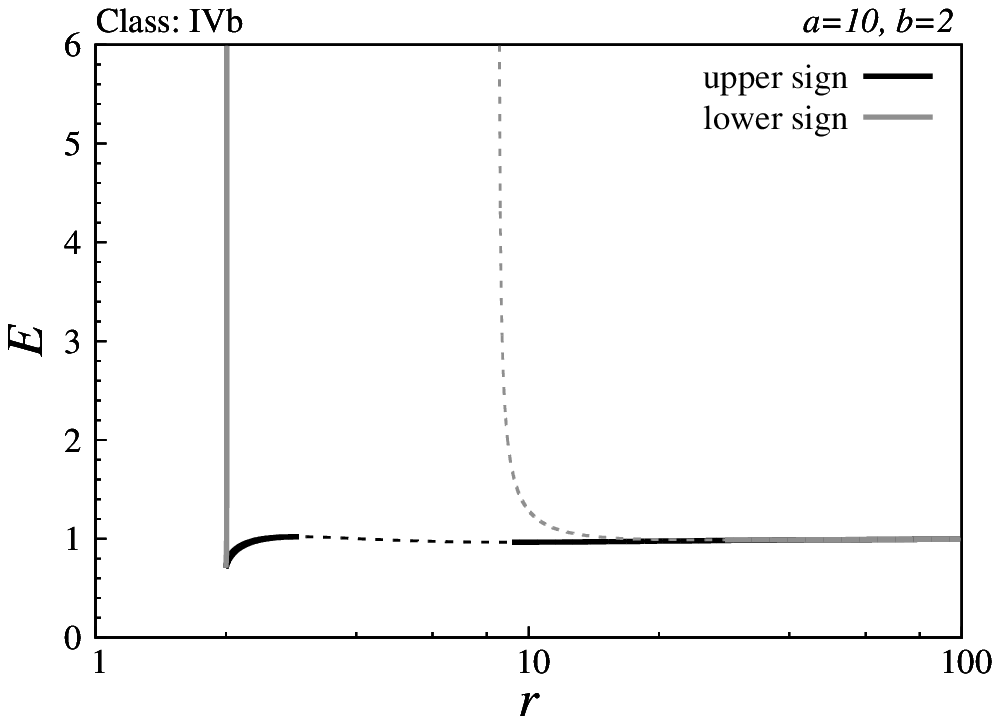}
\end{minipage}\hfill
\begin{minipage}{.33\linewidth}
\includegraphics[width=1\linewidth]{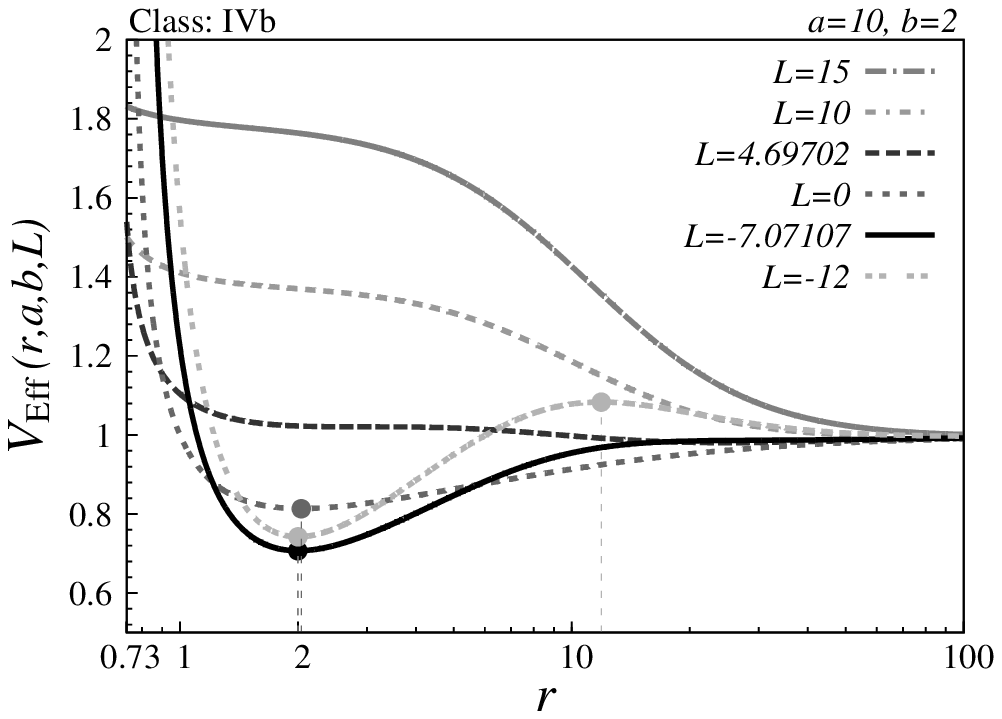}
\end{minipage}
\captionof{figure}{\label{ClIVb} $L$, $E$ and effective potential for class IVb.} 
\end{center}
\end{widetext}

\subsubsection{Class Va}

Class of naked singularity spacetimes with no stable or unstable photon circular orbit. These spacetimes are also without ergosphere. 
The~border of the~Class Va region in the~parameter space is given by the line $b=a_{\mathrm{ms(extr)}}$ and line $a=0$, with intersection point (1).  

For both the lower and upper family circular geodesics, the~marginally stable orbits are not defined. The circular geodesics are only stable, the ISCO's are located at $r=b$, as for all the spacetime classes with $b>1$.

\begin{widetext}
\begin{center}
\begin{minipage}{.33\linewidth}
\includegraphics[width=1\linewidth]{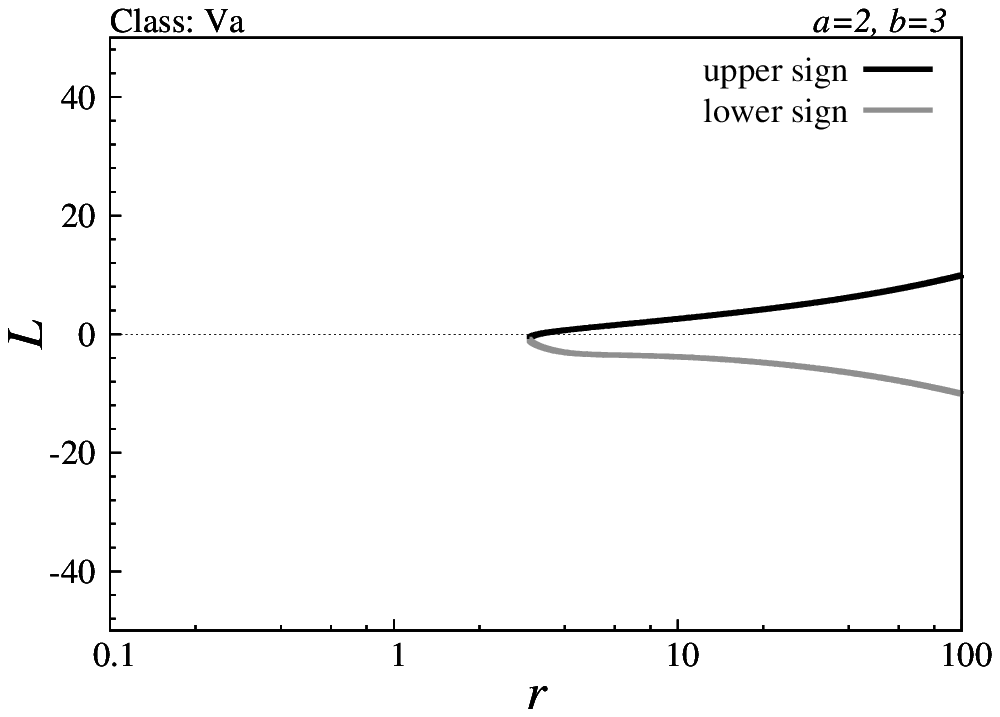}
\end{minipage}\hfill
\begin{minipage}{.33\linewidth}
\includegraphics[width=1\linewidth]{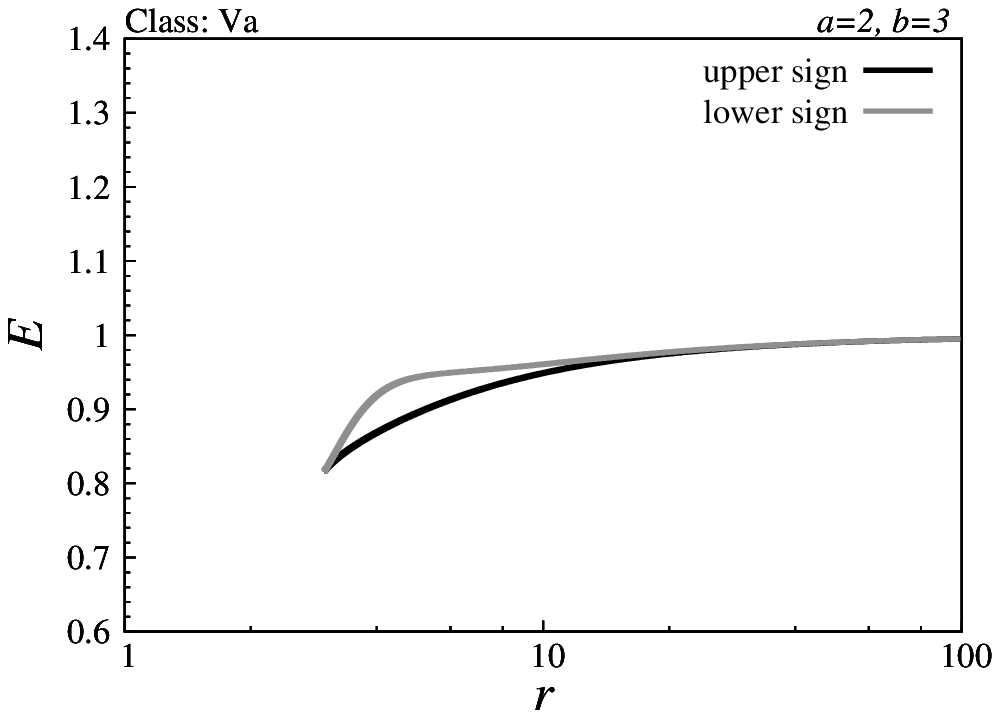}
\end{minipage}\hfill
\begin{minipage}{.33\linewidth}
\includegraphics[width=1\linewidth]{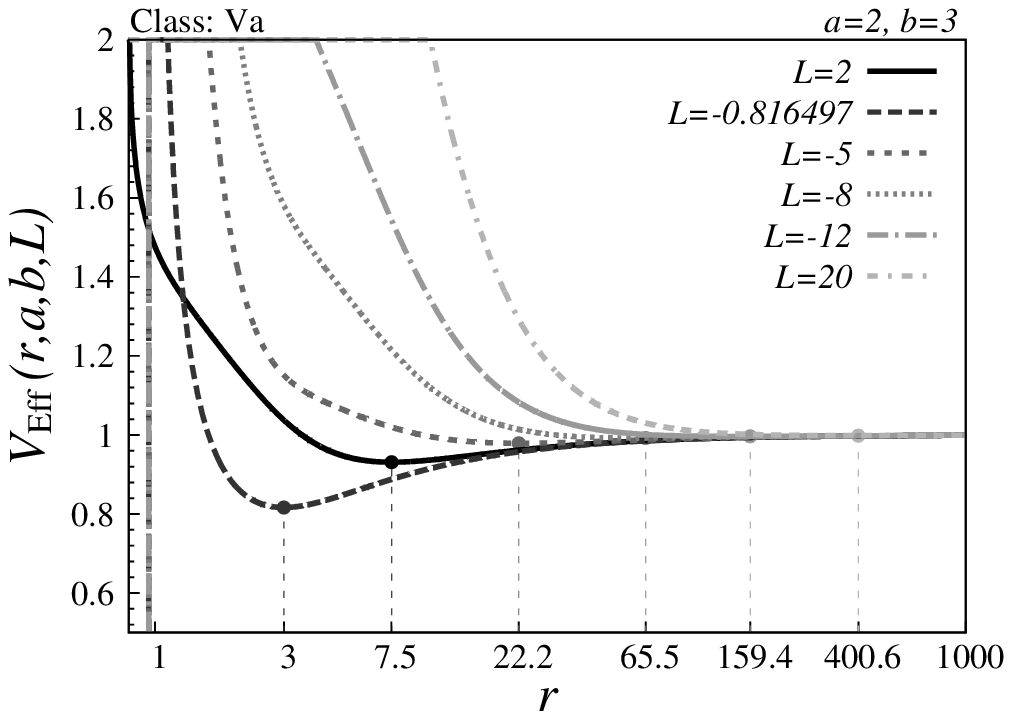}
\end{minipage}
\captionof{figure}{\label{ClVa} $L$, $E$ and effective potential for class Va.} 
\end{center}
\end{widetext}

\subsubsection{Class Vb}

Class of naked singularity spacetime with no stable or unstable photon circular orbit. These spacetimes are also without ergosphere. 
The~border of the~Class Vb region in the~parameter space is given by the line $b=a_{\mathrm{ms(extr)}}$, and line $b=a_{\mathrm{ph}}(4b/3,b)$, with intersection points (1), (3) and (10). The class is infinitely extended in the parameter space.  

The upper family circular geodesics are stable only, finishing at the ISCO located at $r=b$. The marginally stable orbit exists for the~lower family orbits, giving the limit of the standard Keplerian accretion. The lower family orbits continue downwards by sequence of unstable orbits and finally stable orbits finishing at $r=b$.

\begin{widetext}
\begin{center}
\begin{minipage}{.33\linewidth}
\includegraphics[width=1\linewidth]{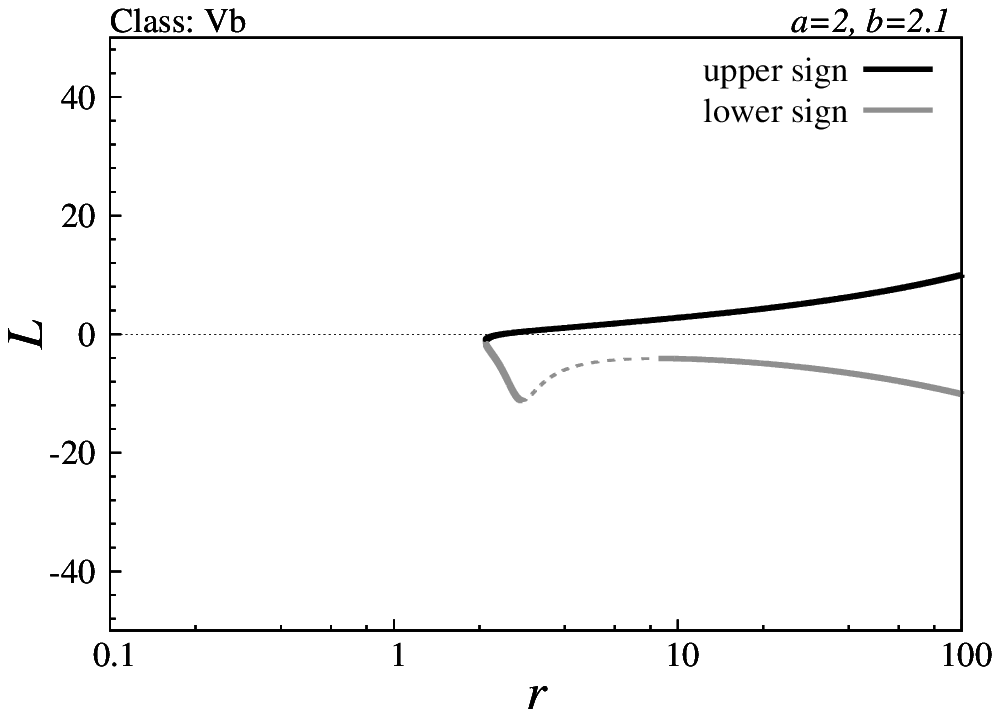}
\end{minipage}\hfill
\begin{minipage}{.33\linewidth}
\includegraphics[width=1\linewidth]{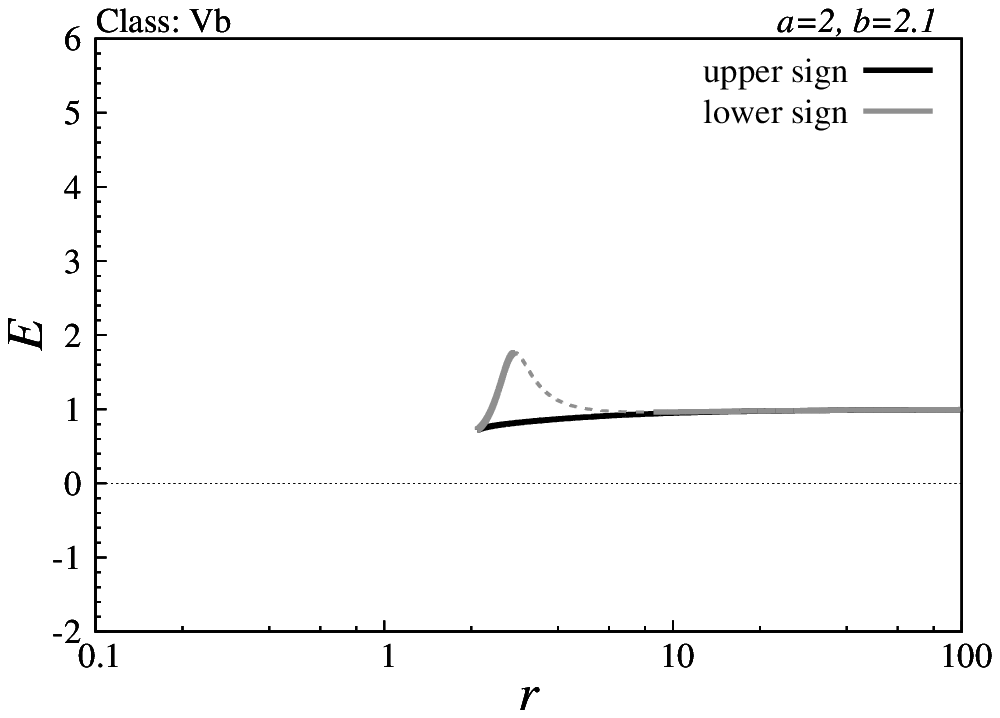}
\end{minipage}\hfill
\begin{minipage}{.33\linewidth}
\includegraphics[width=1\linewidth]{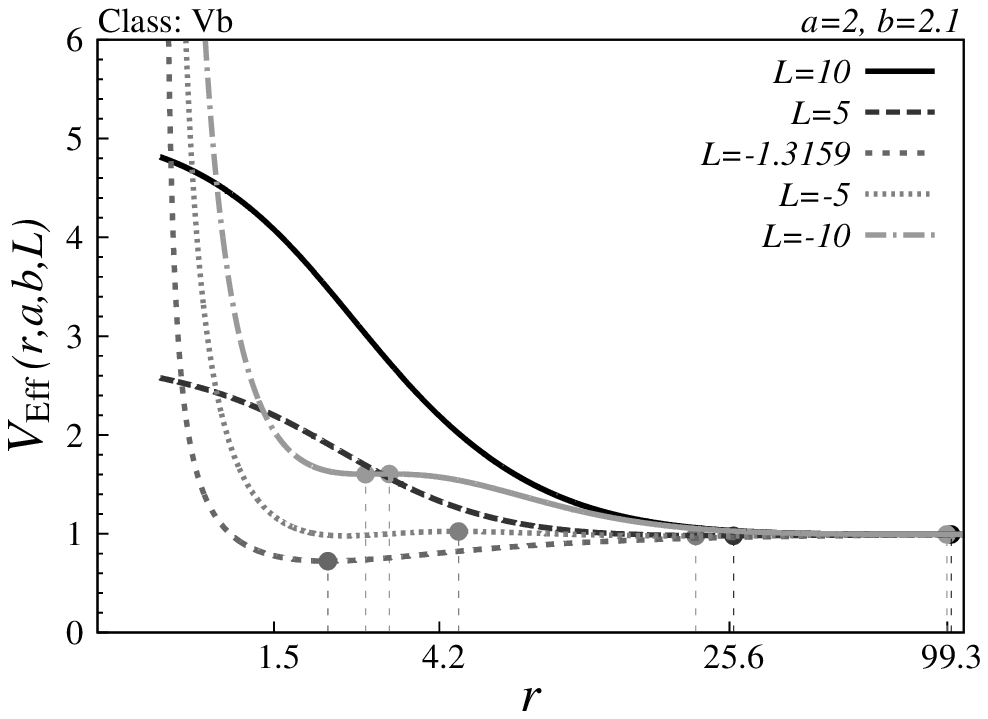}
\end{minipage}
\captionof{figure}{\label{ClVb} $L$, $E$ and effective potential for class Vb.} 
\end{center}
\end{widetext}

\subsubsection{Class Vc}

Class of naked singularity spacetimes with no stable or unstable photon circular orbit. These are again spacetimes without ergosphere. 
In the~parameter space $b-a$ this class disintegrates into two separated areas. The first one is infinitely extended, its border is given by the~line $b=a_{\mathrm{ph-ex}}(r=4b/3,b)$ and line $b=a_{\mathrm{ms(extr)}}$, with the intersection point (10). The second area is finite and its border is given by same line $b=a_{\mathrm{ph-ex}}(r=4b/3,b)$, line $b=a_{\mathrm{ms(extr)}}$ and line $a=0$, with intersection points (1), (2) and (3). 

The marginally stable orbit exists for both the~lower and upper family circular geodesics, giving thus the standard limit of the Keplerian accretion (classic). Both the lower and upper family orbits continue downwards by sequence of unstable orbits and finally stable orbits finishing at $r=b$. 

\begin{widetext}
\begin{center}
\begin{minipage}{.33\linewidth}
\includegraphics[width=1\linewidth]{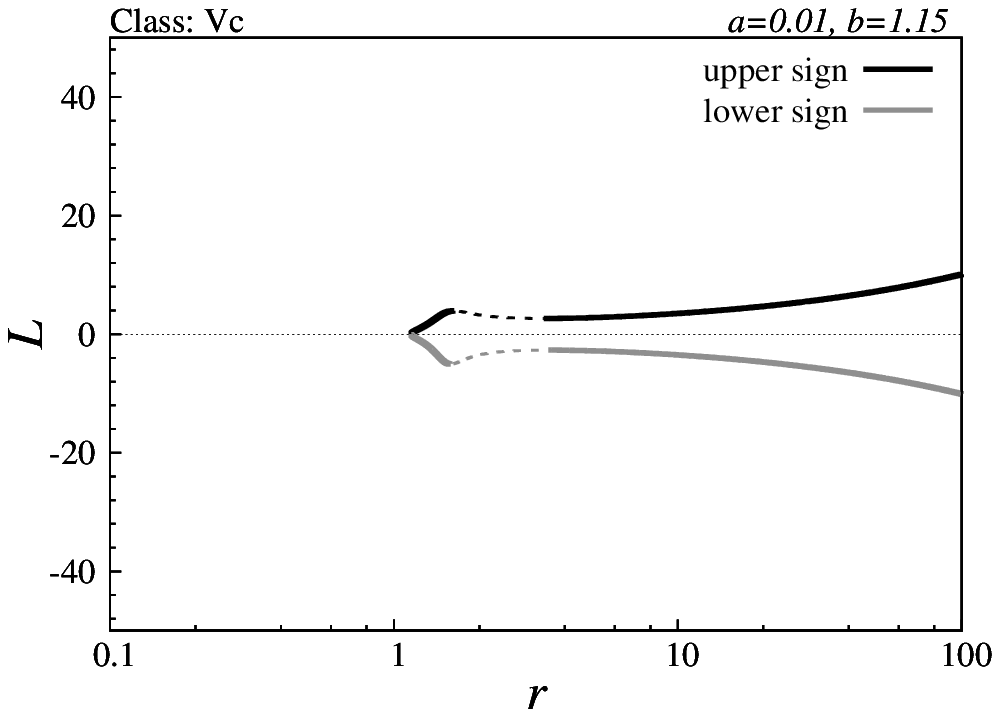}
\end{minipage}\hfill
\begin{minipage}{.33\linewidth}
\includegraphics[width=1\linewidth]{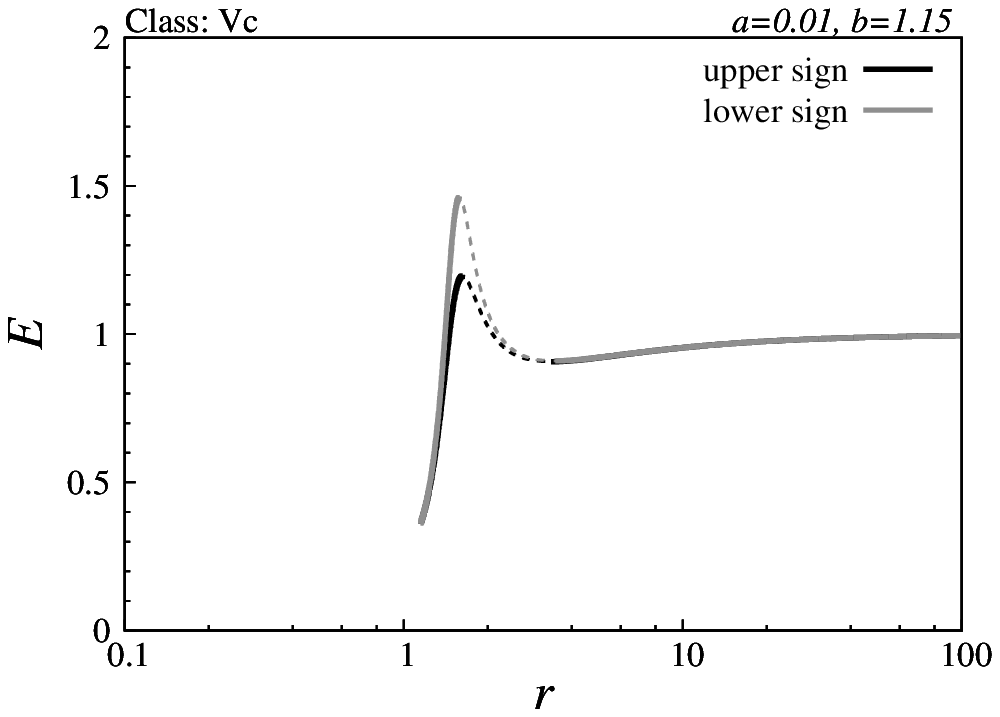}
\end{minipage}\hfill
\begin{minipage}{.33\linewidth}
\includegraphics[width=1\linewidth]{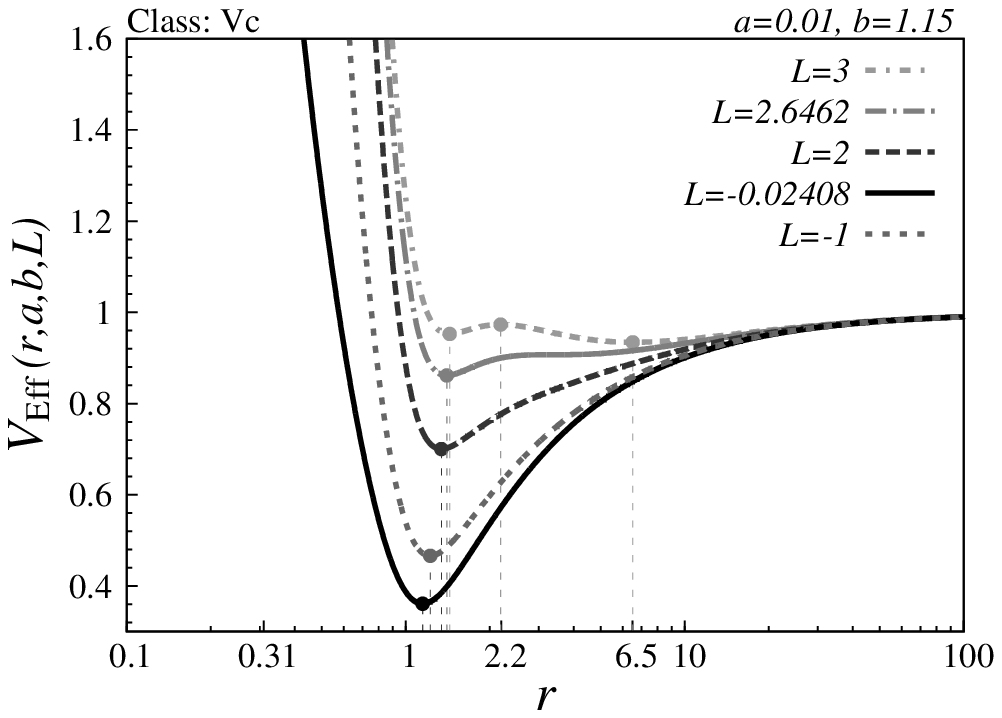}
\end{minipage}
\captionof{figure}{\label{ClVc} $L$, $E$ and effective potential for class Vc.} 
\end{center}
\end{widetext}

\subsubsection{Class VI}

Class of naked singularity spacetimes with two stable and two unstable photon circular orbits and ergosphere. 
In the~parameter space the area of the~Class VI spacetimes has the boundary given by the~line $b=a_{\mathrm{ph-ex}}(r=4b/3,b)$, line $b=1-a^2$ and the~line $b=1$, with the~intersection points (4) and (6). 

For both the lower and upper family of circular geodesics the~marginally stable orbit exists, giving thus the~inner edge of the standard Keplerian accretion. The~upper family orbits have also a very narrow region of stable circular orbits near the~radius $r=b$, starting at the~stable circular photon orbit. 

\begin{widetext}
\begin{center}
\begin{minipage}{.33\linewidth}
\includegraphics[width=1\linewidth]{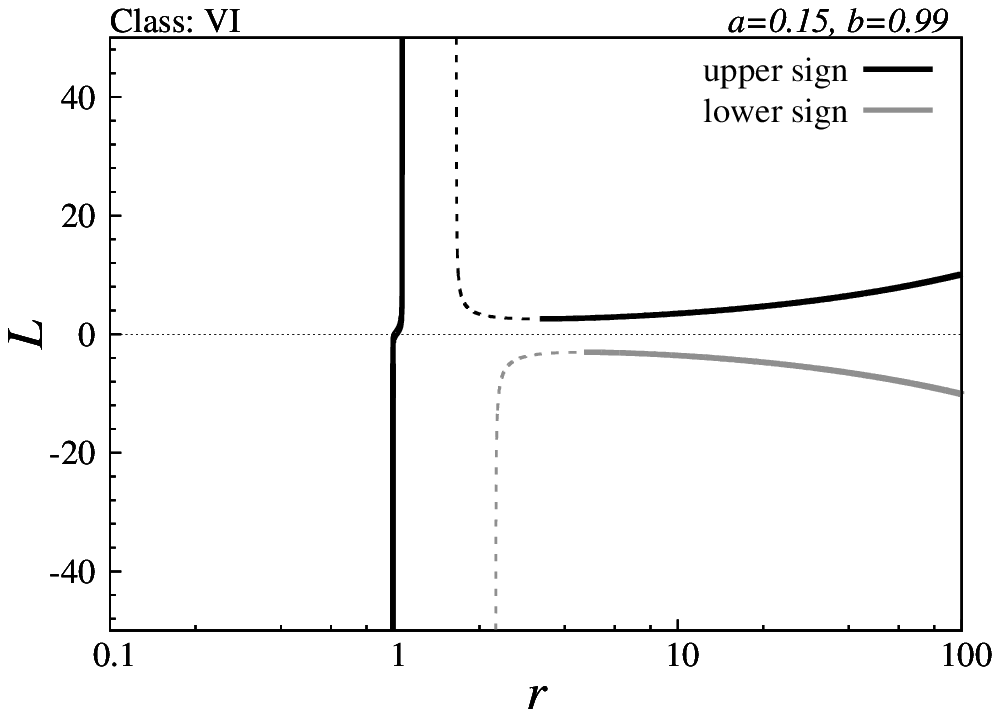}
\end{minipage}\hfill
\begin{minipage}{.33\linewidth}
\includegraphics[width=1\linewidth]{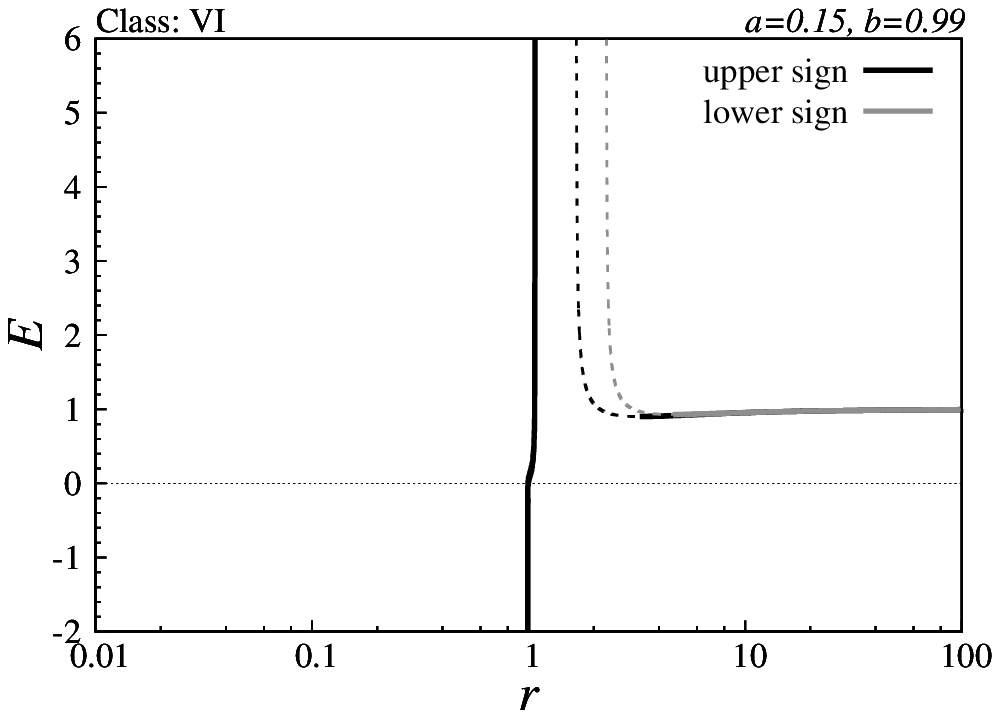}
\end{minipage}\hfill
\begin{minipage}{.33\linewidth}
\includegraphics[width=1\linewidth]{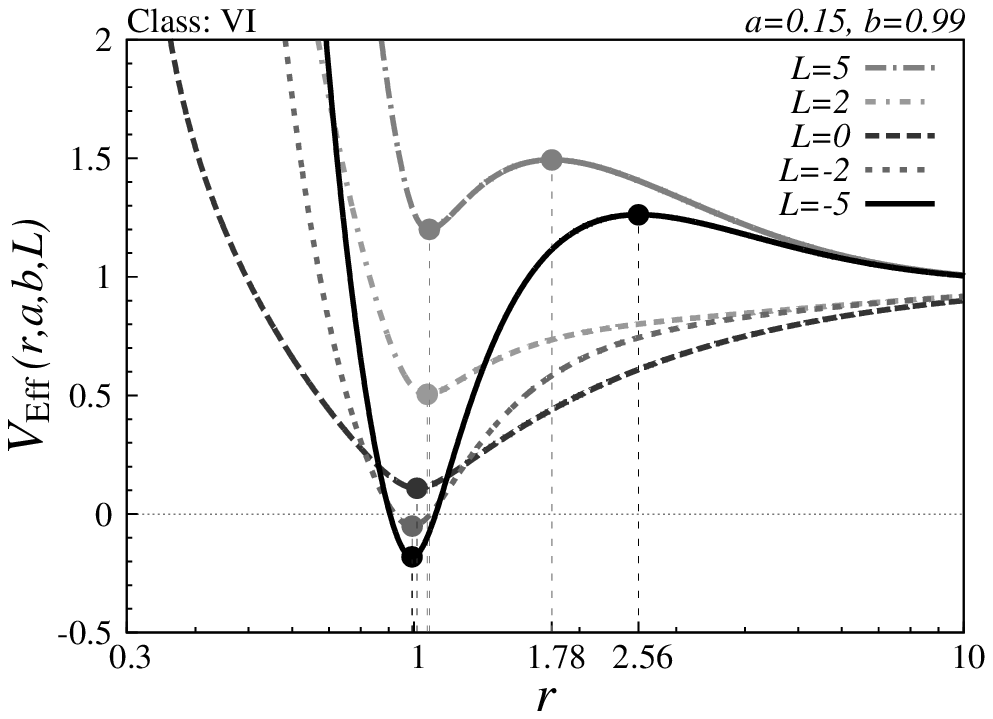}
\end{minipage}
\captionof{figure}{\label{ClVI} $L$, $E$ and effective potential for class VI.} 
\end{center}
\end{widetext}

\subsubsection{Class VII}

Class of naked singularity spacetimes with two stable and two unstable photon circular orbits, having no ergosphere. 
In the~parameter space the~area of the~Class VII spacetimes has the~boundary given by the~line $b=a_{\mathrm{ph-ex}}(r=4b/3,b)$, line $a=0$ and the~line $b=1$, with the~intersection points (2) and (4). 

For both the lower and upper family circular geodesics, the~marginally stable orbit exists giving thus the~edge of the standard Keplerian accretion. Further, both the lower and upper family orbits have the~ISCO at $r=b$ where sequence of stable orbits finishes, starting for each family at the~related photon circular geodesic. 

\begin{widetext}
\begin{center}
\begin{minipage}{.33\linewidth}
\includegraphics[width=1\linewidth]{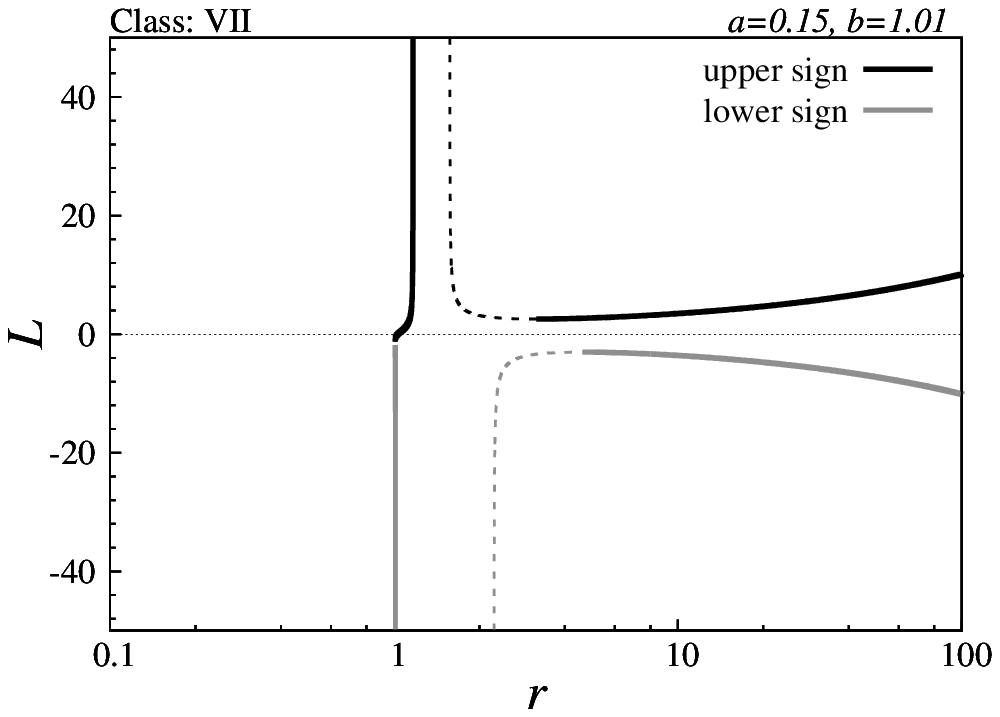}
\end{minipage}\hfill
\begin{minipage}{.33\linewidth}
\includegraphics[width=1\linewidth]{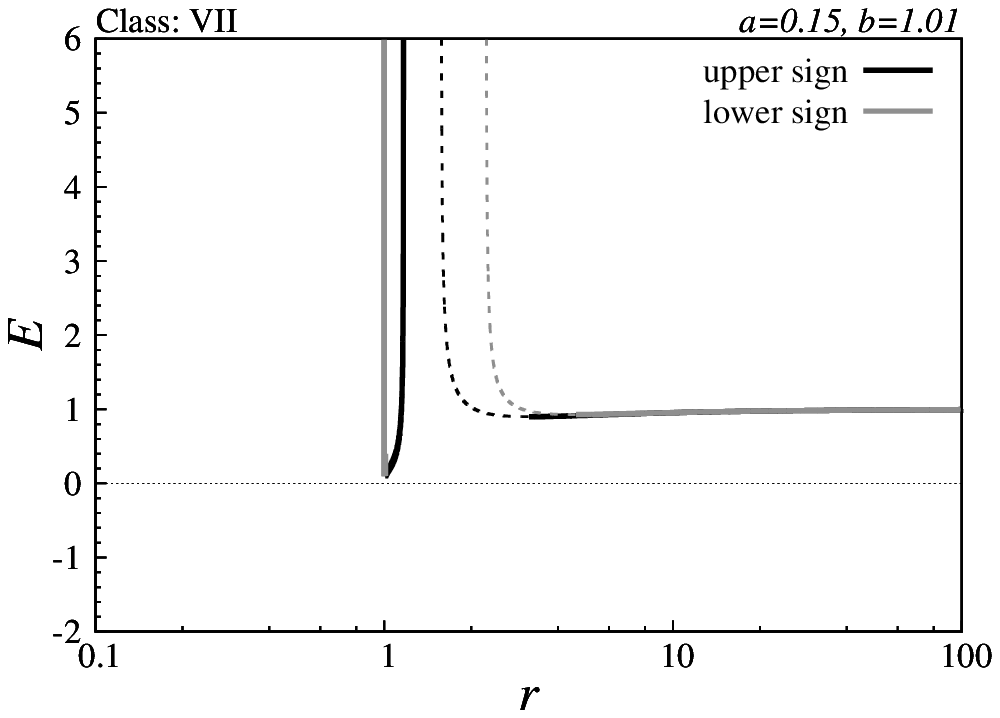}
\end{minipage}\hfill
\begin{minipage}{.33\linewidth}
\includegraphics[width=1\linewidth]{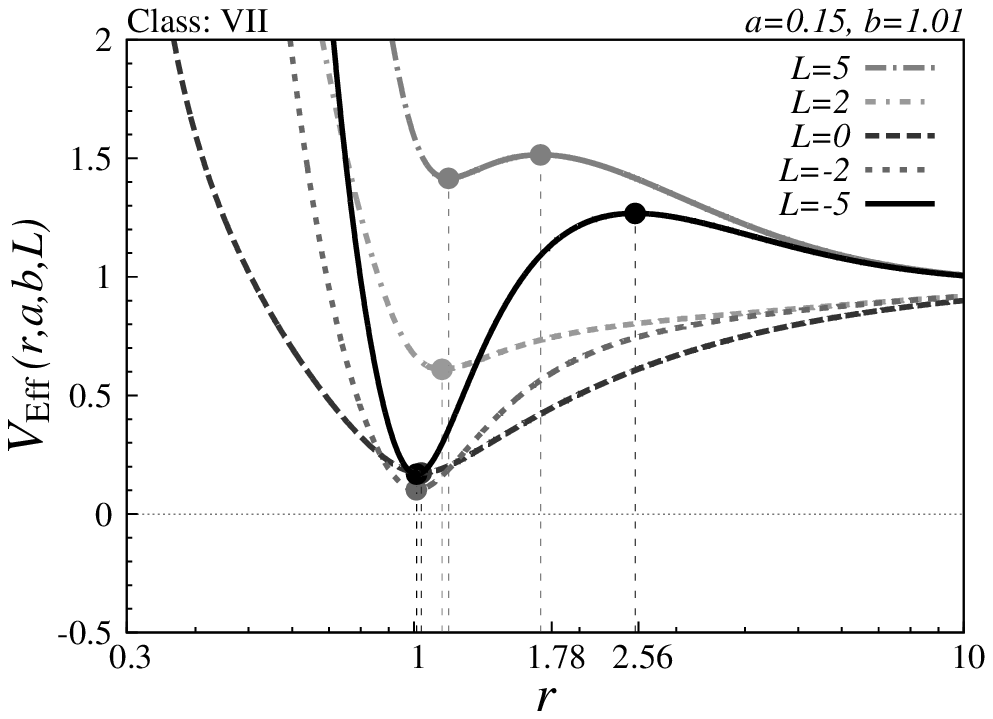}
\end{minipage}
\captionof{figure}{\label{ClVII} $L$, $E$ and effective potential for class VII.} 
\end{center}
\end{widetext}

\subsubsection{Class VIII}

Class of black hole spacetimes with negative braneworld tidal charge parameter $b$ that has only one horizon located at $r>0$, two unstable photon circular orbits and ergosphere. 
In the~parameter space $b-a$, boundary of the~region related to this class is given by the~line $b=-a^2\, ,$ and the~line $a=0$. 

For both the lower and upper family circular geodesics the~marginally stable orbit exist, determining the~inner edge of the~standard Keplerian disk. We obtained thus the standard situation typical for Kerr black holes, but no geodesic structure occurs at $r>0$ under the~event horizon. 

\begin{widetext}
\begin{center}
\begin{minipage}{.33\linewidth}
\includegraphics[width=1\linewidth]{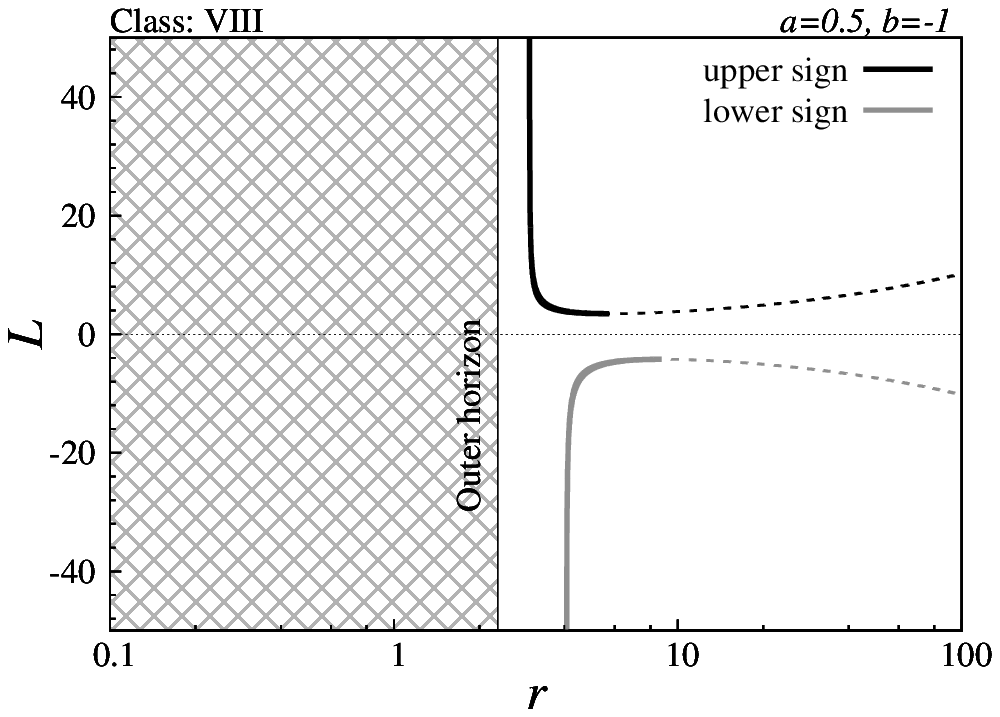}
\end{minipage}\hfill
\begin{minipage}{.33\linewidth}
\includegraphics[width=1\linewidth]{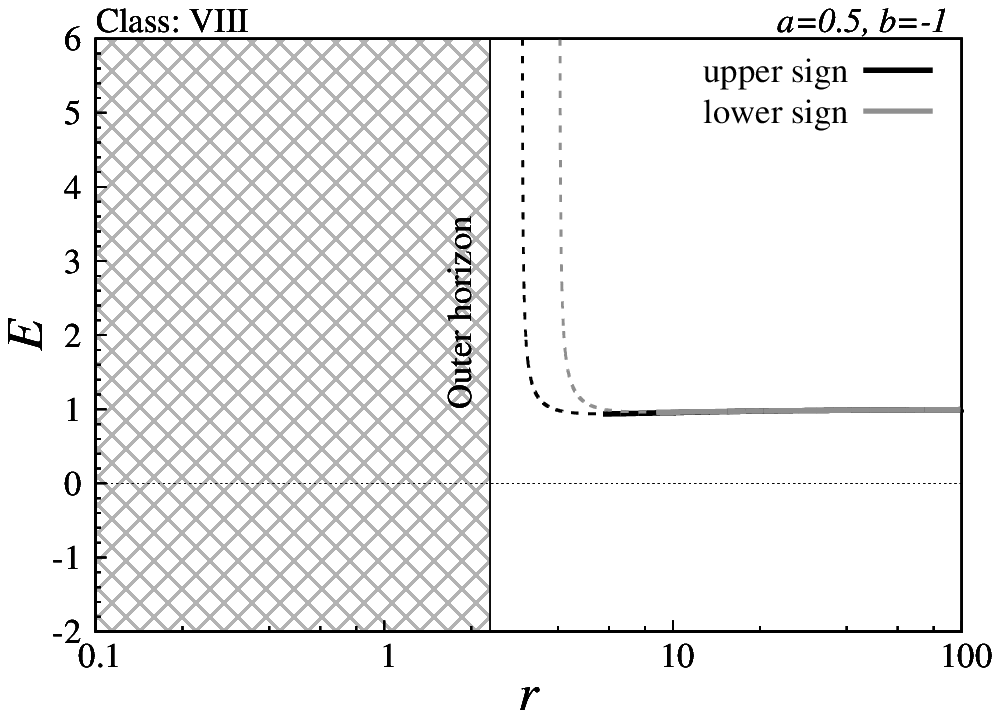}
\end{minipage}\hfill
\begin{minipage}{.33\linewidth}
\includegraphics[width=1\linewidth]{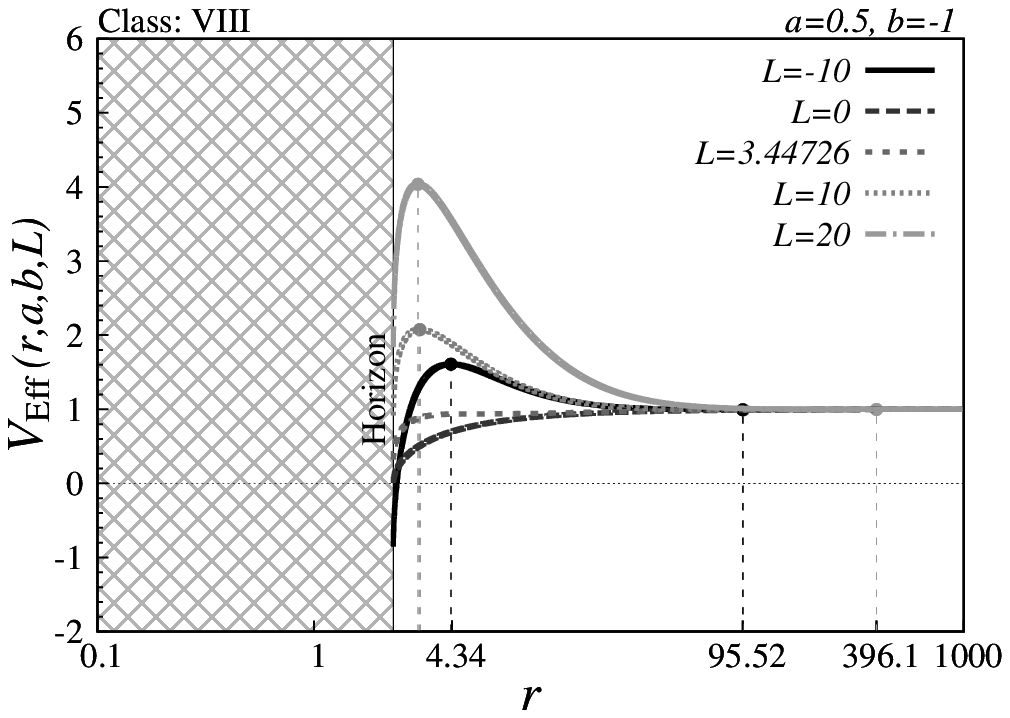}
\end{minipage}
\captionof{figure}{\label{ClVIII} $L$, $E$ and effective potential for class VIII.} 
\end{center}
\end{widetext}

\subsubsection{Class IX}

Class of black hole spacetimes with negative braneworld parameter $b$ having two horizons, three unstable photon circular orbits and ergosphere. 
Border of the~related region of the spacetime parameter space is given by the~line $b=1-a^2\, ,$ line $b=-a^2\, ,$ and the~line $b=0$. 

For both the lower and upper circular orbits the~marginally stable orbit exists, giving in standard way the~inner edge of the~Keplerian accretion. Unstable orbits exist under the~inner horizon. 

\begin{widetext}
\begin{center}
\begin{minipage}{.33\linewidth}
\includegraphics[width=1\linewidth]{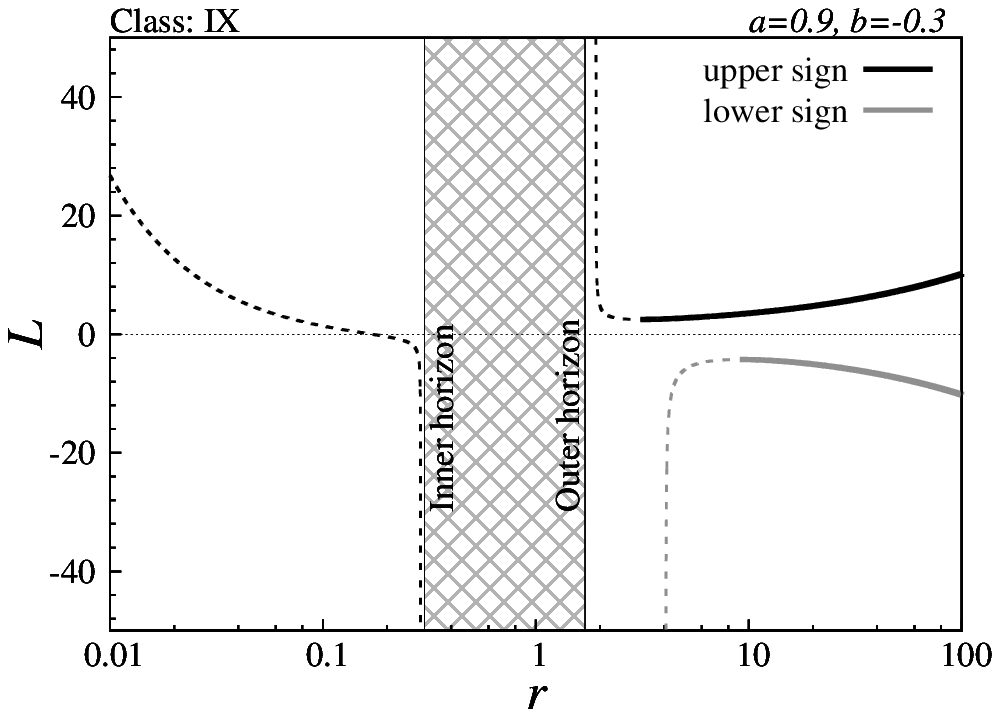}
\end{minipage}\hfill
\begin{minipage}{.33\linewidth}
\includegraphics[width=1\linewidth]{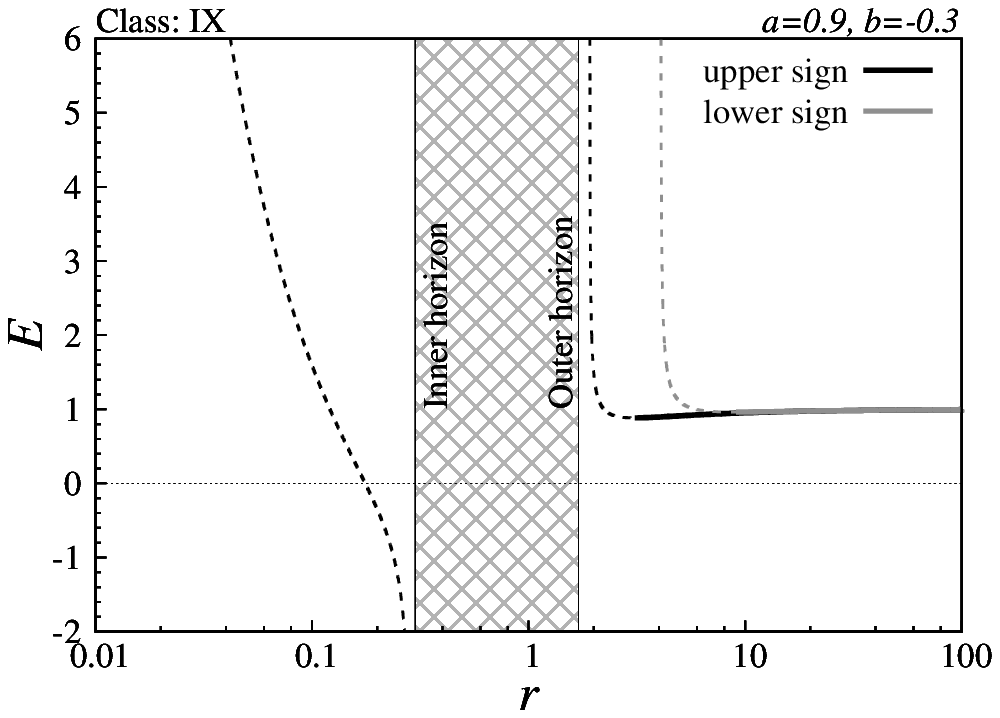}
\end{minipage}\hfill
\begin{minipage}{.33\linewidth}
\includegraphics[width=1\linewidth]{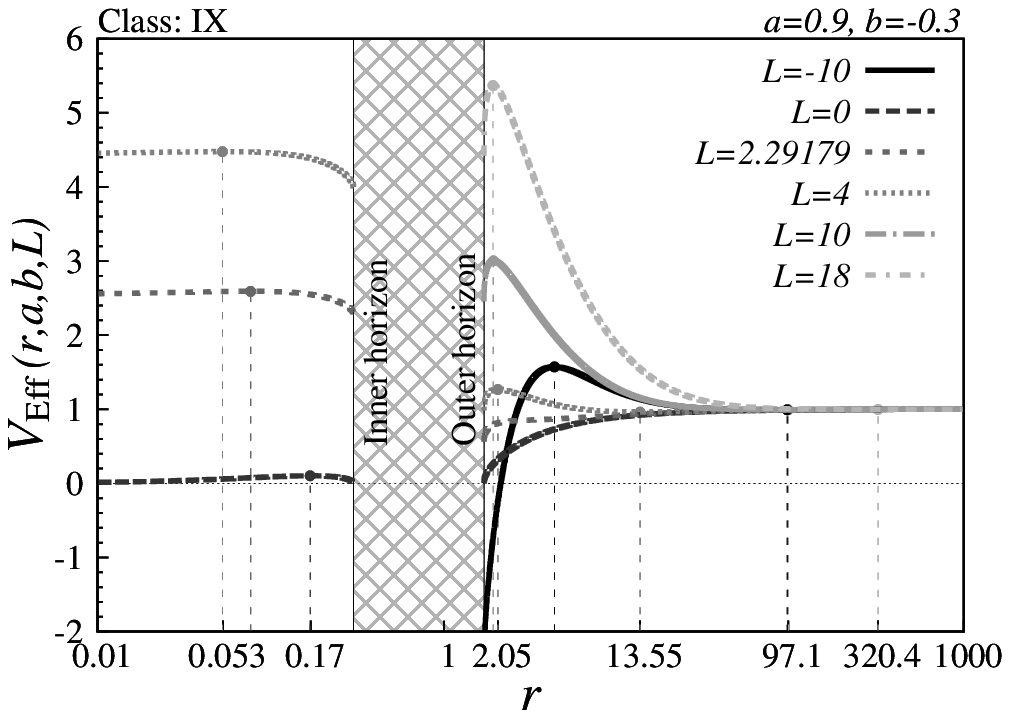}
\end{minipage}
\captionof{figure}{\label{ClIX} $L$, $E$ and effective potential for class IX.} 
\end{center}
\end{widetext}

\subsubsection{Class X}

Class of naked singularity spacetimes with negative brane parameter $b$ having one unstable photon circular orbit and ergosphere. Border of the~related region of the~parameter space is given by the~line $b=1-a^2\, ,$ and the~line $b=0$.

In these naked singularity spacetimes the~marginally stable orbit exists for both the lower and upper family of circular geodesics, representing thus in both cases the~inner edge of the~Keplerian accretion disks. Under the~marginally stable orbits only unstable orbits exist for both families. From the~point of view of the~geodesic structure, the naked singularity spacetimes of Class X resemble the~standard Kerr naked singularity spacetimes. 

\begin{widetext}
\begin{center}
\begin{minipage}{.33\linewidth}
\includegraphics[width=1\linewidth]{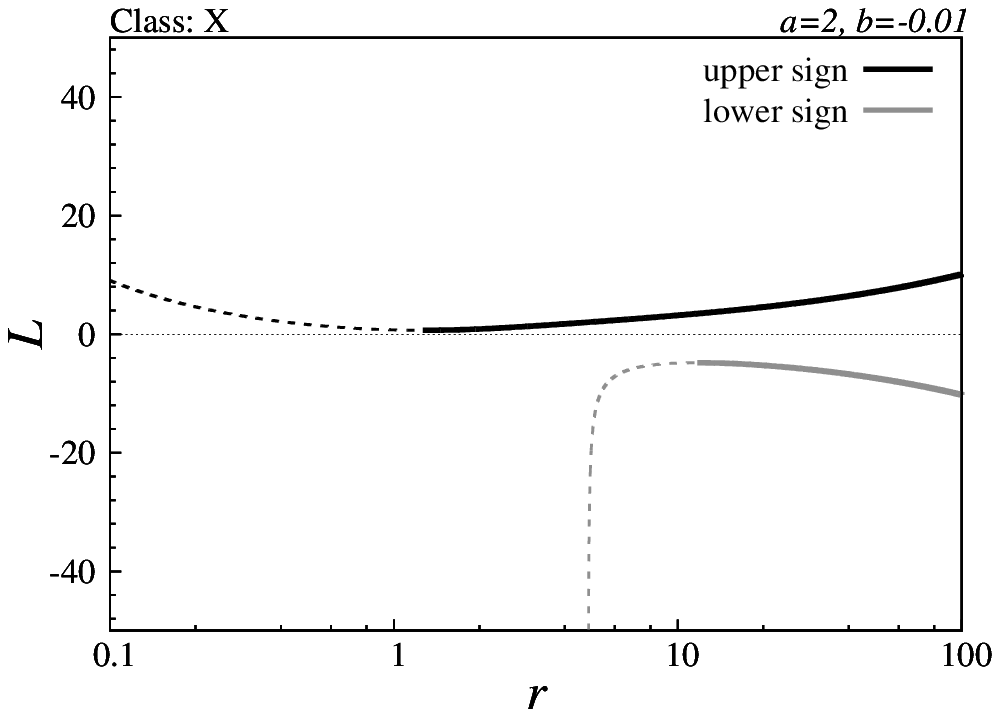}
\end{minipage}\hfill
\begin{minipage}{.33\linewidth}
\includegraphics[width=1\linewidth]{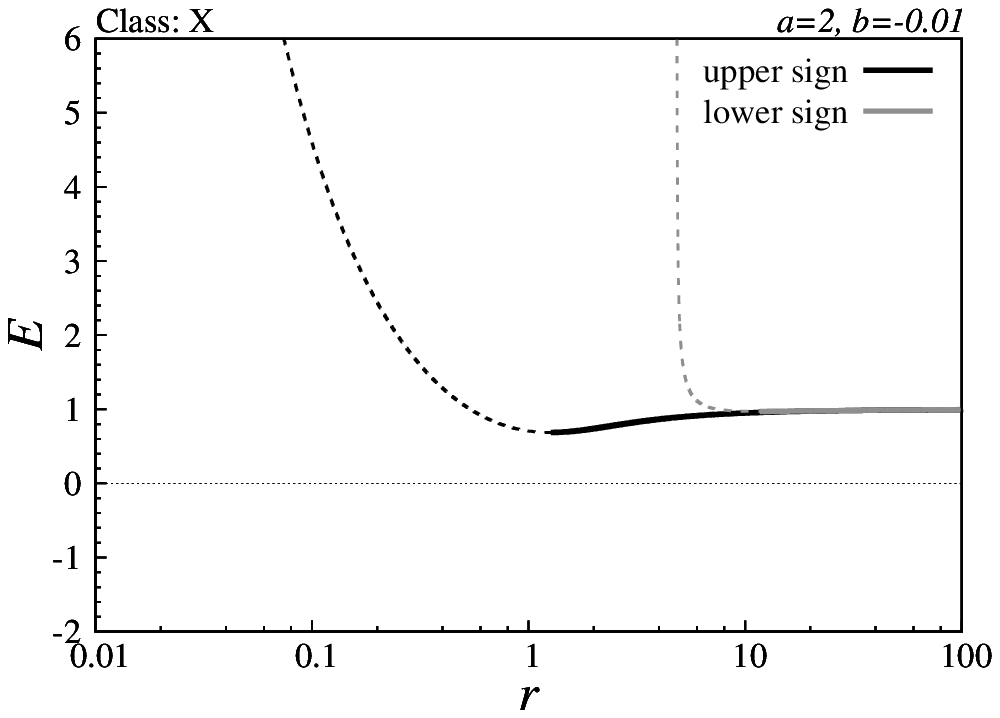}
\end{minipage}\hfill
\begin{minipage}{.33\linewidth}
\includegraphics[width=1\linewidth]{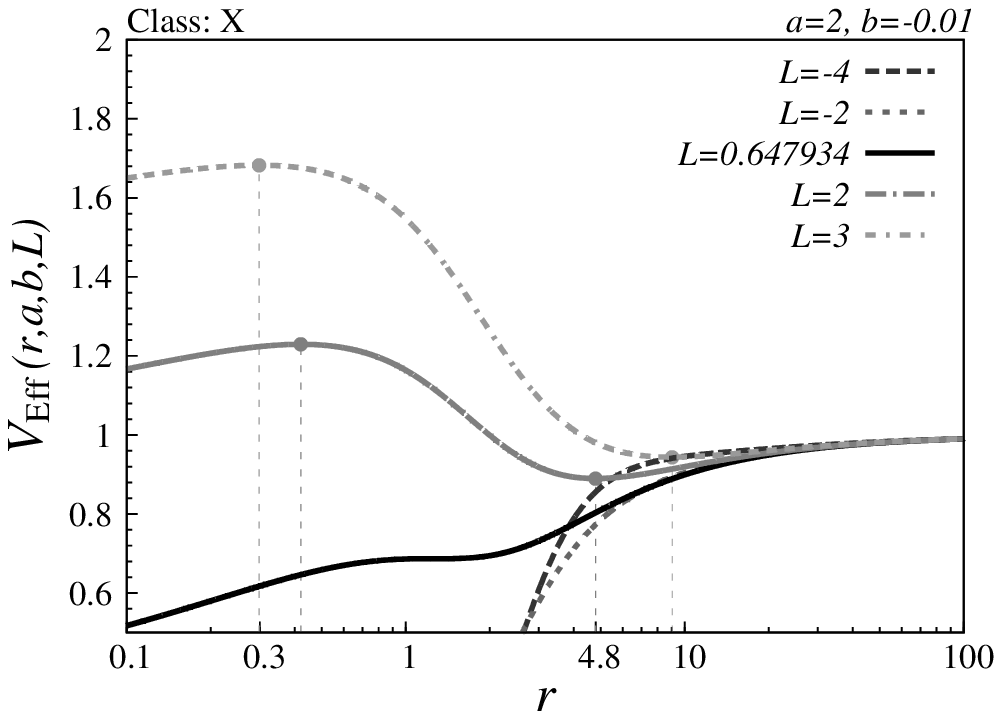}
\end{minipage}
\captionof{figure}{\label{ClX} $L$, $E$ and effective potential for class X.} 
\end{center}
\end{widetext}

\section{Efficiency of the Keplerian accretion}

Now we are able to determine the~energetic efficiency of the~Keplerian accretion. From the~astrophysical point of view, the~standard Keplerian accretion is relevant in the~regions enabling starting of accretion at large distance (infinity) and its finishing at the~first inner edge that can be approached by a~continuous accretion process. We determine efficiency of the~Keplerian accretion for all the~classes of the~braneworld Kerr--Newman spacetimes for the~standard Keplerian accretion. In some of these spacetimes there exists also an~inner region where the~Keplerian accretion could work due to the~decline of both energy and angular momentum with decreasing radius. However, these regions are not related to the~standard notion of Keplerian accretion and will not be considered here for calculations of the~accretion efficiency. Moreover, there could exist also complexities of the~Keplerian accretion process related to the~behaviour of the~angular velocity that are described in detail in \cite{Stu-Sche:2014:CLAQG:} -- we shall not discuss these subtleties in the~present paper. 

We concentrate our attention in determining the~efficiency for the~Keplerian accretion following the~upper family circular geodesics, when the~efficiency can be very high, being in some cases even unlimitedly high (formally). In the~case of the~upper family Keplerian accretion, the~efficiency is discontinuous when transition between the~naked singularity with sufficiently high dimensionless spin and the~related extreme black hole state is considered. The critical value of the spin, and the~related critical tidal charge reads 
\begin{equation}
a_{\mathrm{cr}}=\frac{1}{\sqrt{2}}\, ,\quad b_{\mathrm{cr}}=\frac{1}{2}\, .
\end{equation}
We have to stress that the~efficiency of the~Keplerian accretion in the~near-extreme naked singularity spacetimes exceeds significantly efficiency in the~extreme black hole spacetimes. On the other hand, the~efficiency of the Keplerian accretion in the~upper family regime is fully continuous in the case of transition of the naked singularity to extreme black hole spacetime with sufficiently low spin, $a<a_{\mathrm{cr}}$, and for all the braneworld KN spacetimes in the~case of the~Keplerian accretion in the~lower family accretion regime. Generally, the~efficiency of the~Keplerian accretion is substantially smaller in comparison to the~upper family regime in a~given Kerr--Newman spacetime.  

\begin{figure*}[tp]
\begin{center}
\includegraphics[width=0.8\linewidth]{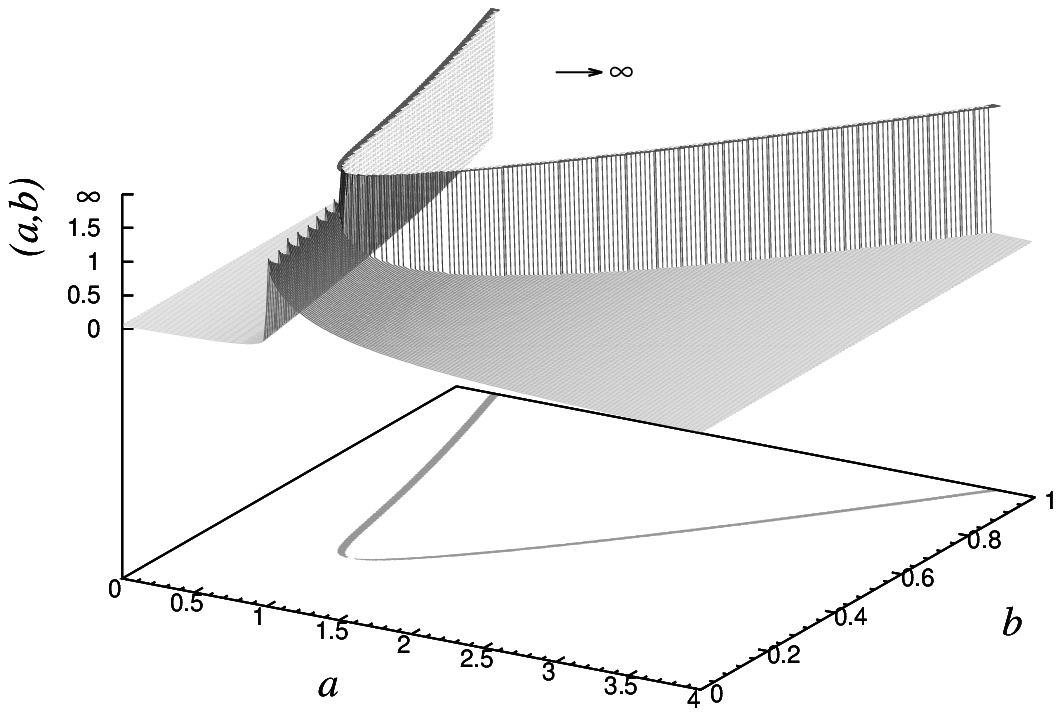}
\includegraphics[width=0.7\linewidth]{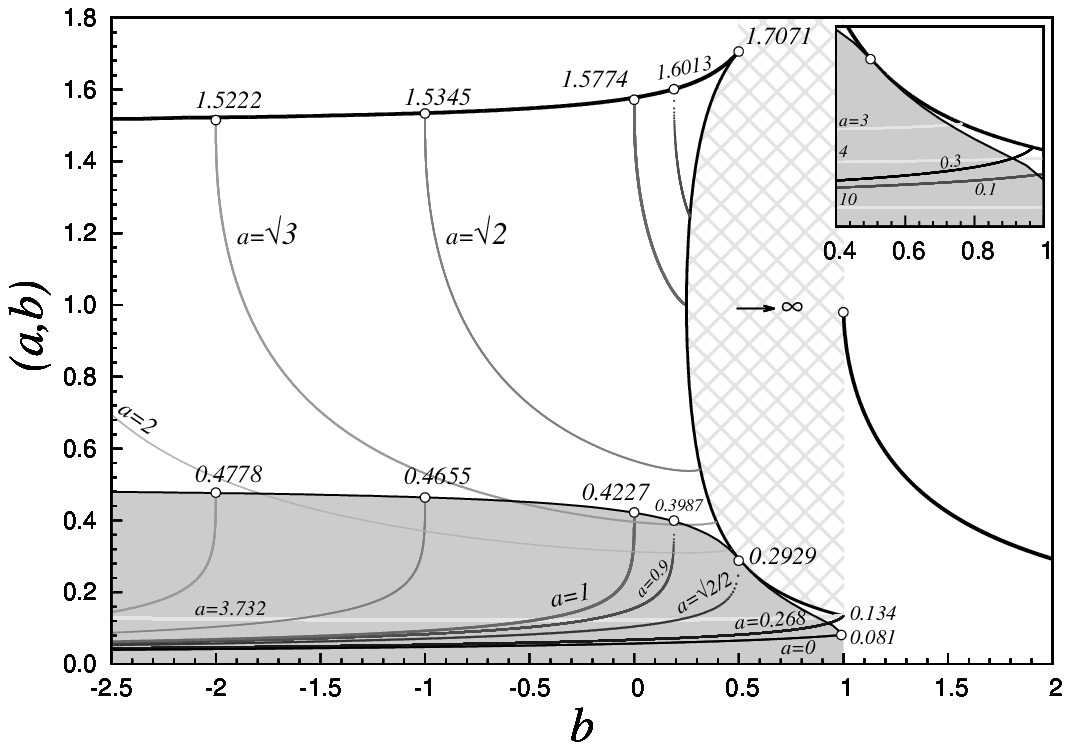}
\caption{\label{graph} Energetic efficiency $\eta(a,b)$ of the~Keplerian accretion following the~upper family circular geodesics is given in dependence on the~spacetime dimensionless tidal charge parameter $b$ and the~spin parameter $a$. 3D diagram is reflecting the~position of the~special class of the~mining unstable Kerr--Newman spacetimes of Class IIIa in the~plane of the spacetime parameters. Due to a complex character of the~efficiency function we give also the~characteristic $a=\mathrm{const}$ sections in the $\eta - b$ plane.} 
\end{center}
\end{figure*}

\begin{figure}[t]
\begin{center}
\includegraphics[width=\linewidth]{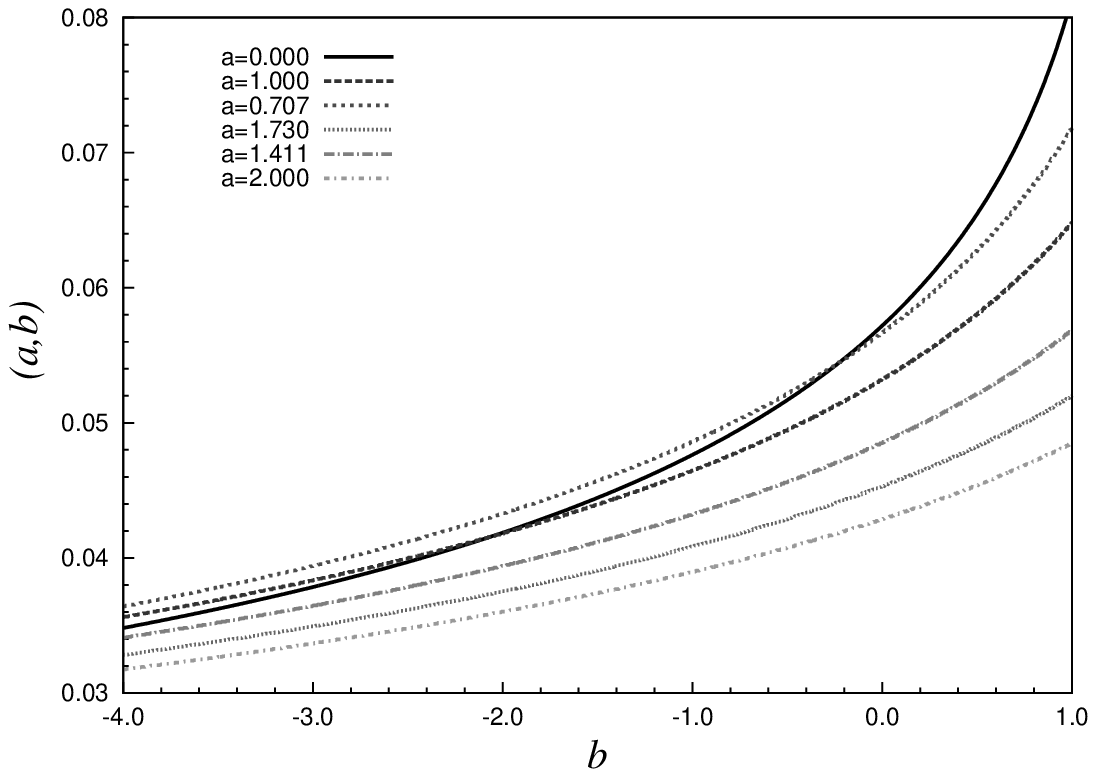}
\caption{\label{graph2} Energy efficiency of the Keplerian accretion following the~lower family circular geodesics is given in dependence on the spacetime dimensionless tidal charge parameter $b$ for characteristic values of the spin parameter $a=\mathrm{const}$.} 
\end{center}
\end{figure}

The~efficiency of the~accretion for the~geometrically thin Keplerian disks governed by the~circular geodesics is defined by the~relation  
\begin{equation}
\eta(a,b) = 1 - E(r_{\mathrm{edge}},a,b)\, ,
\end{equation} 
where $r_{\mathrm{edge}}$ denotes location of the~inner edge of the~standard Keplerian accretion disks. For the~Keplerian disks following the~lower family circular geodesics, the~inner edge of the~disk is always located at the~marginally stable geodesic giving thus always the~scenario of the Keplerian accretion in the~Kerr spacetimes. On the~other hand, for the~upper family Keplerian disks, the~situation is more complex, as follows from the~classification of the~braneworld Kerr--Newman spacetimes. There can occur three qualitatively different cases in dependence on the~combinations of dimensionless spacetime parameters $a$ and $b$.

In the~first family of classes of the~Kerr--Newman spacetimes, the $r_\mathrm{edge}$ is simply located at the~marginally stable geodesic, giving thus the~scenario of the~Keplerian accretion onto Kerr black holes -- this case includes all the~braneworld Kerr--Newman black hole spacetimes. 

In the~second family of the Kerr--Newman classes, the~inner edge of the~Keplerian disk is located at the~radius $r=b$, giving thus the~special case discovered at first for the~Reissner--Nordstr{\" o}m naked singularity spacetimes \cite{Stu-Hle:2002:ActaPhysSlov:,Pug-Que-Ruf:2011:PHYSR4:}. In all classes with $b>1$ (IV,V) the~efficiency of the~Keplerian accretion along the~upper family of circular geodesics is independent of the~spin parameter $a$ being defined by the~simple relation \footnote{Interestingly, the Keplerian efficiency relation for the~lower family of geodetics is much more complex and depends on the~spin parameter $a$.}
\begin{equation}
\eta(b)=1-\sqrt{1-\frac{1}{b}}\, .
\end{equation}
The efficiency goes slowly to 0\% for $b\rightarrow \infty$ -- see Figure (\ref{graph}). 

In the~third and most interesting family of the~Kerr--Newman spacetime classes, $r_\mathrm{edge}$ corresponds to the~radius of the~stable photon orbit approached by particles with specific energy $E \to -\infty$ and specific angular momentum $L \to -\infty$ -- notice that the~limiting photon circular geodesic is a corotating one as the~impact parameter $\lambda = L/E > 0$. In the~third case, the~Keplerian accretion efficiency approaches (theoretically) infinity. This effect occurs explicitly in the~Class IIIa Kerr--Newman spacetimes as clearly demonstrated in Figure (\ref{ClIIIa}). For this class the~tidal charge parameter $b \in (1/4,1)$ and the~dimensionless spin $a \in (2\sqrt{b}-\sqrt{b(4b-1)},2\sqrt{b}+\sqrt{b(4b-1)})$. 

The~Keplerian accretion efficiency is given for the~upper family of circular geodesics in Figure \ref{graph}, and for the~lower family circular orbits in Figure \ref{graph2}. Because of its complexity, we represent the~case of the~upper family accretion regime by a 3D figure with addition of the~figure representing the relevant sections $a=\mathrm{const}$. In the~case of the~lower family accretion regime, the representative $a=\mathrm{const}$ sections are sufficient to clearly demonstrate the~character of the~efficiency of the~Keplerian accretion. 

For the~Keplerian accretion along the~lower family circular geodesics the~situation is quite simple and the~efficiency is always continuously matched between the~naked singularity and the~extreme black hole states. The efficiency of the~lower family regime accretion for fixed dimensionless spin $a$ of the~braneworld Kerr--Newman spacetimes always decreases with decreasing tidal charge parameter $b$. Moreover, for fixed tidal charge $b$, the~efficiency decreases with increasing spin $a$. 

In order to understand the~upper family Keplerian accretion regime and its efficiency, the~dependences $\eta\left(b,a=\mathrm{const}\right)$ are most instructive. They are governed by two crucial families of curves. First, the~efficiency of the~Keplerian accretion in the~extreme braneworld KN black hole spacetimes, and the~related near-extreme braneworld KN naked singularity spacetimes is given by the~relation 
\begin{equation}
\displaystyle \eta_{\mathrm{jump}}(b)= 1\pm \frac{1}{\sqrt{4-\frac{1}{1-b}}}\, ,
\end{equation}
where the~$+$ sign corresponds to the~efficiency in the~near-extreme naked singularity spacetimes, while the~$-$ sign corresponds to the~related extreme black holes. Naturally, this formula is relevant in the~interval of the~tidal charge $b\in\left(-\infty,0.5\right)$, i.e., up to the~critical value of the~tidal charge. Second crucial curve is given by the~efficiency of the~accretion in the~limiting spacetimes governed by the~boundary of Class IIIa spacetimes, $\eta_{\mathrm{mining}}(b,a_{\mathrm{mining}{\pm}}(b))$, where $b\in(1/4,1)$, and $a_{\mathrm{mining}}(b)=2\sqrt{b} \pm \sqrt{b(4b-1)}$. 

The~results can be summarized in the~following way. For whole the~braneworld spacetimes with the~negative tidal charge parameter $b<0$, and for those with positive charge parameter $0 \leq b \leq 1/2$, the~large jump of the~efficiency in transition between the+naked singularity to the~related extreme black hole state occurs. Such a jump was observed for the~first time in the case of transition between the~Kerr naked singularity and the~extreme Kerr black hole ($b=0$) where $\eta \sim 1.57$ goes down to $\eta \sim 0.43$ \cite{Stu:1980:BAC:}. For the~braneworld KN extreme black holes (related near-extreme naked singularities), the~efficiency slightly increases (decreases) with negatively valued tidal charge increasing in its magnitude, so the~efficiency jump slightly decreases from its maximal Kerr value. On the~other hand, for $b \in (0,1/2)$, the~efficiency for the~extreme black holes (near-extreme naked singularities) decreases (increases), and the~jump fastly increases -- for $b=1/2$, the~efficiency jumps from $\eta \sim 1.707$ down to $\eta \sim 0.293$. 

For the~naked singularities with the~tidal charge in the~interval $1/4 < b < 1$, and the~dimensionless spin in the~interval $a \in (2\sqrt{b}-\sqrt{b(4b-1)}, 2\sqrt{b}+\sqrt{b(4b-1)})$ the formally defined efficiency of the~Keplerian accretion is unlimited. At the~boundary of this region, the~efficiency is given by the~limit governed by the~regular Keplerian accretion finished at the~marginally stable orbit. 

For the~naked singularities with $b<1/2$, the efficiency of the~Keplerian accretion, starting at the~near-extreme state and keeping spin $a=\mathrm{const}$, decreases with increasing tidal charge down to the~curve $\eta_{\mathrm{mining}}(b,a_{\mathrm{mining}{\pm}}(b))$. Further increase of $b$ causes entrance to the~region of unlimited efficiency. The curve $\eta(b,a=1)$ starts at $b=0$ and finishes at $b=1/4$, giving the~related efficiency $\eta=1$. For spin in the~interval $1>a>1/\sqrt{2}$, the curves $\eta(b,a=1)$ start at the~extreme state and finish at the~state with $0<b<1/2$ and efficiency $\eta > 1$. For spin approaching $a = 1/\sqrt{2}$, the curve $\eta(b,a=\mathrm{const})$ degenerates at the~point with $b=1/2$, and efficiency approaching $\eta \sim 1.707$. For higher values of spin, $a \in (1,2+\sqrt{3})$, the~efficiency curves $\eta(b,a=\mathrm{const})$ decrease to the curve $\eta_{\mathrm{mining}}(b,a_{\mathrm{mining}{\pm}}(b))$ with $b$ increasing in the~interval $b \in (1/4,1)$ and the~efficiency decreasing down to the~limiting value of $\eta(b=1,a=2+\sqrt{3}) \sim 0.134$. 

For tidal charge $b \in (1/2,1)$, the~efficiency of the~Keplerian accretion at the~transition between the~extreme black hole and the~related near-extreme naked singularity is continuously matched. The~efficiency of $\eta \sim 0.134$ is reached for the~Kerr--Newman spacetime with $b=1$ and $a=2-\sqrt{3}$. For values of the~spin in the~interval of $a \in (2-\sqrt{3},1/\sqrt{2})$, the~transition of the~function $\eta(b,a=\mathrm{const})$ between the~black hole and naked singularity states, obtained due to increasing tidal charge $b$, is still continuous, and the~curve $\eta_{\mathrm{mining}}(b,a_{\mathrm{mining}{\pm}}(b))$ is reached at values $\eta < 0.293$. 

With increasing spin $a$, the~efficiency of the~Keplerian accretion decreases. It is interesting that for naked singularities having spin $a$ higher than $\sim 4.97$ and appropriately valued tidal charge $b$, the~efficiency reaches values smaller than those corresponding to the~Schwarzschild black holes $\left(\eta \sim 0.057\right)$. 

Note that the~results of the~Keplerian accretion analysis for the~braneworld Kerr--Newman spacetimes can be directly applied also for the~Keplerian accretion in the~standard Kerr--Newman spacetimes, if we make transformation $b \to Q^2$ where $Q^2$ represent the~electric charge parameter of the~Kerr--Newman background. 

\section{Conclusions} 
In the~present paper the~circular geodesics of the~braneworld Kerr--Newman black hole and naked singularity spacetimes have been studied and classification of these spacetimes according to character of the~circular geodesic structure has been presented. The circular geodesics have been separated into two families -- the~lower family containing only the~counterrotating circular geodesics, and the~upper family with corotating geodesics at large distance, but possible transformation to counterrotating geodesics in vicinity of the~naked singularity. It has been demonstrated that fourteen different classes of the~Kerr--Newman spacetimes can exist, mainly due to the~properties of the~upper family of circular geodesics. Implications of the~geodesic structure to the~Keplerian accretion have been given, and efficiency of the~Keplerian accretion have been determined. The accretion efficiency is continuously matched between the~naked singularity and extreme black hole spacetimes for the~Keplerian accretion along the~lower family circular geodesics. On the~other hand, there is a strong discontinuity occuring in the~transition between the~naked singularities and the~extreme black holes for the~Keplerian accretion along the~upper family circular geodesic, if the~dimensionless spin of the~Kerr--Newman spacetime is sufficiently high ($a>1/\sqrt{2}$) -- the~energy efficiency of the~Keplerian accretion is then substantially higher for the~naked singularity spacetimes. The~accretion efficiency could then go up to the~value of $\eta \sim 1.707$ for Kerr--Newman near-extreme naked singularity spacetimes with $b \sim 1/2$ and $a \sim 1/\sqrt{2}$. 

For the~Keplerian accretion along the~lower family circular geodesics, the~inner edge of the~disk has to be always located at the~marginally stable circular geodesic corresponding to an~inflexion point of the~effective potential of the~motion in accord with the~scenario of Keplerian accretion onto Kerr black holes and naked singularities. 

It has been shown that the Keplerian accretion along the~upper family geodesics can give three different scenarios. It can finish at the~inner edge located at the~marginally stable circular geodesic -- this is the~standard accretion scenario present in the~black hole spacetimes. However, two other scenarios could occur in the~naked singularity spacetimes. The~inner edge of the~Keplerian accretion could occur at $r=b$ that is the~special limit on existence of the~circular geodesics. For $b>1$ the~efficiency of the~upper family Keplerian accretion is independent of the~naked singularity spin. The~most interesting is the~third scenario, related to the~Kerr--Newman naked singularity spacetimes of Class IIIa having an~infinitely deep gravitational potential of the~upper family Keplerian accretion. Then the~inner edge of the~Keplerian accretion could occur even at the~stable photon circular geodesic, and the~accretion efficiency could be formally unlimited, making such naked singularity spacetimes unstable relative to accretion \uvozovky{mining}. 

The~mining instability of the~Kerr--Newman naked singularity spacetimes (or related superspinars) is a~classical instability that could imply interesting consequences that we plan to study in future. Nevertheless, it is clear that the~mining instability have to be restricted at least by validity of the~test particle approximation used in the~present paper for the~Keplerian accretion. 

The~other classical instability, related to the~conversion of Kerr naked singularities to extreme black holes due to the~Keplerian accretion \cite{Stu:Hle:Tru:2011:}, has to be treated in future work, but this instability has necessarily much more complex character in the~Kerr--Newman naked singularity spacetimes in comparison to the~relatively simple Kerr case, due to the~complexities related to the~mining instability and the~influence of the~tidal charge. 

Interesting phenomena could be expected in the~mining unstable Kerr--Newman spacetimes (Class IIIa) even if it will represent astrophysically more acceptable concept of Kerr--Newman superspinar, with the~inner boundary of the~Kerr--Newman spacetime located at least at the~outer boundary of the~causality violation region \cite{Gim-Hor:2009:PhysLetB:,Stu-Sche:2010:CLAQG:,Stu-Sche:2012:CLAQG:}. We can expect observations of extremely high energy coming out of collisions in vicinity of such an superspinar, enabled by non-existence of the~event horizon and fast rotation of the~superspinar, or inversely, strong magnification of incoming radiation \cite{Stu-Sche:2013:CLAQG:,Stu-Sche-Abd:2014:PHYSR4:}. We could expect a hot doughnut-shaped configuration of accreting matter surrounding the~superspinar, as discussed in \cite{Stu-Sche:2014:CLAQG:}. 

\section*{Acknowledgements}

ZS and MB has been supported by the~Albert Einstein Centre for Gravitation and Astrophysics financed by the~Czech Science Agency Grant No. 14-37086G and by the~Silesian University at Opava grant SGS/14/2016. MB thanks Petr Slan\'{y} for several discussions.  


\end{document}